\documentclass[3p, sort&compress]{elsarticle}
\usepackage[colorlinks=true]{hyperref}
\usepackage{amsmath}
\usepackage{amssymb}
\usepackage{enumitem}
\usepackage{comment}

\newcommand{\lt}{\left}
\newcommand{\rt}{\right}
\newcommand{\pa}{\partial}
\newcommand{\ve}{\varepsilon}
\newcommand{\br}{\mathbf{r}}
\newcommand{\bx}{\mathbf{x}}
\newcommand{\bh}{\mathbf{h}}
\newcommand{\bhr}{\tilde{\mathbf{h}}}
\newcommand{\bE}{\mathbf{E}}
\newcommand{\hr}{\tilde{h}}
\newcommand{\bq}{\mathbf{q}}
\newcommand{\bk}{\mathbf{k}}
\newcommand{\be}{\mathbf{e}}
\newcommand{\ur}{\tilde{u}}
\newcommand{\chir}{\tilde{\chi}}
\newcommand{\lambdar}{\tilde{\lambda}}
\newcommand{\mur}{\tilde{\mu}}
\newcommand{\Gammar}{\tilde{\Gamma}}

\newcommand{\etab}{\bar{\eta}}
\newcommand{\betab}{\bar{\beta}}
\newcommand{\Op}{\mathcal{O}}
\newcommand{\bR}{\mathbf{R}}
\newcommand{\bt}{\mathbf{t}}

\begin{document}
\begin{frontmatter}

\title{Scale without conformal invariance in membrane theory}

\author[mainaddress]{Achille Mauri\corref{correspondingauthor}}
\cortext[correspondingauthor]{Corresponding author}
\ead{a.mauri@science.ru.nl}
\author[mainaddress]{Mikhail I. Katsnelson}
\ead{m.katsnelson@science.ru.nl}
\address[mainaddress]{Radboud University, Institute for Molecules and Materials,
Heyendaalseweg 135, 6525 AJ Nijmegen, The Netherlands}

\begin{abstract}
We investigate the relation between dilatation and conformal symmetries in the
statistical mechanics of flexible crystalline membranes.
We analyze, in particular, a well-known model which describes the fluctuations of a
continuum elastic medium embedded in a higher-dimensional space.
In this theory, the renormalization group flow connects a non-interacting ultraviolet 
fixed point, where the theory is controlled by linear elasticity, to an interacting 
infrared fixed point.
By studying the structure of correlation functions and of the energy-momentum tensor, we 
show that, in the infrared, the theory is only scale-invariant: the dilatation symmetry 
is not enhanced to full conformal invariance.
The model is shown to present a non-vanishing virial current which, despite being 
non-conserved, maintains a scaling dimension exactly equal to $D - 1$, even in presence of 
interactions.
We attribute the absence of anomalous dimensions to the symmetries of the model under 
translations and rotations in the embedding space, which are realized as shifts of phonon 
fields, and which protect the renormalization of several non-invariant operators.
We also note that closure of a symmetry algebra with both shift symmetries and conformal 
invariance would require, in the hypothesis that phonons transform as primary fields, 
the presence of new shift symmetries which are not expected to hold on physical grounds.
We then consider an alternative model, involving only scalar fields, which describes 
effective phonon-mediated interactions between local Gaussian curvatures.
The model is described in the ultraviolet by two copies of the biharmonic theory, which 
is conformal, but flows in the infrared to a fixed point which we argue to be only 
dilatation-invariant.
\end{abstract}
\end{frontmatter}

\section{Introduction}

Asymptotic scale invariance plays a crucial role in quantum field theory, from 
statistical mechanics to models of fundamental interactions.
In several cases, the asymptotically-emergent scaling symmetry is enlarged to full 
conformal invariance, which opens the way to powerful techniques such as bootstrap 
equations~\cite{poland_rmp_2019, rychkov_epfl_2017} or, in two dimensions, methods based 
on the infinite Virasoro algebra~\cite{di-francesco_cft}.
These approaches give access to high-precision non-perturbative calculations and, in some 
cases, even to exact solutions.
Understanding the conditions under which conformal symmetry arises is thus of great 
importance, and has motivated extensive investigations~\cite{nakayama_pr_2015}.

Particularly general results were established for two- and four-dimensional field 
theories assuming unitarity, or, in Euclidean space, the corresponding property of 
reflection positivity~\cite{zamolodchikov_jetp_1986, polchinski_npb_1988, 
cardy_conformal, rychkov_epfl_2017}.
In the two-dimensional case, Zamolodchikov and Polchinski proved that unitary 
scale-invariant field theories are always conformal under two mild assumptions: the 
existence of a well-defined energy-momentum tensor and the discreteness of the spectrum of 
operator dimensions~\cite{zamolodchikov_jetp_1986, polchinski_npb_1988}.
In four-dimensional space, a similar result is expected to hold~\cite{nakayama_pr_2015}, 
as indicated by perturbative proofs to all orders~\cite{jack_npb_1990, luty_jhep_2013, 
fortin_jhep_2013} and corroborated by non-perturbative evidences~\cite{nakayama_pr_2015, 
luty_jhep_2013, dymarsky_jhep_2015, dymarsky_jpa_2015, dymarsky_jhep_2016}.
Some analogue derivations were argued to be applicable to unitary theories in any  
even dimension $D = 2n$~\cite{nakayama_prd_2020}.

These arguments, however, cannot be extended straightforwardly to arbitrary dimensions 
(possibly odd or non-integer) or to models lacking unitarity or reflection positivity.
In addition, several derivations break down when the energy-momentum tensor and its 
two-point function are not well defined, which can happen in sigma models relevant for 
string theories~\cite{polchinski_npb_1988, hull_npb_1986, arutyunov_npb_2016}.
Models with scale but without conformal invariance, in fact, exist and have been 
explicitly identified~\cite{nakayama_pr_2015, hull_npb_1986, polchinski_npb_1988, 
riva_plb_2005, ho_jhep_2008, el-showk_npb_2011, nakayama_prd_2013, arutyunov_npb_2016, 
nakayama_prd_2017, oz_epjc_2018}, or indirectly conjectured based on holographic 
analyses~\cite{nakayama_prd_2017b, nakayama_prd_2017c, li_epjc_2019}.
Although unphysical in the context of fundamental interactions, models defined in 
general dimension $D$ and without unitarity or reflection positivity are recurrent in 
statistical mechanics.
Analyses of the relation between scale and conformal invariance in more general classes of
theories are thus crucial for several physical applications (see 
Refs.~\cite{schafer_jpa_1976, brown_aop_1980, paulos_npb_2016, polchinski_npb_1988, 
delamotte_pre_2016, de-polsi_jsp_2019, rychkov_epfl_2017, meneses_jhep_2019} for some of 
the results and methods).

If we try to consider, roughly speaking, how likely it is for a scale-invariant model to 
exhibit conformal symmetry, we can often run into a dilemma.
On the one hand, dilatation invariance is not a sufficient condition for the extended 
conformal invariance and, therefore, a generic scale-invariant theory can be expected to 
lack conformal symmetry.
On the other hand, there exist arguments suggesting that, for \emph{interacting} 
field theories, scale invariance should imply conformal invariance 
generically~\cite{el-showk_npb_2011, rychkov_epfl_2017, meneses_jhep_2019, 
nakayama_prd_2017}.
A formulation of this reasoning starts from the structure of the energy-momentum tensor 
$T_{\alpha \beta}$ and its trace $T_{\alpha}^{\alpha}$.
In local and scale-invariant theories, dilatation symmetry implies that 
$T_{\alpha}^{\alpha} = \pa_{\alpha}V^{\alpha}$, where $V^{\alpha}$ is a local field, the 
'virial current'.  
Conformal invariance arises instead whenever $V^{\alpha} = j^{\alpha} + \pa_{\beta} 
L^{\alpha \beta}$ where $j^{\alpha}$ is conserved ($\pa_{\alpha}j^{\alpha} = 0$) and 
$L^{\alpha \beta}$ is a tensor field~\cite{polchinski_npb_1988}.
Although the requirements for conformal symmetry are stronger and not automatically 
satisfied a priori, possible candidates for the virial current are constrained, 
because $V^{\alpha}$ must have a scaling dimension exactly equal to $\{V^{\alpha}\} = 
D-1$ in order to match the dimensions of the energy-momentum tensor $\{T_{\alpha \beta}\} 
= D$~\cite{el-showk_npb_2011, rychkov_epfl_2017, meneses_jhep_2019, 
nakayama_prd_2017}\footnote{
More precisely, the change of a symmetric energy-momentum tensor under infinitesimal 
dilatations reads $i[S, T_{\alpha \beta}] = x^{\rho}\pa_{\rho} T_{\alpha \beta} + D 
T_{\alpha \beta} + \pa^{\sigma} \pa^{\rho} Y_{\alpha \sigma \beta \rho}$ where $Y_{\alpha 
\sigma \beta \rho} = -Y_{\sigma \alpha \beta \rho} = Y_{\beta \rho \alpha 
\sigma}$~\cite{polchinski_npb_1988, dymarsky_jhep_2015}.
The first two terms, $x^{\rho}\pa_{\rho} T_{\alpha \beta} + D T_{\alpha \beta}$ describe 
the scaling law of an eigenoperator with dimension $D$, while the third, inhomogeneous 
term is generated by renormalization.
In scale-invariant theories, where $T_{\alpha}^{\alpha} = \pa_{\alpha}V^{\alpha}$, the 
scaling law for the virial current must read, therefore, $i[S, V^{\alpha}] = 
x^{\rho}\pa_{\rho} V^{\alpha} + (D-1) V^{\alpha} + l^{\alpha} + \pa^{\rho} Y^{\sigma 
\alpha}_{~~~~\sigma \rho}$, with $\pa_{\alpha} l^{\alpha} = 0$ (see also 
Ref.~\cite{dymarsky_jhep_2015}).
The inhomogeneous terms $l^{\alpha} + \pa^{\rho} Y^{\sigma \alpha}_{~~~~\sigma \rho}$ 
have precisely the form of the combination of a conserved current and a total divergence, 
which are irrelevant to the discussion of scale and conformal invariance.
This justifies considering $V^{\alpha}$ as a scaling operator of dimension $D-1$.
It is usually possible to choose an improved energy-momentum tensor in such way that 
$Y_{\alpha \sigma \beta \rho} = 0$ and the canonical scaling laws holds (see however
Ref.~\cite{dymarsky_jhep_2015} for a more detailed discussion).
}.
All vector currents are usually expected to acquire anomalous dimensions in presence of 
interactions, unless they are conserved.
Consistent candidates for $V^{\alpha}$ in a generic theory can thus be expected to be 
conserved currents, which implies conformal invariance~\cite{el-showk_npb_2011, 
rychkov_epfl_2017, meneses_jhep_2019, nakayama_prd_2017}.

A basis from which we can formulate similar arguments is provided by the results of 
Refs.~\cite{delamotte_pre_2016, de-polsi_jsp_2019, schafer_jpa_1976} which, instead of 
analyzing the energy-momentum tensor, used non-perturbative renormalization group 
techniques.
Refs.~\cite{delamotte_pre_2016, de-polsi_jsp_2019} showed that, for critical scalar and 
O$(N)$ models, scale implies conformal invariance if no vector eigenoperator with scaling 
dimension $-1$ exists\footnote{
Redundant operators, whose insertion is equivalent to an infinitesimal change of 
variables, are allowed: even if their dimension is exactly equal to $-1$, they do not 
destroy conformal invariance but, rather, modify the transformation of fields under the 
elements of the conformal group~\cite{de-polsi_jsp_2019}.
This is consistent with the fact that the scaling dimension of redundant operators can 
actually be chosen at will, by suitable design of the specific renormalization group 
transformation~\cite{dietz_jhep_2013}.
The dimensions of non-redundant operators are, instead, intrinsic quantities, invariant 
under redefinitions of the RG.
}.
This vector quantity plays a role analogue to the space integral of the virial current.
Ref.~\cite{schafer_jpa_1976}, instead, used a generalization of Wilson's renormalization 
group to argue that, for a general fixed point theory, two- and three-point functions are 
consistent with the constraints imposed by conformal invariance provided that (i) there 
exists no vector eigenoperator with dimension $-1$, (ii) interactions are sufficiently 
local, (iii) the real parts of operator dimensions are bounded from below, and (iv) some 
surface effects are negligible\footnote{
In Ref.~\cite{schafer_jpa_1976} the vector operator dimension is reported as $+1$, 
because length units are used instead of inverse-length units.
Similarly, the lower bound in the real part of operator dimensions is expressed there as 
an upper bound.}.
With the same logic used for the virial current, the existence of vectors with dimension  
tuned to $-1$ appears to be unlikely in generic interacting field theories, suggesting 
that scale implies conformal invariance in a broad class of models.
The argument can actually be improved further by a reasoning based on continuity: even if 
a vector happens by coincidence to have scaling dimension $-1$ in $D$-dimensional space, 
conformal invariance can still be inferred by continuation from neighbouring dimensions $D 
+ \delta D$.
A scenario without conformal invariance thus requires the existence of a vector presenting 
dilatation eigenvalue exactly equal to $-1$ throughout a continuous interval of dimensions 
in the neighbourhood of $D$, which seems even more unlikely~\cite{delamotte_pre_2016}.

Although genericity arguments hint at a general explanation of conformal invariance, 
they cannot set a fully definite answer.
The same reasonings, for example, could be read from a different point of view: it might 
be the case that scale without conformal invariance is recurrent in several field 
theories, and vectors with dimension $-1$ or currents with dimension $D-1$ are not 
unlikely as a first expectation suggests.
With this reversed perspective, the arguments could be regarded as proofs that these 
vectors are common even in interacting theories.
Moreover, in some classes of theories there exist mechanisms ensuring the 
non-renormalization of some vector fields: for example, this can happen in presence of 
BRST invariance~\cite{nakayama_prd_2017}.
In these models, a non-trivial virial current without anomalous dimensions arises 
naturally, without the need of a fine tuning.

For given field theories, it is usually not necessary to argue from genericity.
For example, in the Ising and in the O$(N)$ model, the presence of conformal invariance 
can be proved by setting bounds on the dilatation spectrum~\cite{delamotte_pre_2016, 
de-polsi_jsp_2019, meneses_jhep_2019}.
Also, powerful tools are available to analyze perturbative theories 
explicitly~\cite{parisi_plb_1972, polchinski_npb_1988, nakayama_pr_2015, jack_npb_1990, 
luty_jhep_2013, brown_aop_1980, paulos_npb_2016}.

It is interesting, however, to explore the genericity arguments in more depth.
In this direction, Ref.~\cite{nakayama_prd_2017} identified and analyzed an interacting 
scale invariant model which is not conformal: the theory of SU$(N)$ gauge fields coupled 
to massless fermions at the Banks-Zaks fixed point.
As it was shown, the model is conformal when regarded as a gauge theory, but presents a 
nontrivial virial current $V^{\alpha}$ when gauge fixed.
The scaling dimension of $V^{\alpha}$ was shown to be exactly equal to $\{V^{\alpha}\} = 
D-1$, to all orders in perturbation theory, which was traced to BRST invariance of the 
theory.
Other scale-invariant but nonconformal theories were identified in the context of 
turbulence~\cite{oz_epjc_2018}, sigma models~\cite{hull_npb_1986, polchinski_npb_1988, 
arutyunov_npb_2016, ho_jhep_2008}, topologically-twisted 
theories~\cite{nakayama_prd_2017b, nakayama_prd_2017c}, Wess-Zumino models with 
scale-invariant renormalization-group trajectories~\cite{nakayama_prd_2013}, 
or were recognized by holographic analysis~\cite{nakayama_pr_2015, li_epjc_2019, 
nakayama_prd_2017b, nakayama_prd_2017c}.
Finally, we note that Ref.~\cite{pajer_jhep_2019} recognized the presence of 
scale-invariance without conformal symmetry in an analysis at classical level of 
symmetric superfluids characterized by shift-invariant actions.

In this paper, we analyze the relation between scale and conformal symmetry in the 
statistical mechanics of fluctuating crystalline membranes, a theory which is 
relevant for biological layers and for free-standing samples of atomically-thin 
two-dimensional materials such as graphene~\cite{nelson_statistical, bowick_pr_2001, 
katsnelson_graphene, nelson_jpf_1987, david_epl_1988, aronovitz_prl_1988, 
aronovitz_jpf_1989, guitter_jpf_1989, kownacki_pre_2009, gazit_pre_2009, bowick_prb_2017,
le-doussal_aop_2018, saykin_aop_2020, mauri_npb_2020, coquand_pre_2020}.
The theory of two-dimensional solids in three dimensions, or more generally, of 
$D$-dimensional crystalline membranes embedded in $d$-dimensional space has been 
studied extensively.
For temperatures lower than a transition temperature $T_{c}$, these membranes present a 
'flat phase' where the embedding-space O$(d)$ symmetry is spontaneously broken and the 
state of the system is macroscopically planar~\cite{nelson_jpf_1987,david_epl_1988, 
aronovitz_prl_1988, aronovitz_jpf_1989, guitter_jpf_1989, kownacki_pre_2009}.
As it was crucially recognized, in this broken-symmetry phase, the large-distance 
behavior of fundamental degrees of freedom, the phonon fluctuations, is controlled by an 
interacting scale-invariant theory~\cite{aronovitz_prl_1988, guitter_jpf_1989, 
coquand_pre_2020}.

Here, we show that the asymptotic infrared behavior of the flat phase presents only 
scale invariance, and not the full conformal symmetry.
In particular, we verify that the theory generates a virial current $V^{\alpha}$ 
which cannot be reduced to a combination of a conserved current and a total derivative.
Despite being non-conserved, the $V^{\alpha}$ is shown to have scaling eigenvalue 
$\{V^{\alpha}\} = D - 1$ to all orders in perturbation theory, without anomalous 
dimensions.
This absence of renormalization is traced to the fact that $V^{\alpha}$ is not invariant 
under the spontaneously-broken embedding-space translations and rotations, which are 
realized as shifts of the phonon fields.
A similar result is found for the 'GCI model' in dimension $D = 4 - \ve$, a distinct 
field theory which is expected, however, to become equivalent to the conventional model at 
the physical dimensionality $D = 2$~\cite{mauri_npb_2020}.
Even for this alternative theory, the infrared behavior is shown to be scale invariant but 
nonconformal.
A consequence of our analysis is that methods of conformal field theory (CFT), such as 
the conformal bootstrap, cannot be straightforwardly applied to the flat phase of 
crystalline membranes.

The membrane models analyzed in this work can be viewed as a generalization of the 
linearized theory of elasticity, a model which was identified by Riva and Cardy as an 
example of scale-invariant but non-conformal field theory~\cite{riva_plb_2005, 
el-showk_npb_2011, nakayama_aop_2016}.
The main difference is that the Riva-Cardy model describes an elastic medium confined in 
$D$ dimensions, while solid membranes are allowed to flucutate in an embedding space with 
higher dimension $d > D$.
While linearized elasticity is a Gaussian, non-interacting theory, transverse 
fluctuations in the additional $d - D$ space dimensions make membrane theory an 
anharmonic model, which realizes scale invariance via an interacting RG fixed point.
The presence of interactions makes membrane theory an interesting platform to test the 
genericity arguments on scale and conformal invariance.

\section{Scaling and renormalization in crystalline membranes}
\label{s:model-RG}

This section introduces one of the two membrane models analyzed in this work and 
describes its renormalization within the $\ve$-expansion.
In addition to methods based on dimensional regularization, which were often used in the 
literature~\cite{aronovitz_prl_1988, guitter_jpf_1989, mauri_npb_2020, coquand_pre_2020}, 
in Sec.~\ref{s:RG-cutoff} we discuss an approach based on bare 
renormalization group equations, expressing the response of the theory to variations of 
an ultraviolet cutoff.

\subsection{Model}
\label{s:model}

Analyses in this work focus on a well-known theory for the flat phase of crystalline 
membranes~\cite{nelson_statistical, bowick_pr_2001, katsnelson_graphene, 
nelson_jpf_1987, aronovitz_prl_1988, aronovitz_jpf_1989, guitter_jpf_1989, 
le-doussal_aop_2018, saykin_aop_2020, coquand_pre_2020}.
This theory can be viewed as the most general membrane model which, with the scaling 
properties characteristic of the flat phase, is renormalizable by power counting in the 
$\ve$-expansion.

For a derivation, it is convenient to start from a more accurate model and to obtain the 
effective theory by dropping all irrelevant interactions~\cite{aronovitz_prl_1988}.
We thus start from a general description of a continuum $D$-dimensional crystalline 
membrane embedded in a higher-dimensional space.
Introducing a coordinate $\bx \in \mathbb{R}^{D}$ to label mass elements of the elastic 
medium, fundamental degrees of freedom in the theory are coordinates $\br(\bx) \in 
\mathbb{R}^{d}$ specifying the location of all elements, identified by $\bx$, in the 
$d$-dimensional embedding space.
At leading order in powers of deformations and their gradients, the configuration energy 
can be written as~\cite{aronovitz_prl_1988, david_epl_1988, aronovitz_jpf_1989, 
guitter_jpf_1989}
\begin{equation} \label{H}
H = \frac{1}{2} \int {\rm d}^{D}x [\kappa (\pa^{2}\br)^{2} + \lambda (\bar{U}_{\alpha 
\alpha})^{2}+ 2 \mu \bar{U}_{\alpha \beta}\bar{U}_{\alpha \beta}]~.
\end{equation}
Here
\begin{equation}
\bar{U}_{\alpha \beta} = \frac{1}{2}\lt(\pa_{\alpha}\br \cdot \pa_{\beta} \br - 
\delta_{\alpha 
\beta}\rt)
\end{equation}
is the strain tensor, a measure of the local deviation of the metric $g_{\alpha \beta} 
= \pa_{\alpha}\br \cdot \pa_{\beta}\br$ from the Euclidean metric $\delta_{\alpha 
\beta}$.
At zero temperature, 'ground states' of the model are given by $\br = 
x_{\alpha}\be_{\alpha}$, where $\be_{\alpha}$ are any set of $D$ mutually orthogonal unit 
vectors in $d$-dimensional space.
These states spontaneously break the embedding-space translational and rotational 
symmetries~\cite{coquand_prb_2019}.
For $T > 0$, statistical properties such as correlation functions are calculated by 
functional integration with the Gibbs weigth ${\rm e}^{-H/T}$ through a partition function
\begin{equation}
Z[\mathbf{J}] = \int [{\rm d}\br] {\rm e}^{-H/T + \int {\rm d}^{D}x \mathbf{J}\cdot \br}~.
\end{equation}
We only focus on the flat phase\footnote{A crucial prediction of the theory is that the 
flat phase is stable in a finite window of temperatures $0 < T < T_{c}$ even in dimension 
$D = 2$.
This is possible because the system violates the assumptions of the Mermin-Wagner 
theorem~\cite{nelson_jpf_1987, david_epl_1988, aronovitz_jpf_1989}.} and, in particular, 
on the limit of small temperatures $T \to 0$.
In this broken-symmetry phase, as in the zero-temperature case, the system is 
macroscopically planar and extended: the thermal average of coordinates is $\langle 
\br(\bx)\rangle =  \xi x_{\alpha} \be_{\alpha}$.
A stretching factor $\xi < 1$ in general appears due to a 'hidden area' effect: due to 
transverse fluctuations in the out-of-plane direction, the projected in-plane area is 
smaller than its curvilinear size.
Equivalently, $\xi$ can be viewed as a renormalization of the order parameter for the 
flat phase: thermal fluctuations reduce the degree of order in the 
layer~\cite{david_epl_1988, guitter_jpf_1989, aronovitz_jpf_1989,
burmistrov_prb_2016, burmistrov_aop_2018, saykin_aop_2020}.

To study fluctuations, it is convenient to expand the coordinates $\br(\bx)$ as $\br = \{
[\xi x_{\alpha} + T u_{\alpha}/(\xi \kappa)]\be_{\alpha} + \sqrt{T/\kappa}~\bh\}$, where 
$u_{\alpha}$ and $\bh$ are, respectively, in-plane and out-of-plane phonon 
displacement fields\footnote{
This definition of displacement fields differs by a rescaling from the conventions of 
elasticity theory.
In particular, the units of measurements of the fields are ${\rm dim}(u_{\alpha}) = 
(D-3)$ and ${\rm dim}(\bh) = (D-4)/2$ in terms of inverse-length units.}.
Defining
\begin{equation}
U_{\alpha \beta} = \frac{1}{2} \lt(\pa_{\alpha}u_{\beta} + \pa_{\beta}u_{\alpha} + 
\pa_{\alpha}\bh \cdot \pa_{\beta}\bh + \frac{T}{\xi^{2} \kappa}\pa_{\alpha}u_{\gamma} 
\pa_{\beta} 
u_{\gamma}\rt)~,
\end{equation}
the reduced Hamiltonian $H' = H/T$ takes the form, up to an overall energy shift,
\begin{equation} \label{Hb}
H' = \frac{H}{T} = \frac{1}{2} \int {\rm d}^{D} x \lt[(\pa^{2} \bh)^{2} + 
\frac{T}{\xi^{2}\kappa} (\pa^{2} u_{\gamma})^{2} + \lambda_{0} U_{\alpha \beta} U_{\alpha 
\beta} +  2 \mu_{0} U_{\alpha \beta} U_{\alpha \beta} + 2\sigma_{0} U_{\alpha 
\alpha}\rt]~,
\end{equation}
where $\lambda_{0} = T \lambda/\kappa^{2}$, $\mu_{0} = T \mu/\kappa^{2}$, $\sigma_{0} = 
(D \lambda + 2 \mu)(\xi^{2}-1)/(2\kappa)$.

An analysis of tree-level propagators and canonical dimensions of interactions shows that 
the theory has $D = 4$ as upper critical dimension~\cite{aronovitz_prl_1988, 
guitter_jpf_1989}.
This implies, in analogy with theories of critical behavior, that the perturbative 
expansion is well defined (free of infrared divergences) only for $D\geq 4$ or in $D = 4 
- \ve$ for $\ve$ infinitesimal, that is within the framework of an 
$\ve$-expansion~\cite{zinn-justin_qft, parisi_sft}.
For any finite $\ve$, instead, the perturbation theory in $D < 4$ develops infrared 
problems~\cite{parisi_sft} at an order $\approx 2/\ve$.
At the same time, power counting shows that near $D = 4$, the terms in Eq.~\eqref{Hb} of 
the type $(\pa^{2} u_{\gamma})^{2}$, $(\pa u)^{4}$, and $(\pa u)^{2} (\pa \bh)^{2}$ are 
irrelevant in the sense of canonical dimensional analysis\footnote{The power-counting 
dimensions of displacement fields are determined by the small-momentum behavior of their 
propagator: if the Gaussian two point function scales with momentum $k^{-\sigma}$, the 
dimension is $(D-\sigma)/2$.
The power-counting dimensions are respectively $\{u_{\alpha}\} = (D-2)/2$, $ \{\bh\} = 
(D-4)/2$.
Note that $\{u_{\alpha}\}$ is different from the naive units of measurements, 
because $\lambda_{0}$ and $\mu_{0}$ are themselves dimensionful.
}.

Similarly to critical phenomena~\cite{zinn-justin_qft}, universal exponents controlling 
the leading scaling behavior can be captured within the $\ve$-expansion by an effective 
renormalizable field theory where all canonically-irrelevant interactions are dropped.
For the flat phase of crystalline membranes, the corresponding effective theory can be 
shown~\cite{aronovitz_prl_1988, guitter_jpf_1989} to be\footnote{
The coefficients $\lambda_{0}$, $\mu_{0}$, and $\sigma_{0}$ in the effective theory are 
different, in general, from the corresponding parameters in Eq.~\eqref{Hb} because they 
get renormalized by neglected irrelevant interactions~\cite{zinn-justin_qft}.
We use the same symbols, however, to lighten the notation.
The neglected nonlinearities are suppressed by one power of $T$ in the limit $T \to 0$, 
so the quantitative difference between the two sets of constants is small in the 
low-temperature region.}
\begin{equation} \label{H1}
{\cal H} = \frac{1}{2} \int {\rm d}^{D}x [(\pa^{2}\bh)^{2} + \lambda_{0} (u_{\alpha 
\alpha})^{2} + 2 \mu_{0} u_{\alpha \beta} u_{\alpha \beta} + 2 \sigma_{0} u_{\alpha 
\alpha}]~,
\end{equation}
where $u_{\alpha \beta} = (\pa_{\alpha} u_{\beta} + \pa_{\beta} u_{\alpha} + 
\pa_{\alpha}\bh\cdot \pa_{\beta} \bh)/2$ is a linearized version of the strain tensor.
Eq.~\eqref{H1} differs in form from Eq.~\eqref{Hb} by the neglection of $(\pa^{2} 
u_{\gamma})^{2}$ and by the replacement $U_{\alpha \beta} \to u_{\alpha \beta}$ in all 
terms of the Hamiltonian\footnote{
The replacement $U_{\alpha \beta} \to u_{\alpha \beta}$ is performed not only in 
interaction terms, but also in the term linear in strain $\sigma_{0} u_{\alpha\alpha}$.
At first it could seem that this replacement neglects a contribution to the Gaussian part 
of the energy functional proportional to $\sigma_{0}\pa_{\alpha}u_{\gamma} \pa_{\alpha} 
u_{\gamma}$ which is not irrelevant, but formally marginal by power counting.
Actually, substituting  $U_{\alpha \alpha} \to u_{\alpha \alpha}$ in \emph{all} terms is 
necessary, in order to preserve the invariance ot the theory under the symmetry 
transformations~\eqref{deformed-rotations}, which represent linearized versions of the 
underlying invariance under O$(d)$ transformations in the embedding space.
In practical calculations, it is possible to start with $\sigma_{0} = 0$ at tree level 
and to calculate $\sigma_{0}$ order by order in perturbation theory as a counterterm to 
quadratic ultraviolet divergences via the 'renormalization condition' $\langle 
\pa_{\alpha}u_{\alpha}\rangle = 0$.
Since ${\cal H}$ is invariant under~\eqref{deformed-rotations}, the counterterm generated 
is proportional to $u_{\alpha\alpha}$ and not to $U_{\alpha \alpha}$.
}.

The effective theory~\eqref{H1} is one of the two main models investigated in this work 
and is assumed as a starting point in all further discussions.
In the past, it has been the subject of extensive investigations (see for 
example~\cite{nelson_statistical, bowick_pr_2001, katsnelson_graphene, nelson_jpf_1987, 
david_epl_1988, aronovitz_prl_1988, aronovitz_jpf_1989, guitter_jpf_1989, 
kownacki_pre_2009, gazit_pre_2009, bowick_prb_2017, le-doussal_aop_2018, saykin_aop_2020, 
mauri_npb_2020, coquand_pre_2020}).

To renormalize the theory, Ward identities associated with rotational invariance in the 
embedding $d$-dimensional space play a crucial role~\cite{guitter_jpf_1989, 
aronovitz_prl_1988}.
In the transition from Eq.~\eqref{Hb} to Eq.~\eqref{H1}, the procedure of neglecting 
non-renormalizable interactions has broken the original O$(d)$ symmetry of the model 
explicitly.
However, the underlying rotational symmetry is still presents in a deformed, linearized 
form: the effective theory~\eqref{H1} is, in fact, invariant under the continuous 
transformations defined, for any set of $D$ vectors $\mathbf{A}_{\alpha}$ in 
$(d-D)$-dimensional space, by
\begin{equation} \label{deformed-rotations}
\begin{split}
\bh &\to \bh + \mathbf{A}_{\alpha}x_{\alpha}~,\\
u_{\alpha} & \to u_{\alpha} - (\mathbf{A}_{\alpha} \cdot \bh) - \frac{1}{2} 
(\mathbf{A}_{\alpha}\cdot \mathbf{A}_{\beta}) x_{\beta}~.
\end{split}
\end{equation}
These transformations can be recognized as deformed versions of the broken 
embedding-space rotations.
In addition, the model is manifestly invariant under rigid translations in the embedding 
space ($\bh \to \bh + \mathbf{B}$, $u_{\alpha} \to u_{\alpha} + B_{\alpha}$), in-plane 
rotations ($\bh \to \bh$, $u_{\alpha} \to u_{\alpha} + \omega_{\alpha \beta}x_{\beta}$ 
with $\omega_{\alpha \beta} = - \omega_{\beta \alpha}$) and O$(d-D)$ rotations of the 
field $\bh$.

\subsection{Feynman rules, doubly-soft Goldstone modes and cancellation of tadpole 
diagrams}
\label{s:perturbative}

\begin{figure}[h]
\centering 
\includegraphics{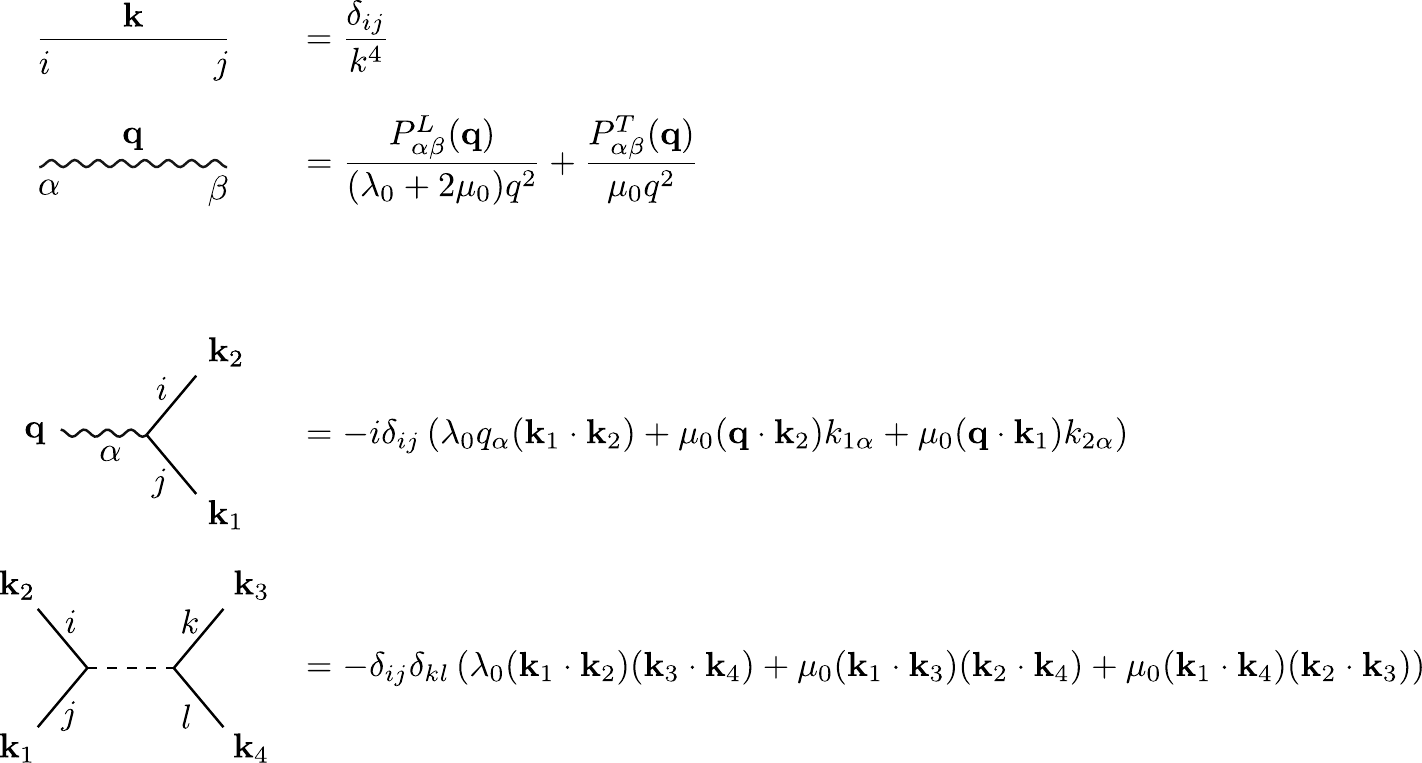}
\caption{\label{feynman-rules} Feynman rules for the elasticity theory of crystalline 
membranes.
Solid and wiggly lines represent propagators of the $\bh$ and of the $u_{\alpha}$ field 
respectively.
The model has a three-leg vertex, corresponding to interactions of the form $(\pa u)(\pa 
\bh)^{2}$ and a four-leg vertex corresponding to $(\pa \bh)^{4}$.
}
\end{figure}

The effective theory defined in Eq.~\eqref{H1} has a perturbative expansion described by 
the Feynman rules illustrated in Fig.~\ref{feynman-rules}.

\noindent
Let us define more precisely the role of the 'tension' term $\sigma_{0}u_{\alpha \alpha}$ 
in the Hamiltonian.
When free boundary conditions are used (as it is implicit throughout all steps of our 
analysis), the value of $\sigma_{0}$ is only relevant to the discussion of zero modes and 
completely decouples from the behavior of finite-wavelength 
fluctuations~\cite{nelson_statistical, guitter_jpf_1989} and, therefore, from the Feynman 
rules.
Any term linear in the trace of the strain tensor $u_{\alpha \alpha}$, in fact, can be 
removed from the Hamiltonian by a change of variables of the form $u_{\alpha} \to 
u_{\alpha} + \ell x_{\alpha}$.
Physically, the presence of a finite $\sigma_{0}$ describes the 'hidden area' effect, the 
reduction in projected area due to transverse thermal fluctuations~\cite{david_epl_1988, 
guitter_jpf_1989, burmistrov_prb_2016, aronovitz_jpf_1989,  burmistrov_prb_2016, 
saykin_aop_2020}.

Consistently with the derivations of Sec.~\ref{s:model} we can choose to set 
$\sigma_{0}$ as a function of other parameters of the theory in such way that the phonon 
displacement field has $\langle \pa_{\alpha}u_{\alpha}\rangle = 0$.
This choice of $\sigma_{0}$ separates phonon fluctuations from zero-modes associated with 
the macroscopic compression of the projected area, ensuring that $u_{\alpha}$ is only a 
superposition of fluctuations\footnote{
For discussions of the equation of state $\xi = \xi(T)$ and more generally for 
stress-strain relations in presence of applied external tension, see 
Refs.~\cite{aronovitz_jpf_1989, guitter_jpf_1989, burmistrov_prb_2016, 
burmistrov_aop_2018, saykin_aop_2020}.
}.

A convenient feature of this convention is that it the tadpole diagrams

\begin{figure}[h]
\centering
\includegraphics{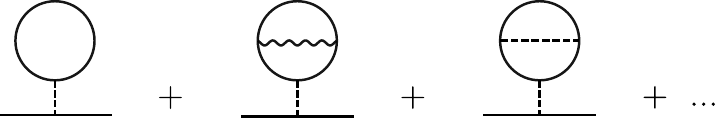}
\end{figure}

\noindent
are precisely cancelled by equal and opposite terms coming from the contribution 
proportional to $\sigma_{0}$ in the bare Hamiltonian, as it can be shown by explicit 
calculation\footnote{The value of $\sigma_{0}$ which ensures 
$\langle\pa_{\alpha}u_{\alpha} \rangle = 0$ can be calculated by arguments analogue to the 
theories in Refs.~\cite{burmistrov_prb_2016, burmistrov_aop_2018, saykin_aop_2020} and 
reads, in the notation adopted here
\begin{equation*}
\sigma_{0} = -\frac{1}{2} (\lambda_{0} + 2 \mu_{0}/D) \langle \pa_{\alpha} 
\bh(\bx) \cdot \pa_{\alpha} \bh(\bx)\rangle~.
\end{equation*}
}.
Tadpoles connected via wiggly lines, instead, are not one-particle irreducible (1PI) and 
should be excluded.
They contribute to the calculation of the minimum of the free-energy, which here is set 
to $\pa_{\alpha} u_{\alpha}= 0$ by definition.

The cancellation of tadpoles reflects the fact that the transverse displacements $\bh$ are 
massless Goldstone fields associated with the spontaneously-broken O$(d)$ invariance in 
the embedding space~\cite{coquand_prb_2019}.
The breakdown of translation and rotation symmetries implies in particular that $\bh$ is 
doubly-soft: not only its inverse propagator vanishes for $\bk \to 0$, but also, it must 
vanish faster than $k^{2}$.
The tree-level inverse propagator and all diagrams of non-tadpole type for the 
$\bh$-field self-energy, in fact, scale as $k^{4}$ up to powers of $k^{-\ve}$ and resum to 
$k^{4 - \eta_{*}}$ for $k \to 0$~\cite{nelson_statistical, nelson_jpf_1987, 
aronovitz_prl_1988, le-doussal_aop_2018}, preserving the softness of the infrared 
behavior.
Only tadpole diagrams could give a 'mass', by generating contributions proportional to 
$k^{2}$ in the self-energy.
Their exact cancellation is, therefore, consistent with the expected infrared physics.

That the self-energy must vanish faster than $k^{2}$ can be derived from Ward identities 
associated with the symmetry 
transformations~\eqref{deformed-rotations}~\cite{guitter_jpf_1989}, or from a direct 
inspection of diagrams.
For any self-energy diagram which is 1PI~\cite{zinn-justin_qft}, and not of the tadpole 
type, all internal wiggly and dashed lines can be replaced by a single non-local 
interaction

\begin{figure}[h]
\centering
\includegraphics{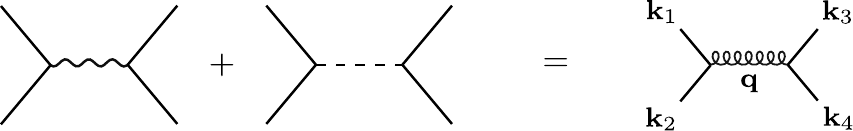}
\begin{equation} \label{effective-interaction}
 = - \lt[\frac{2\lambda_{0}\mu_{0}}{\lambda_{0} + 2 \mu_{0}}P^{T}_{\alpha \beta} 
P^{T}_{\gamma \delta} + \mu_{0} (P^{T}_{\alpha \gamma} P^{T}_{\beta \delta} + 
P^{T}_{\alpha \delta}P^{T}_{\beta \gamma})\rt] k_{1\alpha}k_{2 \beta} k_{3 \gamma} 
k_{4\delta}~,
\end{equation}
\end{figure}

\noindent
where $P^{T}_{\alpha \beta} = \delta_{\alpha \beta} - q_{\alpha}q_{\beta}/q^{2}$ is the 
projector transverse to the momentum transfer $\bq$.
This interaction, can be equivalently derived by integrating out the $u_{\alpha}$ fields 
in favor of an effective theory for $\bh$~\cite{nelson_statistical, nelson_jpf_1987, 
le-doussal_aop_2018, burmistrov_aop_2018}.
Due to transverse projectors, it is always possible to factorize two powers of the 
momentum $\bk$ for each of the two external leg~\cite{le-doussal_aop_2018}, leading to 
diagrams which scale as $k^{4}$ up to powers of $k^{-\ve}$.

\subsection{Renormalization within the dimensional regularization scheme}
\label{s:RG-MS}

For explicit calculation of renormalization-group functions, schemes based on dimensional 
regularization were often used~\cite{aronovitz_prl_1988, guitter_jpf_1989, mauri_npb_2020, 
coquand_pre_2020}.

A convenient feature of this framework is that the counterterm $\sigma_{0} u_{\alpha 
\alpha} $in Eq.~\eqref{H1} is not needed and can be safely set to zero: if $\sigma_{0} = 
0$ at tree level, it remains zero in the renormalized theory~\cite{guitter_jpf_1989}.
This simplification follows from the specific prescriptions of dimensional regularization, 
which automatically remove divergences of power-law type~\cite{zinn-justin_qft}.

We can thus consider a bare Hamiltonian
\begin{equation} \label{H1b}
{\cal H} = \frac{1}{2} \int {\rm d}^{D}x \lt[(\pa^{2} \bh)^{2} + \lambda_{0} 
(u_{\alpha \alpha})^{2} + 2 \mu_{0} u_{\alpha \beta} u_{\alpha \beta}\rt]~.
\end{equation}
All counterterms which can possibly arise in renormalization must be operators invariant 
under the symmetries of the theory, and with relevant or marginal power-counting 
dimensions.
As it can be shown~\cite{guitter_jpf_1989, aronovitz_prl_1988}, Eq.~\eqref{H1b} already 
contains all possible interactions, and the renormalized Hamiltonian, equipped with 
all necessary counterterms, takes the same form up to a redefinition of coefficients:
\begin{equation} \label{H2}
\tilde{{\cal H}}[\bh, u_{\alpha}] = \frac{1}{2} \int {\rm d}^{D}x \lt[Z 
(\pa^{2}\bh)^{2} 
+ M^{\ve} G_{\lambda} (u_{\alpha \alpha})^{2} + 2 M^{\ve} G_{\mu} u_{\alpha \beta} 
u_{\alpha \beta}\rt]~.
\end{equation}
In Eq.~\eqref{H2}, $M$ is an arbitrary wavevector scale, and $Z$, $G_{\lambda}$, and 
$G_{\mu}$ are functions of the dimensionless renormalized coupling constants $\lambdar$, 
$\mur$.
Comparing Eq.~\eqref{H1} and~\eqref{H2} shows that bare and renormalized quantities are 
related as~\cite{aronovitz_prl_1988, guitter_jpf_1989}:
\begin{equation}\label{renormalization-amplitudes-MS}
\bh = \sqrt{Z} \bhr~, \qquad u_{\alpha} = Z \ur_{\alpha}~, \qquad \lambda_{0}  = 
\frac{M^{\ve} G_{\lambda}}{Z^{2}}~, \qquad \mu_{0} = \frac{M^{\ve} G_{\mu}}{Z^{2}}~, 
\qquad \tilde{{\cal H}}[\bhr, \ur_{\alpha}] = {\cal H}[\bh, u_{\alpha}]~.
\end{equation}
Renormalization group equations follow, as usual, from the fact that bare correlation 
functions are independent of $M$~\cite{aronovitz_prl_1988, guitter_jpf_1989}. 
After introduction of the RG functions
\begin{equation}\label{RG-functions}
\eta = \frac{\pa \ln Z}{\pa \ln M}\Big|_{\lambda_{0}, \mu_{0}}~, \qquad
\beta_{\lambda} = \frac{\pa \lambdar}{\pa \ln M}\Big|_{\lambda_{0}, \mu_{0}}~, \qquad 
\beta_{\mu} = \frac{\pa \mur}{\pa \ln M}\Big|_{\lambda_{0}, \mu_{0}}~, 
\end{equation}
renormalization group equations read
\begin{equation}\label{RG-equations}
\lt[\frac{\pa}{\pa \ln M}\Big|_{\lambdar, \mur} + \beta_{\lambda}\frac{\pa}{\pa 
\lambdar}\Big|_{M, \mur} + \beta_{\mu} \frac{\pa}{\pa \mur}\Big|_{M, \lambdar} + 
\lt(\frac{n}{2} + \ell\rt)\eta\rt] \langle \hr_{i_{1}}(\bx_{1}) ..  
\hr_{i_{n}}(\bx_{n}) \ur_{\alpha_{1}}(\bx'_{1}) .. \ur_{\alpha_{n}}(\bx'_{n})\rangle = 0~.
\end{equation}
RG functions at one-loop order have been explicitly calculated in 
Refs.~\cite{aronovitz_prl_1988, guitter_jpf_1989, coquand_pre_2020}, and read:
\begin{equation}\label{RG-functions-1L}
\begin{split}
\beta_{\lambda} & = - \ve \lambdar + \frac{d_{c}}{16 \pi^{2}}\lt(\lambdar^{2} + 
\lambdar \mur + \frac{1}{6}\mur^{2}\rt) + \frac{5}{8 \pi^{2}} \frac{\lambdar \mur 
(\lambdar + \mur)}{\lambdar + 2 \mur}~,\\
\beta_{\mu} & = - \ve \mur + \frac{d_{c}}{96 \pi^{2}} \mur^{2} + \frac{5}{8 
\pi^{2}}\frac{\mur^{2}(\lambdar + \mur)}{\lambdar + 2 \mur}~,\\
\eta & = \frac{5}{16 \pi^{2}} \frac{\mur (\lambdar + \mur)}{\lambdar + 2\mur}~,
\end{split}
\end{equation}
where $d_{c} = d - D$ is the number of components of the $\bh$ field.
An extension to two loops has been recently derived in Ref.~\cite{coquand_pre_2020}.

Scaling behavior emerges at fixed points ($\lambdar_{*}, \mur_{*}$), where 
$\beta_{\lambda} = \beta_{\mu} = 0$.
At these points, RG equations express dilation symmetry of correlation functions, 
characterized by an anomalous dimension $\eta_{*} = \eta(\lambdar_{*}, \mur_{*})$.
In particular, it can be shown that the two-point function of the $\bh$ in momentum space 
scales with the wavevector $\bk$ as $G(\bk) \approx k^{-4 + \eta_{*}}$, while the 
interacting propagator of the field $u_{\alpha}$ behaves as $D_{\alpha \beta}(\bk) 
\approx k^{-6+D+2\eta_{*}}$.
This scaling behavior is often described qualitatively as an infinite stiffening of the 
bending rigidity $\kappa \to \kappa(\bk) \approx \kappa k^{-\eta_{*}}$ and a softening of 
effective elastic moduli $\lambda \to \lambda(\bk)\approx k^{4-D-2\eta_{*}}$, $\mu \to 
\mu(\bk) \approx k^{4-D-2\eta_{*}}$.

\subsection{RG flow and fixed points}
\label{s:fixed-points}

\begin{figure}[h]
\centering
\includegraphics[scale=0.9]{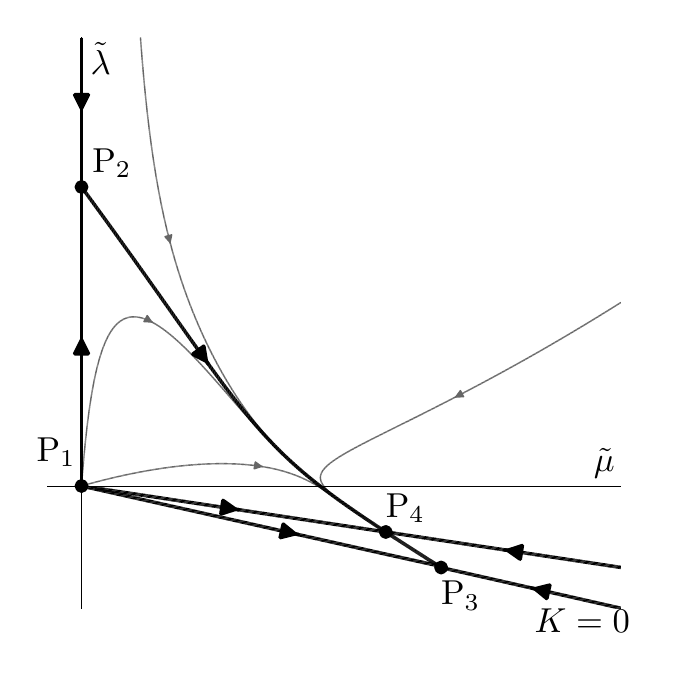}
\caption{\label{f:RG-flow} Renormalization group flow for the flat phase of crystalline 
membranes at one-loop order.}
\end{figure}

The structure of the renormalization group flow is illustrated in Fig.~\ref{f:RG-flow}, 
which portraits the one-loop $\beta$-functions~\eqref{RG-functions-1L}.
For membranes with generic elastic constants, RG trajectories connect the Gaussian 
fixed point P$_{1}$, which is ultraviolet-stable, to an infrared-attractive interacting 
fixed point P$_{4}$~\cite{aronovitz_prl_1988, aronovitz_jpf_1989}.
After extrapolation of the $\ve$-expansion to $D = 4 - \ve \to 2$, this is the case of 
interest for fluctuating two-dimensional materials, and, thus, it is the only case which 
will be analyzed in the rest of this paper.

A different behavior arises for peculiar membranes with either vanishing shear modulus 
($\mu_{0}=0$) or vanishing bulk modulus ($B_{0} = \lambda_{0} +  2 \mu_{0}/D=0$).
In fact, for these special values of the bare elastic constants, the theory presents 
enhanced symmetries~\cite{guitter_jpf_1989}.
For $\mu_{0} = 0$, the model is invariant under the shift $u_{\alpha} \to u_{\alpha} + 
s_{\alpha \beta} x_{\beta}$ for any traceless matrix $s_{\alpha \beta}$.
For vanishing bulk modulus $B_{0}= 0$, the theory is instead invariant under uniform 
compression ($u_{\alpha} \to u_{\alpha} + \ell x_{\alpha}$) or, more, generally under the 
transformation $u_{\alpha} \to u_{\alpha} + \tau_{\alpha}$ for any vector field 
$\tau_{\alpha}$ satifsying the conformal Killing equation $\pa_{\alpha} \tau_{\beta} + 
\pa_{\beta} \tau_{\alpha} = 2 \delta_{\alpha 
\beta} (\pa_{\gamma}\tau_{\gamma})/D$~\cite{guitter_jpf_1989, sun_pnas_2012}\footnote{
This symmetry is not equivalent to the usual notion of conformal invariance intended in 
CFT: the conformal transformation, here, does not act on the coordinates $\bx$, but, 
rather, acts as a shift of the field itself.
In two dimensions with $\lambda_{0} + \mu_{0} = 0$, the linear model of in-plane 
displacement fields is also conformal in the standard CFT sense if $u_{\alpha}$ is 
regarded as a collection of scalars (see Sec.~\ref{s:linear-theory} and 
Ref.~\cite{riva_plb_2005}).
}.
The lines $\mu_{0} = 0$ and $B_{0} = 0$, therefore, cannot be in the basin of attraction 
of P$_{4}$, a fixed point where these enhanced symmetries are absent.
The infrared behavior of membranes with zero shear and zero bulk modulus is instead 
controlled by two different fixed points, P$_{2}$ and P$_{3}$\footnote{The line 
$\mu_{0} = 0$ corresponds, to all orders in perturbation theory, to the line $\mur = 
0$, as it can be verified by inspecting the structure of Feynman diagrams.
The curve in the $(\lambdar, \mur)$ plane corresponding to $B_{0} = 0$, instead, is less
straightforward to express explicitly.
In Ref.~\cite{guitter_jpf_1989}, which used a renormalized bulk modulus as fundamental 
coupling constant, this line corresponds simply to $\tilde{B} = 0$.
However, defining minimal subtraction with $\lambdar$ and $\mur$ as couplings reshuffles 
the parametrization of renormalization constants in a non-trivial way.
At leading order in perturbation theory the curve $B_{0} = 0$ corresponds to the line 
$\lambdar + \mur/2 = 0$.
Already at two loop order, however, the coordinates of the fixed point P$_{3}$ can be 
seen to lie outside of this line.
In Ref.~\cite{coquand_pre_2020}, this was interpreted as an artifact of the 
renormalization scheme.
It is likely in fact that the RG-invariant manifold $B_{0} = 0$ is not a straight line, 
but, rather, a curve $(\lambdar, \mur)$ plane.
}.

\noindent
The lines $\mu_{0} = 0$ and $B_{0} = 0$ mark the boundaries of the overall region of 
stability for the elastic medium: $\mu_{0} \geq 0$, $B_{0} \geq 0$.
Physically, the line $\mu_{0} = 0$ has been proposed to be associated to 
fixed-connectivity fluid membranes~\cite{aronovitz_prl_1988}, or possibly to generic fluid 
membranes~\cite{guitter_jpf_1989}.
A difficulty, however, is that the elastic energy associated with transverse waves is 
exactly zero for vanishing shear modulus, and higher-derivative terms of the form 
$(\pa^{2} u)^{2}$, neglected in the theory, could play a role~\cite{xing_pre_2003}.
The line $B_{0} = 0$, instead, has a physical counterpart, for example, in 
two-dimensional twisted kagome lattices~\cite{sun_pnas_2012}.

Coordinates of fixed points at one-loop order are reported in table~\ref{fixed-points} 
(for results at two-loops order see Ref.~\cite{coquand_pre_2020}).
\begin{table}[h]
\centering
\begin{tabular}{cccc}
\hline
& $\lambdar_{*}$ & $\mur_{*}$  & $\eta_{*}$\\
\hline
P$_{1}$ & 0 & 0 & 0\\
P$_{2}$ & $16\pi^{2} \ve/d_{c}$  & 0 & 0\\
P$_{3}$ & $-48 \pi^{2}\ve/(d_{c}+20)$ & $96 \pi^{2}\ve/(d_{c}+20)$ & $10 \ve/(d_{c}+20)$\\
P$_{4}$ & $-32 \pi^{2} \ve/(d_{c}+24)$ & $96 \pi^{2} \ve/(d_{c}+24)$  & $12 
\ve/(d_{c}+24)$\\
\hline
\end{tabular}
\caption{\label{fixed-points} Coordinates of fixed points and corresponding anomalous 
dimensions at leading order in the $\ve$-expansion.}
\end{table}

\subsection{Bare renormalization group equations}
\label{s:RG-cutoff}

To derive an alternative set of RG equations, we can introduce a cutoff scale $\Lambda$ 
and consider the Hamiltonian
\begin{equation}\label{H-Lambda}
{\cal H}_{\Lambda} = \frac{1}{2} \int {\rm d}^{D}x \lt[(\pa^{2}\bh)^{2} 
+ \frac{c_{1}}{\Lambda^{2}}\lt(\pa_{\alpha}\pa^{2}\bh\rt)^{2} + 
\frac{c_{2}}{\Lambda^{4}}\lt(\pa^{2}\pa^{2}\bh\rt)^{2} + \lambda \Lambda^{\ve} 
\lt(u_{\alpha \alpha}\rt)^{2} + 2 \mu\Lambda^{\ve} u_{\alpha \beta}u_{\alpha \beta} + 2 
\sigma \Lambda^{2} u_{\alpha \alpha}\rt]~.
\end{equation}
Eq.~\eqref{H-Lambda} is almost identical to the model discussed in Sec.~\ref{s:RG-MS}, 
with three differences.
The propagator of the $\bh$ field, $G_{0}(\bk) = 1/k^{4}$, is replaced here by a cutoff 
propagator $G_{0\Lambda}(\bk) = 1/(k^{4} + c_{1}k^{6}/\Lambda^{2} + c_{2} 
k^{8}/\Lambda^{4})$.
This is sufficient to regularize all ultraviolet divergences in perturbation theory, both 
in dimension four and in dimension $D = 4-\ve$ within the framework of the 
$\ve$-expansion\footnote{In analogy with theories of critical 
phenomena~\cite{zinn-justin_qft}, we define the $\ve$-expansion as a simultaneous (double 
series) expansion in $\ve$ and in the perturbative coupling constant.
At any finite order in this expansion, propagators and vertices behave with the 
same scaling of corresponding tree-level functions up to powers of $\ln k$, where $k$ is 
the momentum scale.
From the point of view of power counting and UV divergences, the $\ve$-expansion 
is thus identical to the theory in dimension $D = 4$.
}.
A second difference is in the normalization of couplings: in Eq.~\eqref{H-Lambda} all 
dimensionful interactions are expressed by factorizing corresponding powers of the cutoff 
scale, in such way that the coefficients $c_{1}$, $c_{2}$, $\lambda$, $\mu$, and $\sigma$ 
are dimensionless\footnote{
Despite the different normalization, we use the same symbols for elastic 
coefficients $\lambda$ and $\mu$ in order to lighten the notation.}.
Finally, the 'tension' term $\sigma u_{\alpha\alpha}$, which vanishes in dimensional 
regularization, is non-zero in general, and has been reintroduced in the expression of the 
Hamiltonian (effects of $\sigma$ have been discussed in Sec.~\ref{s:perturbative}).

To study scaling behavior, we can write bare RG equations~\cite{zinn-justin_qft} 
expressing the equivalence between changes of the cutoff and renormalizations of coupling 
constants:
\begin{equation} \label{bare-RG}
 \lt[\frac{\pa}{\pa \ln \Lambda}\Big|_{\lambda, \mu} + \betab_{\lambda} \frac{\pa}{\pa 
\lambda}\Big|_{\Lambda, \mu} + \betab_{\mu} \frac{\pa}{\pa \mu}\Big|_{\Lambda, \lambda} + 
\lt(\frac{n}{2} + \ell\rt)\etab\rt] \langle h_{i_{1}}(\bx_{1}) .. h_{i_{n}}(\bx_{n}) 
u_{\alpha_{1}}(\bx'_{1})..u_{\alpha_{\ell}}(\bx'_{\ell})\rangle = 0~,
\end{equation}
or, for 1PI correlation functions with $n$ external $\bh$ lines
\begin{equation} \label{bare-RG-1PI}
\lt[\frac{\pa}{\pa \ln \Lambda}\Big|_{\lambda, \mu} + \betab_{\lambda} \frac{\pa}{\pa 
\lambda}\Big|_{\Lambda, \mu} + \betab_{\mu} \frac{\pa}{\pa \mu}\Big|_{\Lambda, 
\lambda} - \lt(\frac{n}{2} + \ell\rt)\etab\rt] \Gamma^{(n, \ell)}_{i_{1}.. 
i_{n} \alpha_{1} .. \alpha_{\ell}}(\bk_{1}, .., \bk_{n}, \bk'_{1}, .., \bk'_{\ell}) = 0~.
\end{equation}
Eqs.~\eqref{bare-RG} and~\eqref{bare-RG-1PI} are a consequence of the perturbative 
renormalizability of the $\ve$-expansion, which follows from 
power-counting arguments in analogy with other field theories~\cite{zinn-justin_qft}.
As usual, the RG functions $\betab_{\lambda}$, $\betab_{\mu}$, and $\etab$ cannot depend 
on $\Lambda$, because they are dimensionless and $\Lambda$ is the only scale in the 
problem.
It follows that $\betab_{\lambda}$, $\betab_{\mu}$, and $\etab$ depend only on the 
dimensionless bare couplings $\lambda$ and $\mu$, and, implicitly, on the specific form of 
regularization, expressed via the coefficients $c_{1}$ and $c_{2}$ (parameters which we 
choose to keep fixed as the cutoff is lowered).

In this setting, perturbative RG equations are closely analogue to Wilson's exact 
renormalization group equations.
The main difference is that in most formulations of Wilson's RG the lowering of an UV 
cutoff is compensated by the flow of coupling constants exactly.
Here, instead, after a change of $\Lambda$ and subsequent renormalizations, the physics 
is preserved up to small corrections which vanish roughly as $\Lambda^{-2}$ in the limit 
of $\Lambda$ large.
In more detail, adapting an analogue result for the critical scalar field 
theory~\cite{zinn-justin_qft}, we expect that 1PI correlation functions behave for large 
$\Lambda$ as
\begin{equation}\label{double-series-expansion}
\mathbf{\Gamma}^{(n, \ell)}  = k_{1\beta_{1}} .. k_{n\beta_{n}} k'_{1 \gamma_{1}} .. 
k'_{\ell \gamma_{\ell}} \sum_{\substack{m, p, q \geq 0\\ m + p + q = L-1 + \frac{n}{2} + 
\ell}} \mathbf{\Gamma}_{Ls, mpq}^{(n, \ell) \beta_{1} .. \beta_{n} \gamma_{1} .. 
\gamma_{\ell}} \mu^{m} \lambda^{p} \lt(\frac{\lambda \mu}{\lambda + 2 \mu}\rt)^{q} 
\ve^{s}~,
\end{equation}
where, schematically,
\begin{equation}
\mathbf{\Gamma}^{(n, \ell)}_{Ls, mpq} = \sum_{k = 0}^{N(n, \ell, L, s)}  
\mathbf{\Gamma}^{(n, \ell)}_{Ls, mpq, k} (\ln \Lambda)^{k} + \Lambda^{-2} \times 
(\text{powers of }\ln \Lambda) + ...
\end{equation}
Perturbative renormalizability implies that the bare RG equations~\eqref{bare-RG} 
and~\eqref{bare-RG-1PI} are exact for the part which does not vanish in the limit $\Lambda 
\to \infty$~\cite{zinn-justin_qft}.
As a result, fixed points and anomalous dimensions of the perturbative renormalization 
group describe exactly the exponent of the leading scaling behavior, and only misses 
corrections due to strongly-irrelevant operators, separated by a large gap in the 
dilatation spectrum.

\subsection{Comment on reflection positivity}
\label{s:reflection-positivity}

Although we could not develop a detailed derivation, we expect that the membrane model 
discussed in this section is not reflection-positive.
In the ultraviolet limit, where interactions can be neglected, the theory reduces to
\begin{equation}
{\cal H}_{\rm UV} = \frac{1}{2} \int {\rm d}^{D}x \lt[(\pa^{2} \bh)^{2} + 
(\lambda_{0}+ \mu_{0})(\pa_{\alpha}u_{\alpha})^{2} + \mu_{0} \pa_{\alpha}u_{\beta} 
\pa_{\alpha}u_{\beta}\rt]~,
\end{equation}
the combination of $d_{c}$ copies of a higher-derivative scalar theory and a Gaussian 
vector model.
These non-interacting theories were analyzed in Refs.~\cite{arici_jmp_2018, riva_plb_2005, 
el-showk_npb_2011} and were shown to lack reflection positivity or, equivalently, 
unitarity in Minkowski space.
We find it likely, therefore, that the also the full interacting model is not 
reflection-positive.
A conclusive result requires, however, an analysis of the infrared 
region~\cite{pisarski_prd_1983}.
We leave this question to further investigations.

\section{Gaussian-curvature interactions}
\label{s:GCI}

In addition to the theory of elasticity, we discuss the relation between scale and 
conformal invariance in an alternative model, discussed in detail in 
Ref.~\cite{mauri_npb_2020} (see also Ref.~\cite{coquand_pre_2020}).
The starting point in the derivation of this model is the observation that the 
Hamiltonian depends on the in-plane displacement fields $u_{\alpha}$ quadratically.
As a result, integration over $u_{\alpha}$ can be computed analytically, and gives an 
effective interaction of a form already introduced in 
Eq.~\eqref{effective-interaction}~\cite{nelson_statistical, nelson_jpf_1987, 
le-doussal_aop_2018}.
In the case $D = 2$, which is the dimension of interest physically, the geometrical 
structure of the effective interaction simplifies because, due to the presence of a 
single transverse direction, $P^{T}_{\alpha \beta} P^{T}_{\gamma \delta} = P^{T}_{\alpha 
\gamma} P^{T}_{\beta \delta}=P^{T}_{\alpha \delta} P^{T}_{\beta \gamma}$.
As a result, the interaction becomes separable~\cite{gazit_pre_2009}, and can be 
decoupled by introducing a scalar field via a Hubbard-Stratonovich 
transformation~\cite{mauri_npb_2020}.

It follows that, for $D = 2$, the physics of $\bh$-field fluctuations can be captured by 
an alternative local field theory:
\begin{equation}\label{gci}
{\cal H} = \int {\rm d}^{D}x \lt[\frac{1}{2} (\pa^{2}\bh)^{2} + \frac{1}{2Y_{0}} 
(\pa^{2} \chi)^{2} + i \chi K\rt]~,
\end{equation}
where $\chi$ is a scalar field mediating interactions and
\begin{equation}
K(\bx) = - \frac{1}{2} (\delta_{\alpha \beta} \pa^{2} - \pa_{\alpha}\pa_{\beta}) 
(\pa_{\alpha} \bh \cdot \pa_{\beta} \bh) = \frac{1}{2} [(\pa^{2} \bh \cdot \pa^{2} \bh) - 
(\pa_{\alpha}\pa_{\beta}\bh \cdot \pa_{\alpha} \pa_{\beta}\bh)]~.
\end{equation}
As it can be shown, $K(\bx)$ is an approximate version of the Gaussian curvature of the 
membrane~\cite{nelson_jpf_1987}.
Eq.~\eqref{gci} thus expresses, qualitatively, a theory for membrane fluctuations with 
long-range interactions between Gaussian curvatures.
In the following, Eq.~\eqref{gci} will be referred to as the 'Gaussian curvature 
interaction', or 'GCI' model.

\begin{figure}[h]
\centering 
\includegraphics{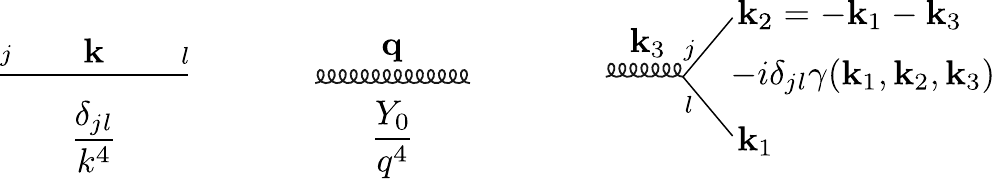}
\begin{equation*}
~ \qquad \gamma(\bk_{1}, \bk_{2}, \bk_{3}) = k_{1}^{2} k_{2}^{2} - (\bk_{1}\cdot 
\bk_{2})^{2} = 
k_{2}^{2} k_{3}^{2} - (\bk_{2}\cdot \bk_{3})^{2} = k_{3}^{2} k_{1}^{2}- (\bk_{3} \cdot 
\bk_{1})^{2}
\end{equation*}
\caption{\label{feynman-gci} Feynman rules for the effective model.
Solid and curly lines represent, respectively, propagators of the $\bh$ and of the $\chi$ 
field.}
\end{figure}

The theory is controlled by a single coupling, the Young modulus $Y_{0} = 4 \mu_{0} 
(\lambda_{0} + \mu_{0})/(\lambda_{0} + 2 \mu_{0})$, which is proportional to both the 
shear coefficient $\mu_{0}$ and the two-dimensional bulk coefficient $\lambda_{0} + 2 
\mu_{0}/D = \lambda_{0} + \mu_{0}$.
Perturbative expansions can be computed from the Feynman rules in Fig.~\ref{feynman-gci}.

\noindent
As discussed in Ref.~\cite{mauri_npb_2020}, the long-wavelength behavior of the theory 
can be studied by perturbative techniques within an $\ve$-expansion near $D = 4$, 
dimension in which the model is renormalizable.

Renormalization is particularly simple because, as an analysis of power counting shows, 
there are only two primitive divergences: the amplitude and the coupling constant 
renormalization~\cite{mauri_npb_2020, le-doussal_aop_2018}.
The vertex function, instead, is superficially UV-convergent.
These properties follow directly from the special form of the vertex function 
$\gamma(\bk_{1}, \bk_{2}, \bk_{3})$, which, in any 1PI diagram, allows to factorize two 
powers of each external momentum, reducing the degree of divergence.
We note that a similar result emerges in Galileon theories, which can include terms of 
the same form of the interaction $i \chi K$ in Eq.~\eqref{gci}.
Also in these theories, vertex non-renormalization plays a crucial 
role~\cite{goon_jhep_2016}.

Due to the considerations above, the renormalized action can be written as
\begin{equation}
\tilde{{\cal H}} = \int {\rm d}^{D}x \lt[\frac{Z}{2} (\pa^{2} \bh)^{2} + \frac{1}{2 
Z_{Y}Y M^{\ve}} (\pa^{2}\chi)^{2} + i \chi K\rt]~,
\end{equation}
where $M$ is an arbitrary scale, $Y$ is the dimensionless renormalized coupling, and 
$Z$, $Z_{Y}$ are divergent factors.
After introduction of
\begin{equation}
\beta(Y) = \frac{\pa Y}{\pa \ln M}\Big|_{Y_{0}}~, \qquad \eta(Y) = \frac{\pa \ln Z}{\pa 
\ln M}\Big|_{Y_{0}}~,
\end{equation}
the relations between bare and renormalized quantities
\begin{equation} \label{gci-renormalization}
\bh = \sqrt{Z} \bhr~, \qquad \chi = Z^{-1} \chir~, \qquad Y_{0} = 
M^{\ve}\frac{Z_{Y} Y}{Z^{2}}
\end{equation}
imply the RG equations
\begin{equation} \label{RG-gci}
\lt[\frac{\pa}{\pa \ln M} + \beta(Y) \frac{\pa}{\pa Y} + \lt(\frac{n}{2} - 
\ell\rt)\eta(Y)\rt] \langle \hr_{i_{1}}(\bx_{1}) .. \hr_{i_{n}}(\bx_{n}) \chir (\bx'_{1}) 
.. \chir (\bx'_{\ell})\rangle = 0~.
\end{equation}
For the renormalized 1PI functions with $n$ external $\bh$ legs and $\ell$ external 
$\chi$ lines in momentum space, the corresponding RG relations read
\begin{equation}
\lt[\frac{\pa}{\pa \ln M} + \beta(Y) \frac{\pa}{\pa Y} - 
\lt(\frac{n}{2} - \ell\rt)\eta(Y)\rt]\Gammar^{(n, \ell)} = 0~.
\end{equation}
The $\beta$ function presents, in the $\ve$-expansion, an infrared-stable interacting 
fixed point at $Y = Y_{*}$ with $Y_{*} = {\rm O}(\ve)$~\cite{mauri_npb_2020, 
coquand_pre_2020}.
This fixed point controls the asymptotic infrared behavior.
In particular, the propagator of the $\bh$ field behaves as $G(\bk) = [\Gamma^{(2, 
0)}(\bk)]^{-1}\approx k^{-4 + \eta_{*}}$, and the two-point function of the mediator 
field as $D(\bk) = [\Gamma^{(0, 2)}(\bk)]^{-1}\approx k^{-D-2\eta_{*}}$.
More generally $\Gamma^{(n, \ell)}$ behaves with overall momentum scale as $ 
\Gamma^{(n, \ell)} \approx k^{D + \ell \eta_{*} + n(\ve -\eta_{*})/2}$.
The exponent has been calculated at two-loop order in Refs.~\cite{mauri_npb_2020, 
coquand_pre_2020} and reads
\begin{equation}
\eta_{*} = \frac{2\ve}{d_{c} + 4} - \frac{d_{c}(2-d_{c})}{6 (d_{c}+4)^{3}}\ve^{2} + {\rm 
O}(\ve^{3})~.
\end{equation}
This exponent differs from anomalous dimensions of all fixed points in 
table~\ref{fixed-points}~\cite{coquand_pre_2020}.
The GCI model, although equivalent to Eq.~\eqref{H1} for $D=2$, becomes a distinct theory 
in generic dimension, and provides a separate dimensional continuation to $D = 4 - \ve$.

Finally, let us discuss the shift symmetries of the GCI model.
The Hamiltonian density is invariant under the transformations $\bh \to \bh + \mathbf{A} + 
\mathbf{B}_{\alpha} x_{\alpha}$, where $\mathbf{A}$ and $\mathbf{B}_{\alpha}$ are vectors 
in $d_{c}$-dimensional space.
The theory is also invariant under the shifts $\chi \to \chi + A' + 
B'_{\alpha}x_{\alpha}$, which change the energy density by a total derivative.

To conclude, we note that, the GCI model behaves in the UV as two copies of the biharmonic 
theory, which is not reflection-positive~\cite{arici_jmp_2018}.
Thus, we find it likely that the full theory will also lack reflection positivity.

\section{Energy-momentum tensor in scale-invariant and conformal field theories}
\label{s:energy-momentum}

Let us briefly discuss the relation between scale, conformal invariance, and the 
structure of the energy-momentum tensor.
In any local Euclidean-invariant model, rotational symmetry implies the existence of an 
energy-momentum tensor $T_{\alpha \beta}$ which is symmetric and 
conserved~\cite{di-francesco_cft, rychkov_epfl_2017, polchinski_npb_1988}.
As shown in Ref.~\cite{polchinski_npb_1988}, scale invariance requires that the trace is 
expressible as a total divergence,
\begin{equation}\label{scale-condition}
T_{\alpha \alpha} = \pa_{\alpha} V_{\alpha}~,
\end{equation}
where $V_{\alpha}$ is a local 'virial current' without explicit coordinate dependence.
Conformal invariance requires instead a stronger condition~\cite{polchinski_npb_1988}: 
that
\begin{equation}\label{conformal-condition}
T_{\alpha \alpha} = \pa_{\alpha}\pa_{\beta}L_{\alpha \beta}~,
\end{equation}
or, equivalently, that the virial current $V_{\alpha}$ can be expressed as $V_{\alpha} = 
J_{\alpha} + \pa_{\beta} L_{\alpha \beta}$, where $J_{\alpha}$ is a conserved current 
(with $\pa_{\alpha} J_{\alpha} = 0$).
In dimension $D = 2$, two alternatives should be distinguished: if $L_{\alpha \beta} = 
\delta_{\alpha \beta} L$  the system displays invariance under the full 
infinite-dimensional group of local conformal maps.
If, instead, $T_{\alpha \beta} = \pa_{\alpha}\pa_{\beta} L_{\alpha \beta}$ but 
$L_{\alpha \beta}$ is not expressible as $\delta_{\alpha \beta} L$ the theory is 
invariant under the global conformal group (it is 'M\"{o}bius invariant'), but not under 
the infinite Virasoro symmetry~\cite{nakayama_aop_2016, polchinski_npb_1988}.

A remark is that in scale- and conformally-invariant theories the 
relations~\eqref{scale-condition} and~\eqref{conformal-condition} are usually not 
satisfied identically, but only up to operators which can be identified as generators of 
infinitesimal field redefinitions~\cite{paulos_npb_2016, brown_aop_1980, 
de-polsi_jsp_2019}.
Examples of such operators are $\bE \cdot \bh$, $E_{\alpha}u_{\alpha}$, 
$\bE \cdot \pa_{\beta}\bh$ and $E_{\alpha}\pa_{\beta} u_{\alpha}$, where $\bE(\bx) = 
\delta {\cal H}/\delta \bh(\bx)$ and $E_{\alpha}(\bx) = \delta {\cal H}/\delta 
u_{\alpha}(\bx)$ are variational derivatives of the action, defining the equations of 
motion.
When inserted in correlation functions, these operators produce contact terms and 
generate local changes of field variables which contribute to the transformation law of 
fields under scale and conformal maps~\cite{paulos_npb_2016, brown_aop_1980, 
de-polsi_jsp_2019}.
Further, when referring to the operators $\bE$ and $E_{\alpha}$, we will tell simply 
``equation of motion $E$'' instead of ``$E$ is the variational derivative of the action 
such that $E$ = 0 is the equation of motion''.

\section{Scale vs. conformal invariance in linear elasticity theories}
\label{s:linear-theory}

Before analyzing the complete theories, let us examine the membrane and the GCI model at 
the level of a non-interacting, free-field approximation.

For membrane theory, starting from the Hamiltonian defined in Eq.~\eqref{H1b} and 
neglecting all interactions between the fields $\bh$ and $u_{\alpha}$ we obtain:
\begin{equation}
H'' = \frac{1}{2} \int {\rm d}^{D}x \lt[(\pa^{2}\bh)^{2} + (\lambda_{0} + \mu_{0}) 
(\pa_{\alpha}u_{\alpha})^{2} + \mu_{0} \pa_{\alpha} u_{\beta} \pa_{\alpha} u_{\beta}\rt]~.
\end{equation}

Fluctuations of $\bh$ are thus described by the free bi-harmonic model $H^{(\bh)} = \int 
{\rm d}^{D}x (\pa^{2} \bh)^{2}/2$.
It is a well-known result that this model is conformally-invariant in general 
dimension~\cite{nakayama_aop_2016, brust_jhep_2017}.
An explicit calculation, in fact, shows that the theory admits a symmetric energy-momentum 
tensor with trace
\begin{equation}
T^{(\bh)}_{\alpha \alpha} = \frac{1}{2}(4-D)(\pa^{2}\bh)^{2} = \frac{1}{2}(4-D) \lt[ 
\bh \cdot \pa^{2}\pa^{2} \bh + \pa_{\alpha}\pa_{\beta} L_{\alpha \beta}\rt]~,
\end{equation}
and $L_{\alpha \beta} = 2 (\pa_{\alpha} \bh \cdot \pa_{\beta} \bh) - \delta_{\alpha 
\beta} (\pa_{\gamma}\bh \cdot \pa_{\gamma}\bh) - \delta_{\alpha \beta} (\bh \cdot 
\pa^{2}\bh)$.
This form is consistent with that expected for a conformal 
theory~\cite{nakayama_aop_2016}: the trace can be reduced to a total second derivative, 
up to the term $\bh \cdot \pa^{2}\pa^{2}\bh$, which vanishes with the equation of motion 
$\pa^{2} \pa^{2} \bh=0$ and can be identified as the generator of local field rescaling.
Since $L_{\alpha \beta} \neq \delta_{\alpha \beta} L$, the biharmonic theory in dimension 
$D = 2$ is invariant under the global conformal group but not under the infinite Virasoro 
symmetry~\cite{nakayama_aop_2016}.

The theory for $u_{\alpha}$ fluctuations,
\begin{equation}\label{H-linear-elasticity}
H^{(u)} = \frac{1}{2} \int {\rm d}^{D}x 
[(\lambda_{0} + \mu_{0}) (\pa_{\alpha}u_{\alpha})^{2} + \mu_{0} \pa_{\alpha}u_{\beta} 
\pa_{\alpha}u_{\beta}]
\end{equation}
is the well-known theory of linear isotropic elastic media.
As it was shown in Refs.~\cite{riva_plb_2005, el-showk_npb_2011}, this model provides a
physical realization of a scale-invariant but nonconformal field theory.

The lack of conformal invariance can be seen by showing that $u_{\alpha}$ cannot be a 
primary field nor a descendant~\cite{el-showk_npb_2011}.
That $u_{\alpha}$ is not primary follows from the fact that its two-point function is 
inconsistent with constraints imposed by conformal invariance.
Any primary vector field $y_{\alpha}$ of dimension $\Delta_{y}$ in a CFT, in fact, 
presents a propagator with a specific tensor structure~\cite{el-showk_npb_2011, 
rychkov_epfl_2017, nakayama_prd_2017}:
\begin{equation}\label{vector-primary}
\langle y_{\alpha}(\bx) y_{\beta}(\bx') \rangle = \frac{A}{|\bx-\bx'|^{2\Delta_{y}}} 
\lt(\delta_{\alpha \beta} -2 
\frac{(x_{\alpha}-x'_{\alpha})(x_{\beta}-x'_{\beta})}{|\bx-\bx'|^{2}}\rt)
\end{equation}
in real space and 
\begin{equation}\label{vector-primary-Fourier}
\langle y_{\alpha}(\bk) y_{\beta}(-\bk)\rangle = \frac{A'}{k^{D-2\Delta_{y}}} 
\lt(\delta_{\alpha \beta} + \frac{(D-2\Delta_{y})}{(\Delta_{y}-1)} \frac{k_{\alpha} k 
_{\beta}}{k^{2}}\rt)
\end{equation}
in momentum space.
Explicit calculation of the propagator of $u_{\alpha}$, which has dimension $\Delta_{u} = 
(D-2)/2$, shows that its two-point function is inconsistent with 
Eq.~\eqref{vector-primary-Fourier}, unless elastic constants are tuned in such way that 
$D\lambda_{0} + (D + 4) \mu_{0}= 0$.
That $u_{\alpha}$ is not descendant follows from a simple dimensional analysis: for $D 
\geq 2$, $u_{\alpha}$ is the field with lowest possible dimension, and there exists no 
candidate operator with dimension $\{u_{\alpha}\} -1$ of which $u_{\alpha}$ could be a 
derivative.
The conclusion is therefore that the theory is scale invariant but lacks conformal 
symmetry~\cite{el-showk_npb_2011}.

In $D=2$, the field dimension becomes $\Delta_{u} = 0$, and the propagator behaves as 
$\ln |\bx - \bx'|$, but it can still be shown that the theory lacks conformal 
invariance~\cite{riva_plb_2005}.

These results are confirmed by an inspection of the energy-momentum tensor: the theory 
admits an improved symmetric energy-momentum tensor $T_{\alpha \beta}$ with trace
\begin{equation} \label{T-linear-elasticity}
T_{\alpha \alpha} = \pa_{\alpha}V_{\alpha}~, \qquad 
V_{\alpha} = \frac{1}{2} (D \lambda_{0} + (D+2) \mu_{0}) u_{\alpha} 
\pa_{\gamma}u_{\gamma} - \frac{1}{2} (D-2) \mu_{0} u_{\gamma}\pa_{\alpha} u_{\gamma} - 
\mu_{0} u_{\gamma} \pa_{\gamma} u_{\alpha}~,
\end{equation}
up to terms which vanish with the equations of motion.
For generic $\lambda_{0}$ and $\mu_{0}$, the virial current cannot be reduced to the form 
$V_{\alpha} = j_{\alpha} + \pa_{\beta} L_{\alpha \beta}$, with $\pa_{\alpha} j_{\alpha} = 
0$, implying the absence of conformal invariance.

Conformal symmetry is only recovered in special cases.
When $D \lambda_{0} + (D+4)\mu_{0} = 0$, the virial current reduces to the form 
$V_{\alpha} = \pa_{\beta} L_{\alpha \beta}$, and the theory becomes conformal with 
$u_{\alpha}(\bx)$ as a primary field.
The corresponding model is unphysical as an elasticity theory, being outside of the 
stability region $\mu_{0} \geq 0$, $B_{0} = \lambda_{0}+2\mu_{0}/D\geq 0$, but it is 
relevant for gauge-fixed electrodynamics~\cite{el-showk_npb_2011}.

For $\lambda_{0} + \mu_{0} = 0$ another, 'twisted', form of conformal invariance 
appears.
In this case, the symmetry of the theory is enhanced from O$(D)$ to O$(D) 
\times$O$(D)$, and we can choose to regard $u_{\alpha}$ as a set of scalar fields rather 
than a vector field~\cite{riva_plb_2005, nakayama_prd_2017}.
The Hamiltonian is identical to $D$ copies of free scalar field theory, and is, therefore, 
conformal\footnote{
The virial current in Eq.~\eqref{T-linear-elasticity} no longer holds for this twisted 
theory.
In fact, Eq.~\eqref{T-linear-elasticity} was derived by including improvement terms 
needed to make $T_{\alpha \beta}$ symmetric.
If $\lambda_{0} + \mu_{0}=0$ and $u_{\mu}$ is assumed to transform as a scalar, $T_{\alpha
\beta}$ is already symmetric and the improvement must not be 
performed~\cite{riva_plb_2005}.
}.
The possibility to consider $u_{\alpha}$ as a collection of scalars, however, is destroyed 
in the full membrane model, which breaks O$(D) \times$O$(D)$ symmetry even for 
$\lambda_{0} + \mu_{0} = 0$.

As mentioned in Sec.~\ref{s:fixed-points}, a form of embedding-space conformal 
invariance appears for zero bulk modulus $B_{0} = \lambda_{0} + 2 \mu_{0}/D = 0$.
In this case, Eq.~\eqref{H-linear-elasticity} is invariant under the shift of 
displacement fields $u_{\alpha} \to u_{\alpha} + \tau_{\alpha}$, where $\tau_{\alpha}$ is 
a conformal Killing vector with $\pa_{\alpha} \tau_{\beta} + \pa_{\beta} \tau_{\alpha} = 
2 \delta_{\alpha \beta} (\pa_{\gamma} \tau_{\gamma})/D$~\cite{guitter_jpf_1989, 
sun_pnas_2012}.
This symmetry differs from the usual definition of conformal invariance in CFT, because 
transformations act as shifts of the fields and not as shifts of the coordinates $\bx$.

Finally, Ref.~\cite{nakayama_aop_2016} showed that in two dimensions the elasticity model 
for \emph{any} choice of $\lambda_{0}$ and $\mu_{0}$ presents a hidden conformal symmetry 
which emerges when displacement fields are represented as gradients of scalar potentials: 
$u_{\alpha} = \pa_{\alpha} \phi + \epsilon_{\alpha \beta}\pa_{\beta} 
\omega$, where $\phi$ and $\omega$ are respectively a scalar and a pseudoscalar field.
This representation maps Eq.~\eqref{H-linear-elasticity} to two copies of the biharmonic 
theory, which is conformal in general dimension.

The GCI model defined in Eq.~\eqref{gci}, similarly, reduces to two decoupled biharmonic 
theories in the non-interacting limit $Y_{0} \to 0$.

\section{Scale vs. conformal invariance in membrane theory}
\label{s:T-membrane-theory}

\subsection{Inconsistency between vector two-point function and conformal selection rules}
\label{selection-rules}
To analyze whether conformal invariance holds in membrane theory, let us examine the 
two-point function $D_{\alpha \beta}(\bk)$ of the vector field $u_{\alpha}$ in momentum 
space\footnote{We are grateful to S. Rychkov for attracting our attention to the 
advantage of such analysis.}.
If we choose a renormalization scale $M \simeq |\bk|$ of the order of the magnitude of 
a given momentum of interest, the renormalized propagator $\tilde{D}_{\alpha \beta}(\bk) 
= Z^{-2} D_{\alpha \beta}(\bk)$ is accurately captured by renormalized perturbation 
theory and, thus, for $\ve$ small, can be approximated by the corresponding tree-level 
contribution.
After calculation at scales $|\bk| \simeq M$, the result can be rescaled to any 
wavelength via scaling relations.
We thus deduce that the correlation function at an arbitrary $\bk$ in the infrared region 
takes approximately the form
\begin{equation} \label{D-tree-level-scaled}
\tilde{D}_{\alpha \beta}(\bk) \simeq \frac{1}{M^{2 \eta_{*}} k^{6 - D - 2 \eta_{*}}} 
\lt[\frac{P^{L}_{\alpha \beta}(\bk)}{\lambdar_{*} + 2 \mur_{*}} + 
\frac{P^{T}_{\alpha \beta}(\bk)}{\mur_{*}}\rt]~,
\end{equation}
where $P^{L}_{\alpha \beta} = k_{\alpha} k_{\beta}/k^{2}$ and $P^{T}_{\alpha \beta} = 
\delta_{\alpha \beta} - k_{\alpha} k_{\beta}/k^{2}$ are longitudinal and transverse 
projectors.
In particular, the fixed point values of the renormalized couplings can be used to 
estimate, at the leading order in the $\ve$-expansion, the tensor structure of $D_{\alpha 
\beta}(\bk)$.

We can now compare Eq.~\eqref{D-tree-level-scaled} with 
Eq.~\eqref{vector-primary-Fourier}, the special form of the two-point function of a 
primary vector field.
Near $D = 4$, the scaling dimension of $u_{\alpha}$ is $\Delta_{u} = (1 +\eta_{*}-\ve) 
\simeq 1$ and thus, Eq.~\eqref{vector-primary-Fourier} implies that a vector consistent 
with conformal symmetry should have a two-point function which is almost purely 
longitudinal.
In contrast, taking the O$(\ve)$ values of the couplings at the fixed point P$_{4}$, 
$\mur_{*} = 96 \pi^{2} \ve/(d_{c} + 24)$, $\lambdar_{*} = -\mur_{*}/3$, we see that in 
$\tilde{D}_{\alpha \beta}(\bk)$ longitudinal and transverse components have the same 
order of magnitude.
This consideration, in analogy with Ref.~\cite{el-showk_npb_2011} shows that $u_{\alpha}$ 
cannot be a conformal primary field.

\subsection{Analysis of the virial current}
For an alternative analysis, let us consider the structure of the energy-momentum 
tensor.
An explicit calculation gives\footnote{
In order to obtain an improved energy-momentum tensor which is symmetric identically, 
without the use of equations of motion, we define $T_{\alpha \beta}$ as the response of 
the Hamiltonian to the infinitesimal transformation $\bh(\bx) \to \bh'(\bx) = \bh(\bx')$,
$u_{\alpha}(\bx) \to u'_{\alpha}(\bx) = \lt(\delta_{\alpha \beta} + 
(\pa_{\alpha}\epsilon_{\beta} - \pa_{\beta}\epsilon_{\alpha})/2 \rt)u_{\beta}(\bx')$, 
$x_{\alpha} \to x'_{\alpha} = x_{\alpha} + \epsilon_{\alpha}$, including a local rotation 
of $u_{\alpha}$ in reaction to the antisymmetric part of $\pa_{\alpha} \epsilon_{\beta}$.
For this reason the conservation law, Eq.~\eqref{T-conservation-law}, includes the term 
$-\pa_{\alpha}(E_{\beta}u_{\alpha} - E_{\alpha} u_{\beta})/2$, an operator which, 
inserted in correlation functions, acts as a generator for local rotations of the 
$u_{\alpha}$ field.
}
\begin{equation}
\begin{split}
T_{\alpha \beta} & = -\frac{1}{2}\delta_{\alpha \beta} \lt[(\pa^{2}\bh)^{2} + 
\lambda_{0} (u_{\gamma \gamma})^{2} + 2 \mu_{0} u_{\gamma \delta} u_{\gamma \delta}\rt] + 
2 \pa_{\alpha}\pa_{\beta}\bh \cdot \pa^{2}\bh - \pa_{\alpha}\bh \cdot 
\pa_{\beta}\pa^{2}\bh - \pa_{\beta}\bh \cdot \pa_{\alpha}\pa^{2}\bh\\
& +\frac{1}{D-1}\lt[\delta_{\alpha \beta} \pa_{\gamma}\bh \cdot 
\pa_{\gamma}\pa^{2}\bh + \delta_{\alpha \beta} \pa_{\gamma} \pa_{\delta} \bh \cdot 
\pa_{\gamma}\pa_{\delta}\bh + (D-2)\pa_{\gamma}\bh \cdot 
\pa_{\alpha}\pa_{\beta}\pa_{\gamma}\bh - D \pa_{\alpha}\pa_{\gamma}\bh \cdot 
\pa_{\beta}\pa_{\gamma}\bh\rt]\\
& +  2\lambda_{0} u_{\gamma \gamma}u_{\alpha \beta} + 4 \mu_{0} u_{\alpha 
\gamma}u_{\beta \gamma} - \frac{1}{2} (E_{\alpha} u_{\beta} + E_{\beta} u_{\alpha})+ 
\lambda_{0}\pa_{\gamma} \lt[(\delta_{\alpha \beta} u_{\gamma} - \delta_{\beta \gamma} 
u_{\alpha} - \delta_{\alpha \gamma} u_{\beta}) u_{\delta \delta}\rt] \\
&+ 2\mu_{0}\pa_{\gamma} \lt[u_{\gamma} u_{\alpha \beta} -u_{\alpha} 
u_{\beta \gamma} - u_{\beta} u_{\alpha \gamma}\rt]~,
\end{split}
\end{equation}
which is symmetric and locally conserved.
The conservation law for $T_{\alpha \beta}$, in particular, reads 
\begin{equation} \label{T-conservation-law}
\pa_{\alpha}T_{\alpha \beta} = -\bE \cdot \pa_{\beta} \bh - 
E_{\alpha}\pa_{\beta}u_{\alpha} - \frac{1}{2} \pa_{\alpha} (E_{\beta} u_{\alpha} - 
E_{\alpha} u_{\beta})
\end{equation}
where
\begin{equation}\label{equation-motion-dimensional-reg}
\bE = \frac{\delta {\cal H}}{\delta \bh} = \pa^{2}\pa^{2} \bh - \pa_{\alpha} (\lambda_{0} 
u_{\beta \beta} \pa_{\alpha}\bh + 2 \mu_{0} u_{\alpha \beta} \pa_{\beta}\bh)~, \qquad
E_{\alpha} = \frac{\delta {\cal H}}{\delta u_{\alpha}}=-\lambda_{0} \pa_{\alpha}u_{\beta 
\beta} - 2 \mu_{0} \pa_{\beta} u_{\alpha 
\beta}~,
\end{equation}
are equations of motion of the $\bh$ and the $u_{\alpha}$ field.
In contrast with the free-field approximation discussed in Sec.~\ref{s:linear-theory}, 
the theory at finite $\lambda_{0}$ and $\mu_{0}$ is neither conformal nor scale invariant.
The reason is that coupling constants are dimensionful, with dimension $\{\lambda_{0}\} = 
\{\mu_{0}\} = \ve$, and introduce a characteristic length in the problem.
Dilatation symmetry emerges only asymptotically, in the infrared region, when the theory 
becomes controlled by a fixed point.
Adapting a method which was widely used in other field theories~\cite{brown_aop_1980, 
paulos_npb_2016}, we examine this region by expanding $T_{\alpha \alpha}$ on a basis of 
renormalized composite operators, $[(u_{\alpha \alpha})^{2}]$, $[u_{\alpha \beta} 
u_{\alpha \beta}]$, $[u_{\alpha} u_{\beta \beta}]$, and $[u_{\beta} u_{\alpha \beta}]$, 
defined by suitable subtractions in such way that, order by order in perturbation theory, 
their insertion into renormalized correlation functions is ultraviolet-finite (free of 
poles in $\ve$ for $\ve \to 0$).

Detailed derivations, illustrated in appendices~\hyperref[appA]{A} and~\hyperref[appB]{B} 
show that the relation between the bare fields $u_{\alpha \beta}$, $u_{\alpha \alpha}$, 
$(\pa^{2}\bh)^{2}$, $u_{\alpha} u_{\beta \beta}$, $u_{\beta} u_{\alpha \beta}$ and the 
corresponding renormalized operators is almost completely determined by the RG functions 
$\beta_{\lambda}$, $\beta_{\mu}$, $\eta$, and by amplitude and coupling constant 
renormalizations ($Z$, $G_{\lambda}$ and $G_{\mu}$) which can be calculated from 
correlation functions \emph{without} operator insertions.
In particular, we can obtain relations for two distinct types of operators.
A first type is the group of composite fields $\Op_{1} = (\pa^{2}\bh)^{2}/2 + 
\lambda_{0}(u_{\alpha \alpha})^{2} + 2 \mu_{0} u_{\alpha \beta} u_{\alpha \beta}$, 
$\Op_{2} = \lambda_{0} (u_{\alpha \alpha})^{2}/2$, $\Op_{3} = \mu_{0} u_{\alpha \beta} 
u_{\alpha \beta}$, $\Op_{4} = \pa^{2} u_{\alpha \alpha}$, $\Op_{5} = 
\pa_{\alpha}\pa_{\beta} u_{\alpha \beta}$, $\Op_{6} = u_{\alpha \alpha}$, which are 
invariant under all symmetries of the Hamiltonian.
For these operators, the analysis is closely analogue to derivations in 
Ref.~\cite{brown_aop_1980} (see appendix~\hyperref[appA]{A}): we can express, to all 
orders in perturbation theory, the scale-invariance breaking effects in $T_{\alpha 
\alpha}$ in terms of renormalized composite fields multiplied by RG functions.

A second type is constituted by the operators $u_{\alpha} u_{\beta \beta}$ and $u_{\beta} 
u_{\alpha \beta}$, which break the shift symmetry $u_{\alpha} \to u_{\alpha} + B_{\alpha}$ 
and the invariance under the approximate embedding-space rotations defined in 
Eq.~\eqref{deformed-rotations}.
As shown in appendix~\hyperref[appB]{B}, their explicit renormalization relation reads 
(in a non-minimal scheme):
\begin{equation}\label{current-1-renormalization}
u_{\alpha} u_{\beta \beta} = \frac{M^{\ve}(D\lambdar + 2 \mur)}{D \lambda_{0} + 
2\mu_{0}}[u_{\alpha} u_{\beta \beta}] + b_{1}\pa_{\alpha} [\pa_{\beta}\bh\cdot 
\pa_{\beta}\bh] + b_{2} \pa_{\beta}[\pa_{\alpha}\bh \cdot \pa_{\beta}\bh] + b_{3} \pa^{2} 
[u_{\alpha}] + b_{4} \pa_{\alpha}\pa_{\beta}[u_{\beta}]~,
\end{equation}
\begin{equation}\label{current-2-renormalization}
\begin{split}
u_{\beta} u_{\alpha \beta} - \frac{1}{D} u_{\alpha} u_{\beta \beta} & = \frac{M^{\ve} 
\mur}{\mu_{0}} \lt\{[u_{\beta} u_{\alpha \beta}] - \frac{1}{D} [u_{\alpha} u_{\beta 
\beta}]\rt\}\\
& + b'_{1}\pa_{\alpha} [\pa_{\beta}\bh\cdot \pa_{\beta}\bh] + b'_{2} 
\pa_{\beta}[\pa_{\alpha}\bh \cdot \pa_{\beta}\bh] + b'_{3} \pa^{2} [u_{\alpha}] + b'_{4} 
\pa_{\alpha}\pa_{\beta}[u_{\beta}]~.
\end{split}
\end{equation}
where $b_{k}$ and $b'_{k}$, $k=1, .., 4$ are ultraviolet divergent coefficients.
These relations can be interpreted as 'non-renormalizations', in the sense that the 
product of bare couplings with bare operators is equal to the product of renormalized 
couplings and renormalized operators.
Eqs.~\eqref{current-1-renormalization} and~\eqref{current-2-renormalization} are much 
simpler than the general relations expected by symmetry and power counting: counterterms 
with the schematic form $u^{3}$ are absent and mixing of operators 
of the type $u (\pa \bh \cdot \pa \bh)$ and $u \pa u$ is exactly determined in terms of 
the elementary renormalization constants $Z$, $G_{\lambda}$, and $G_{\mu}$.
Although appendix~\hyperref[appB]{B} presents a more complete proof, the particular 
simplicity of the renormalization relations can be directly understood from the structure 
of Feynman rules: in almost any diagram, we can factorize a power of the momentum of each 
external line.
Diagrammatic corrections, therefore, tend to be shift-symmetric even if the inserted 
operators $u_{\alpha}u_{\beta \beta}$ and $u_{\beta} u_{\alpha \alpha}$ are not.
This, in particular, protects the 'diagonal' renormalization (the generation of 
counterterms proportional to the inserted composite fields $u_{\alpha} u_{\beta \beta}$ 
and $u_{\beta} u_{\alpha \beta}$) and implies the simple normalization 
formulas~\eqref{current-1-renormalization} and~\eqref{current-2-renormalization}.
A similar non-renormalization property associated with shift invariance occurs in 
Galileon theories~\cite{goon_jhep_2016}.

For the following analysis, it is also useful to note that the composite operator 
$E_{\alpha} h^{2}$ is not renormalized: $[E_{\alpha} h^{2}] = E_{\alpha}h^{2}$.
In fact, power counting shows that the product $(\bh(\bx) \cdot \bh(\bx))$ at coincident 
points does not generate UV divergences.
As a result $[h^{2}] = \bhr \cdot \bhr = Z^{-1} h^{2}$, where $Z$ is the field-amplitude 
renormalization.
On the other hand, $E_{\alpha}(\bx') h^{2}(\bx')$ is a redundant operator which vanishes 
with equations of motion and acts as the infinitesimal generator of the field redefinition 
$u_{\beta}(\bx) \to u_{\beta}(\bx) - \epsilon \delta_{\alpha \beta} \delta(\bx - \bx') 
h^{2}(\bx)$.
Since $u_{\alpha}$ renormalizes as $\ur_{\alpha} = Z^{-1} u_{\alpha}$, insertion of 
$E_{\alpha}(\bx')h^{2}(\bx')$ can be equivalently represented as the generator of the 
infinitesimal transformation $\ur_{\alpha}(\bx) \to \ur_{\alpha}(\bx) - \epsilon 
\delta_{\alpha \beta} \delta(\bx - \bx') [h^{2}(\bx)]$, which is finite and, thus, does 
not require subtractions.

Collecting results, we obtain the following equivalent expressions for the trace 
$T_{\alpha \alpha}$:
\begin{equation} \label{T-aa-renormalized}
\begin{split}
T_{\alpha \alpha} & = (\ve - \eta) \lt(\frac{1}{2}(\pa^{2} \bh)^{2} + \lambda_{0} 
(u_{\alpha \alpha})^{2} + \mu_{0} u_{\alpha \beta} u_{\alpha \beta}\rt)- E_{\alpha} 
u_{\alpha} + \frac{1}{2} \beta_{\lambda} M^{\ve} \lt[(u_{\alpha \alpha})^{2}\rt] + 
\beta_{\mu} M^{\ve} [u_{\alpha \beta} u_{\alpha \beta}] \\
& + ((D-2) \lambda_{0} + 2 \mu_{0}) \pa_{\alpha}(u_{\alpha} u_{\beta \beta}) - 4 \mu_{0} 
\pa_{\alpha} (u_{\beta} u_{\alpha \beta}) + a_{1} \pa^{2} u_{\alpha \alpha} + a_{2} 
\pa_{\alpha} \pa_{\beta} u_{\alpha \beta}\\
& = -\frac{(\eta - \ve)}{2} \bE \cdot \bh - (1 + \eta - \ve) E_{\alpha} u_{\alpha} + 
\frac{1}{2}\beta_{\lambda} M^{\ve} [(u_{\alpha \alpha})^{2}] + \beta_{\mu} M^{\ve} 
[u_{\alpha \beta} u_{\alpha \beta}] + \pa_{\alpha}V_{\alpha}~,
\end{split}
\end{equation}
with
\begin{equation} \label{Va-unrenormalized}
\begin{split}
V_{\alpha} & = - \frac{1}{4} (\eta - \ve) E_{\alpha} h^{2}  + ((2 - \eta) \lambda_{0} + 2 
\mu_{0}) u_{\alpha} u_{\beta \beta} - 2 (2 +\eta-\ve) \mu_{0} u_{\beta} u_{\alpha \beta} 
\\ 
&  + \frac{1}{2} (\ve - \eta) \pa_{\beta} \lt(- \delta_{\alpha \beta} (\bh \cdot 
\pa^{2} \bh) + \frac{\lambda_{0}}{2} \delta_{\alpha \beta} h^{2} u_{\gamma \gamma} + 
\mu_{0} h^{2} u_{\alpha \beta}\rt)+ a_{1} \pa_{\alpha} u_{\beta \beta} + a_{2} 
\pa_{\beta} u_{\alpha \beta}
\end{split}
\end{equation}
or, after expansion in the basis of renormalized operators $[u_{\alpha} u_{\beta 
\beta}]$, $[u_{\beta} u_{\alpha \beta}]$,
\begin{equation} \label{Va-renormalized}
\begin{split}
V_{\alpha}& = -\frac{1}{4} (\eta - \ve) E_{\alpha} h^{2} +  ((2 - \eta) \lambdar + 2 
\mur) M^{\ve} [u_{\alpha} u_{\beta \beta}] - 2 (2 + \eta - \ve) \mur M^{\ve} [u_{\beta} 
u_{\alpha \beta}] + \pa_{\beta} L_{\alpha \beta}~,\\
L_{\alpha \beta} & = \frac{1}{2}(\ve - \eta) \lt[-\delta_{\alpha \beta} \bh \cdot 
\pa^{2}\bh + \frac{1}{2} \lambda_{0} \delta_{\alpha \beta} h^{2} u_{\gamma \gamma} + 
\mu_{0} h^{2} u_{\alpha \beta}\rt] + b_{1} \delta_{\alpha \beta} (\pa_{\gamma} 
u_{\gamma}) \\ & + b_{2} (\pa_{\alpha} u_{\beta} + \pa_{\beta}u_{\alpha}) + b_{3} 
\delta_{\alpha \beta} (\pa_{\gamma}\bh \cdot \pa_{\gamma}\bh) + b_{4} \pa_{\alpha}\bh 
\cdot \pa_{\beta}\bh~.
\end{split}
\end{equation}
In Eqs.~\eqref{T-aa-renormalized},~\eqref{Va-unrenormalized}, and~\eqref{Va-renormalized}, 
$a_{1}$, $a_{2}$, and $b_{i}$, ($i = 1, .., 4$) are UV-divergent coefficients generated by 
renormalization.

In order to analyze scale and conformal invariance in the asymptotic infrared region, we
assume that all renormalized operators remain finite\footnote{
See Ref.~\cite{paulos_npb_2016} for a related analysis.
} when $\lambdar$ and 
$\mur$ 
approach their fixed point values $\lambdar \to \lambdar_{*}$ and $\mur \to 
\mur_{*}$.
Since $\beta_{\lambda} = \beta_{\mu} = 0$ at the IR fixed point, the scale-invariance 
breaking terms $\beta_{\lambda} M^{\ve} [(u_{\alpha \alpha})^{2}]/2 + \beta_{\mu} M^{\ve} 
[u_{\alpha \beta} u_{\alpha \beta}]$ can be dropped from the expression of $T_{\alpha 
\alpha}$ and the scaling symmetry of the theory, known from RG arguments, becomes 
manifest.
In particular, we can define a dilatation current $S_{\alpha} = x_{\beta} T_{\alpha 
\beta} - V_{\alpha}$~\cite{polchinski_npb_1988} which is locally conserved and presents a 
conservation law
\begin{equation} 
\label{scale-current}
\pa_{\alpha} S_{\alpha} = -x_{\beta} \lt(\bE \cdot \pa_{\beta}\bh + E_{\alpha}\pa_{\beta} 
u_{\alpha}\rt) - \frac{(\eta_{*} - \ve)}{2} \bE \cdot \bh - (1 + \eta_{*} - \ve) 
E_{\alpha} u_{\alpha}~,
\end{equation}
consistent with the form expected for fields of dimension $\Delta_{\bh} = \{\bh\} = 
(\eta_{*} - \ve)/2$ and $\Delta_{u} = \{u_{\alpha}\} = (1 + \eta_{*} 
-\ve)$~\cite{brown_aop_1980, paulos_npb_2016}.
More generally it is possible to show that, for general $\lambdar$ and $\mur$, the Ward 
identity generated by the dilatation current is equivalent to the RG equation (see 
appendix~\hyperref[appA]{A}), similarly to the case of scalar field 
theory~\cite{brown_aop_1980}.

The vanishing of $\beta$ functions, however, is not sufficient to imply the conformal 
invariance of the model due to the presence of the non-zero virial current $V_{\alpha}$.
An algebraic analysis of terms in Eq.~\eqref{Va-unrenormalized} shows that $V_{\alpha}$ 
cannot be written as the total derivative $V_{\alpha} = \pa_{\beta} l_{\alpha \beta}$ of a 
local operator $l_{\alpha \beta}$.
This remains true even in the scale-invariant infrared limit because, as 
Eq.~\eqref{Va-renormalized} shows, contributions proportional to $E_{\alpha} h^{2}$, 
$[u_{\alpha} u_{\beta \beta}]$, $[u_{\beta} u_{\alpha \beta}]$ do not vanish as $\lambdar$ 
and $\mur$ approach their fixed point value.
It follows that it is impossible to construct a conformal current with the 
form~\cite{polchinski_npb_1988}
\begin{equation}
C_{\mu, \alpha} = (2 x_{\mu} x_{\beta} - \delta_{\mu \beta} x^{2}) T_{\alpha \beta} - 2 
x_{\mu} V_{\alpha} + 2 l_{\mu \alpha} 
\end{equation}
and the conservation law
\begin{equation}
\pa_{\alpha} C_{\mu, \alpha} = - (2 x_{\mu}x_{\beta} - \delta_{\mu \beta}x^{2})(\bE 
\cdot \pa_{\beta} \bh + E_{\gamma}\pa_{\beta} u_{\gamma}) + 2x_{\beta} (E_{\beta} u_{\mu} 
- E_{\mu} u_{\beta}) - 2x_{\mu} \lt(\Delta_{\bh} \bE \cdot \bh + \Delta_{u} E_{\alpha} 
u_{\alpha}\rt)
\end{equation}
expected for a scenario in which $\bh$ and $u_{\alpha}$ are conformal primary fields.

It is also impossible to reduce $V_{\alpha}$ to the form $j_{\alpha} + \pa_{\beta} 
l_{\alpha \beta}$ where $j_{\alpha}$ is a conserved current.
If $V_{\alpha} = j_{\alpha} + \pa_{\beta} l_{\alpha \beta}$ was true, the total derivative 
$\pa_{\alpha} V_{\alpha}$ should reduce to a combination $\Op_{\rm red} + \pa_{\alpha} 
\pa_{\beta} l_{\alpha \beta}$, where $\Op_{\rm red}$ is a redundant operator, removable by 
field redefinition.
Working within dimensional regularization, we can assume that the $\Op_{\rm red}$ has the 
form $E_{\alpha} \chi_{\alpha} + \bE \cdot \mathbf{F}$, where $\chi_{\alpha}$ and 
$\mathbf{F}$ are local functionals of the field, and we can neglect contributions arising 
from the Jacobian of the transformation.
The only candidates for $\Op_{\rm red}$ with power-counting dimension $4$ near $D = 4$ 
are then linear combinations of the form $f_{1}(h^{2}) \bE \cdot \bh + f_{2}(h^{2}) 
E_{\alpha} u_{\alpha} + f_{3}(h^{2}) E_{\alpha} (\bh \cdot \pa_{\alpha} \bh)$, where 
$f_{1}(h^{2})$, $f_{2}(h^{2})$, $f_{3}(h^{2})$ are functions of $h^{2}$.
We checked by explicit calculation that $\pa_{\alpha} V_{\alpha}$ cannot be reduced to 
such a combination up to a total second derivative $\pa_{\alpha} \pa_{\beta} l_{\alpha 
\beta}$.

We can thus conclude that the form of the virial current is inconsistent with the 
structure expected in a conformal theory.
Therefore, the theory must exhibit only scale invariance and not the enhanced conformal 
symmetry.
This confirms the result expected from the inconsistency of conformal selection rules 
illustrated in Sec.~\ref{selection-rules} and also excludes the possibility that 
conformal invariance is realized in a more general way, with a transformation law of 
$u_{\alpha}$ differing from that of a primary field.

As a remark, we note that the arguments above rely essentially on the 
'non-renormalization' 
relations~\eqref{current-1-renormalization},~\eqref{current-2-renormalization}, which 
allowed to control contributions to the energy-momentum tensor in the limit $(\lambdar, 
\mur) \to (\lambdar_{*}, \mur_{*})$ via subtracted fields.
In fact, when $\lambdar$ and $\mur$ approach their fixed point values, the bare couplings 
$\lambda_{0}$ and $\mu_{0}$ diverge\footnote{
Since, in absence of a cutoff, the bare couplings are the only scales in the problem, the 
theory can become scale invariant at all wavelengths only if $\lambda_{0}$ and $\mu_{0} 
\to \infty$.
}.
We assume, instead, that subtracted quantities remain finite\footnote{
This is indicated by analogy with the theory of critical phenomena~\cite{zinn-justin_qft, 
parisi_sft}.
We assume that the finiteness of renormalized quantities at the fixed point remains 
valid in the case of composite operators.}.
Differently from the scale-invariance breaking terms, which vanished as 
$\beta_{\lambda}$, $\beta_{\mu} \to 0$, there is no analogue cancellation of 
conformal-breaking terms at the fixed point.

To conclude, we notice that, due to the use of dimensional regularization, the role of 
the 'tension' counterterm $\sigma u_{\alpha\alpha}$ described in 
Sec.~\ref{s:perturbative} remained hidden.
The symmetric energy-momentum tensor corresponding to this term is proportional to $(\pa 
\bh)^{2}$ and thus, breaks the rotational invariance in the embedding space.
The effects of these terms on the relation between scale and conformal invariance can be 
analyzed by generalizing the bare RG equations of Sec.~\ref{s:RG-cutoff} to composite 
operators.

\subsection{Scaling dimension of the virial current}
Having obtained that the membrane theory is not conformal, let us comment on the 
naturalness of having vector operators with dimension exactly equal to $D - 1$.
The absence of anomalous dimensionality is a direct consequence of the 
'non-renormalization' 
relations~\eqref{current-1-renormalization},~\eqref{current-2-renormalization}.
In fact, it can be seen by applying RG equations that both $u_{\alpha} u_{\beta \beta}$ 
and $u_{\beta} u_{\alpha \beta}$ scale at the IR fixed point with the same dimension and 
that the naive dimension $M^{\ve} \{u_{\alpha} u_{\beta \beta}\} = D-1$ remains true in 
the long-wavelength region\footnote{\label{note1}
More rigorously, these terms are not exactly scaling eigenoperators, because they mix 
under renormalization with total derivatives of lower-dimensional fields.
This mixing has a direct connection to a general property of the virial current which, in 
general scales according to a non-canonical current algebra, which 
can include the mixing with total-derivative operators and conserved 
currents~\cite{dymarsky_jhep_2015}.
Here, to simplify the discussion, we describe fields as having dimension $D - 1$ 
meaning that they scale up to total derivatives.
}.
This dimension can also be interpreted as the sum of the infrared dimensions of 
$u_{\alpha}$, which is $1 + \eta_{*} - \ve$ by the RG equations~\eqref{RG-equations}, and 
$u_{\alpha \beta}$ which, as shown in appendix~\hyperref[appA]{A} has dimension $2 - 
\eta_{*}$.
The 'non-renormalization' properties imply that the combination of $u_{\alpha}$ and 
$u_{\alpha \beta}$ into a single operator does not generate any new divergences and, 
therefore, anomalous dimensions.

The existence of non-conserved currents with dimension exactly $D - 1$ can thus be traced 
to the shift-symmetries of the model, which are responsible for the absence of 
renormalizations.

\section{Symmetry argument for the absence of conformal invariance}
\label{s:symmetry}

The derivation in Sec.~\ref{s:T-membrane-theory} suggests an important role of the shift 
symmetries of the model.
In fact, an argument for the absence of conformal invariance can be directly deduced by 
considering the structure of the symmetries.
The Hamiltonian of membrane theory is invariant under translations and rotations of the 
internal coordinates $\bx$ and under embedding-space translations and rotations, which 
are realized as shifts of the $u_{\alpha}$ and $\bh$ fields.
The corresponding generators, written as operators acting on functionals of $\bh$ and 
$u_{\alpha}$, can be written as\footnote{
See also Ref.~\cite{coquand_prb_2019} for a discussion of the symmetries of membrane 
theory.
A detailed analysis of linearly realized symmetries in the biharmonic model and in 
higher-derivative linear theories was given in Ref.~\cite{brust_jhep_2017}.
}
\begin{equation}\label{P}
P_{\alpha} = -i \int {\rm d}^{D}x \lt[\pa_{\alpha} \bh \cdot \frac{\delta}{\delta \bh} + 
\pa_{\alpha}u_{\beta}\frac{\delta}{\delta u_{\beta}}\rt]~,
\end{equation}
\begin{equation}
J_{\alpha \beta} = i \int {\rm d}^{D}x \lt[(x_{\alpha}\pa_{\beta}\bh - 
x_{\beta}\pa_{\alpha}\bh) \cdot \frac{\delta}{\delta \bh} + 
(x_{\alpha}\pa_{\beta}u_{\gamma} - x_{\beta}\pa_{\alpha}u_{\gamma})\frac{\delta}{\delta 
u_{\gamma}} + u_{\beta}\frac{\delta}{\delta u_{\alpha}} - 
u_{\alpha}\frac{\delta}{\delta u_{\beta}}\rt]~,
\end{equation}
\begin{equation}\label{functional-shift}
\bt = -i \int {\rm d}^{D}x  \frac{\delta}{\delta \bh}~, \qquad t_{\alpha} = -i\int 
{\rm d}^{D}x \frac{\delta}{\delta u_{\alpha}}~,
\end{equation}
\begin{equation}\label{functional-rotation-shift}
\bR_{\alpha} = i \int {\rm d}^{D}x \lt[x_{\alpha}\frac{\delta}{\delta \bh} - \bh 
\frac{\delta}{\delta u_{\alpha}}\rt]~, \qquad R_{\alpha \beta} = i \int{\rm d}^{D}x 
\lt[x_{\alpha}\frac{\delta}{\delta u_{\beta}} - x_{\beta}\frac{\delta}{\delta u_{\alpha}} 
\rt]~,
\end{equation}
where bold symbols denote vectors in $d_{c}$-dimensional space.
The generators $P_{\alpha}$ and $J_{\alpha \beta}$ of internal-space transformations 
satisfy the commutation relations of the Euclidean algebra: $[P_{\alpha}, P_{\beta}] = 
0$, $[P_{\gamma}, J_{\alpha \beta}] = i (\delta_{\alpha \gamma}P_{\beta} - \delta_{\beta 
\gamma}P_{\alpha})$, $[J_{\alpha \beta}, J_{\gamma \delta}] = i (\delta_{\alpha \gamma} 
J_{\beta \delta} + \delta_{\beta \delta}J_{\alpha \gamma} - \delta_{\alpha 
\delta}J_{\beta \gamma} - \delta_{\beta \gamma}J_{\alpha \delta})$.
For shift symmetries, we have, instead $[\bt^{i}, \bt^{j}] = [\bt, 
t_{\alpha}] = [\bt, R_{\alpha \beta}] = [t_{\alpha}, t_{\beta}] = [t_{\alpha}, 
\bR_{\beta}] = [t_{\alpha}, R_{\beta \gamma}]= [\bR_{\alpha}, R_{\beta 
\gamma}] = [R_{\alpha \beta}, R_{\gamma \delta}] = 0$, where $\bt^{i}$ denotes the 
$i$ component of the vector $\bt$ in $d_{c}$-dimensional space.
The only nonzero commutators between the embedding-space generators are 
$[\bt^{i}, \bR^{j}_{\alpha}] = i \delta^{ij} t_{\alpha}$ and 
$[\bR^{i}_{\alpha}, \bR^{j}_{\beta}] = i \delta^{ij} R_{\alpha \beta}$.
Mixed commutators between shift generators and internal-space transformations have a 
simple form: in the commutation with $J_{\alpha \beta}$, the generator $\bt$, which has no 
internal-space index, transforms as a scalar, $t_{\alpha}$ and $\bR_{\alpha}$ as vectors, 
and $R_{\alpha \beta}$ as a second-rank tensor.
Commutators between internal translations and shifts read $[P_{\alpha}, \bt] = 
[P_{\alpha}, t_{\beta}] = 0$, $[P_{\alpha}, \bR_{\beta}] = -i \delta_{\alpha \beta}\bt$, 
$[P_{\gamma}, R_{\alpha \beta}] = -i (\delta_{\gamma \alpha}t_{\beta} - 
\delta_{\gamma \beta} t_{\alpha})$.

At the IR fixed point, the theory acquires an additional dilatation symmetry.
We can represent the corresponding generator as 
\begin{equation}\label{dilatation-generator}
D = -i \int {\rm d}^{D}x \lt[\lt(x_{\alpha} \pa_{\alpha}\bh + \Delta_{\bh} \bh \rt) \cdot 
\frac{\delta}{\delta \bh} + \lt(x_{\alpha} \pa_{\alpha}u_{\beta} 
+ \Delta_{u} u_{\beta}\rt)\frac{\delta}{\delta u_{\beta}}\rt]~,
\end{equation}
where $\Delta_{\bh}$ and $\Delta_{u}$ define the scaling dimension of fields.
An analysis of the commutation relations between $D$ and the 
generators~\eqref{P}--\eqref{dilatation-generator} shows that the algebra is not closed 
for general values of $\Delta_{\bh}$ and $\Delta_{u}$.
All commutators are linear combinations of generators a part from one:
\begin{equation}
[D, \bR_{\alpha}] = \int{\rm d}^{D}x \lt[(\Delta_{u} - \Delta_{\bh})\bh 
\frac{\delta}{\delta u_{\alpha}} - (\Delta_{\bh} + 1)x_{\alpha}\frac{\delta}{\delta 
\bh}\rt]~,
\end{equation}
which is not an element of the algebra.
The only way to close the symmetry group without adding new generators is to assume that 
the field dimensions $\Delta_{u}$ and $\Delta_{\bh}$ are related by $\Delta_{u} = 2 
\Delta_{\bh} + 1$, in such way that $[D, \bR_{\alpha}] = i (\Delta_{\bh} + 
1)\bR_{\alpha}$.
This relation, in fact, is satisfied: it is exactly equivalent to the rotational Ward 
identity for scaling exponents~\cite{aronovitz_prl_1988, guitter_jpf_1989, 
gazit_pre_2009} which, in the RG language, arises from the link between 
$\bh$ and $u_{\alpha}$ amplitudes in the renormalization relations $\bh = \sqrt{Z} \bhr$ 
and $u_{\alpha} = Z \ur_{\alpha}$.
In a more conventional notation, $\Delta_{\bh}$ and $\Delta_{u}$ are parametrized by a 
single anomalous dimension as $\Delta_{\bh} = (\eta_{*}-\ve)/2$ and $\Delta_{u} = 1 + 
\eta_{*} - \ve$.

After fixing scaling dimensions as $\Delta_{\bh} = \Delta$ and $\Delta_{u} = 2 
\Delta + 1$, let us suppose that the theory is conformal and that $\bh$ and $u_{\alpha}$ 
are both primary fields.
In this case the symmetry group would contain additional special conformal generators 
whose action can be represented as
\begin{equation}
\begin{split}
K_{\alpha} & = -i \int {\rm d}^{D}x \Big\{\lt[(2 x_{\alpha}x_{\beta} - \delta_{\alpha 
\beta}x^{2}) \pa_{\beta}\bh + 2 \Delta x_{\alpha} \bh\rt] \cdot 
\frac{\delta}{\delta \bh} \\
&+ \lt[(2 x_{\alpha} x_{\beta} - \delta_{\alpha \beta}x^{2}) \pa_{\beta} 
u_{\gamma} + 2 (2\Delta + 1)x_{\alpha} u_{\gamma}\rt]\frac{\delta}{\delta 
u_{\gamma}} + 2 (\delta_{\alpha \gamma}x_{\beta} - \delta_{\alpha 
\beta}x_{\gamma})u_{\beta} \frac{\delta}{\delta u_{\gamma}}\Big\}~.
\end{split}
\end{equation}
The introduction of $K_{\alpha}$, however, breaks the closure of the algebra.
In particular the commutator
\begin{equation}
[K_{\alpha}, \mathbf{t}] = 2\Delta \int {\rm d}^{D}x ~ x_{\alpha} \frac{\delta}{\delta 
\bh}
\end{equation}
requires the introduction of a new generator
\begin{equation}
\bt'_{\alpha} = -i \int {\rm d}^{D}x~ x_{\alpha} \frac{\delta}{\delta \bh}~.
\end{equation}
In turn, $[K_{\alpha}, \bt'_{\beta}]$ requires to add a symmetry under
\begin{equation}
\bt''_{\alpha \beta} = -i \int {\rm d}^{D}x~ \lt[2 (1 + \Delta) x_{\alpha}x_{\beta} - 
\delta_{\alpha \beta}x^{2} \rt] \cdot \frac{\delta}{\delta 
\bh}~.
\end{equation}
For general anomalous dimensionality $\Delta$ the process can be iterated to obtain new 
shift symmetries.
On physical grounds, however, we do not expect these symmetries to hold: shifting $\bh 
\to \bh + \mathbf{B}_{\alpha} x_{\alpha}$ without a compensating shift of $u_{\alpha}$ is 
not a symmetry of the Hamiltonian\footnote{The shift symmetry $\bh \to \bh + 
\mathbf{B}_{\alpha} x_{\alpha}$ is realized, instead, in the 'Gaussian curvature 
interaction' model.} and we do not see reasons why it should emerge in the IR.

This argument indicates, consistently with the analysis in 
Sec.~\ref{s:T-membrane-theory}, 
that $\bh$ can not be interpreted as the primary fields of a conformal field theory.

\section{'Gaussian curvature interaction' model: scale without conformal invariance}

Differently from elasticity theory, the GCI model is exactly conformal in the Gaussian 
approximation, and therefore, in the ultraviolet region.
In fact, the Hamiltonian~\eqref{gci} reduces in the weak-coupling limit $Y_{0} \to 0$ to 
two copies of the biharmonic theory, which is exactly scale and conformal 
invariant~\cite{brust_jhep_2017, nakayama_aop_2016}.
In this section we show that, instead, conformal symmetry is broken in the infrared 
region: the IR fixed point theory is only dilatation-invariant.

With calculations illustrated in appendix~\hyperref[appC]{C} and some further algebraic 
steps, it can be shown that the model admits a symmetric energy-momentum tensor 
$T_{\alpha \beta}$ with trace
\begin{equation}\label{gci-T-renormalized}
\begin{split}
T_{\alpha \alpha} & = -\frac{(\eta-\ve)}{2} \bE\cdot \bh + \eta E \chi - 
\frac{\beta(Y)}{2 Y^{2}}M^{-\ve} [(\pa^{2}\chi)^{2}] + \pa_{\alpha}V_{\alpha}
\end{split}
\end{equation}
where
\begin{equation}
\bE = \pa^{2}\pa^{2} \bh + i (\pa^{2}\chi \pa^{2}\bh - \pa_{\alpha}\pa_{\beta} \chi 
\pa_{\alpha}\pa_{\beta}\bh)~, \qquad E = \frac{1}{Y_{0}} \pa^{2}\pa^{2}\chi + i K
\end{equation}
are, respectively, the equations of motion of the $\bh$ and the $\chi$ field, and 
$[(\pa^{2}\chi)^{2}]$ denotes the renormalized insertion of $(\pa^{2}\chi)^{2}$.
The expression for $T_{\alpha \alpha}$ includes a non-zero 'virial current'
\begin{equation}\label{virial-gci}
\begin{split}
V_{\alpha} & = - \frac{i}{2} \big\{(D-3 + 2 \eta) \pa_{\alpha} \chi (\pa_{\beta} \bh 
\cdot \pa_{\beta}\bh)+ 2(1-\eta)\pa_{\beta} \chi (\pa_{\beta}\bh \cdot \pa_{\alpha}\bh) 
\big\} + \pa_{\beta}L_{\alpha \beta}~,
\end{split}
\end{equation}
where $L_{\alpha \beta}$ is a local tensor field.

At the IR fixed point $Y = Y_{*}$, assuming that the renormalized operator 
$[(\pa^{2}\chi)^{2}]$ remains finite, the term $-\beta(Y) M^{-\ve} [(\pa^{2}\chi)^{2}]/(2 
Y^{2})$ becomes zero due to the vanishing of the $\beta$-function $\beta(Y_{*})=0$.
We can thus introduce a dilatation current $S_{\alpha} = x_{\beta} T_{\alpha \beta} - 
V_{\alpha}$ which is locally conserved:
\begin{equation}
\pa_{\alpha}S_{\alpha} = -x_{\beta} (\bE \cdot \pa_{\beta}\bh + E \pa_{\beta}\chi) - 
\frac{(\eta_{*} - \ve)}{2} \bE \cdot \bh + \eta_{*} E \chi~.
\end{equation}

\begin{figure}[h]
\centering 
\includegraphics[scale=1]{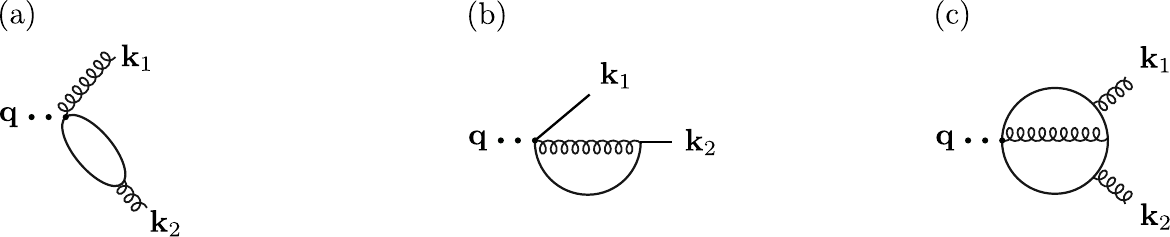}
\caption{\label{P-renormalization} Examples of 1PI diagrams of type (a), (b), and (c).}
\end{figure}

Whether the scaling symmetry is enhanced to the full conformal invariance depends on the 
structure of the virial current.
It is useful, therefore, to examine insertions of the composite field $P_{\mu, \alpha 
\beta} = \pa_{\mu} \chi (\pa_{\alpha}\bh \cdot \pa_{\beta}\bh)$, an elementary building 
block from which the nontrivial terms in Eq.~\eqref{virial-gci} can be constructed.
The renormalization of $P_{\mu, \alpha \beta}$ has a particularly simple form.
In fact, let us consider an arbitrary diagram $\gamma$ for a 1PI correlation function 
with $n$ external $\bh$ lines, $\ell$ external $\chi$ lines, and one insertion of 
$P_{\mu, \alpha \beta}$.
The diagram can be of one of the three types illustrated in Fig.~\ref{P-renormalization}: 
in diagrams of the groups (a) and (b) one of the elementary fields contained in the 
composite operator is directly connected with external lines, while in diagrams of type 
(c) all inserted lines enter as loop propagators.

\noindent 
The Feynman rules of the theory imply that the degree of superficial 
divergence~\cite{zinn-justin_qft} is 
\begin{equation} \label{gci-power-counting}
\delta(\gamma) = 3 + D L - 4 I_{h} - 4 I_{\chi} + 4 v - 2 n - 2 \ell + a_{\gamma}~,
\end{equation}
where $I_{\bh}$ and $I_{\chi}$ denote the number of internal $\bh$ and $\chi$ propagators, 
$v$ the number of vertices, and $L$ the number of loops.
The coefficient $a_{\gamma}$ is $a_{\gamma} = 1$ for diagrams of type (a) and (b) and 
$a_{\gamma}= 0$ for type (c).
Using the topological relations $L = 3 + 2 v - I_{\bh} - I_{\chi} - n- \ell$, $2 I_{\bh} 
+ n = 2 v + 2$, and $2 I_{\chi} + \ell = v + 1$, we see that the degree of divergence in 
the $\ve$ expansion is
\begin{equation}
\delta(\gamma) = 3 -2 n - 2 \ell + a_{\gamma}~.
\end{equation}
It follows that the only counterterms needed for the renormalization of $P_{\mu, \alpha 
\beta}$ have the schematic form $\pa^{3} \chi$, $\pa \chi \pa^{2}\chi$, $\pa \bh 
\cdot \pa^{2}\bh$.
These composite operators can always be represented as total derivatives (see 
Eq.~\eqref{total-derivative}).

We can conclude that insertions of the composite fields $\pa_{\alpha}\chi (\pa_{\beta}\bh 
\cdot \pa_{\beta}\bh)$ and $\pa_{\beta}\chi (\pa_{\alpha}\bh \cdot \pa_{\beta}\bh)$, which 
contribute to the virial current, are finite up to 
total-derivative counterterms.
Therefore, the ``bulk'' of the virial current is unrenormalized: we can set 
$\pa_{\alpha}\chi_{\beta} (\pa_{\beta}\bh\cdot \pa_{\beta}\bh) = 
[\pa_{\alpha}\chi_{\beta} (\pa_{\beta} \bh \cdot \pa_{\beta}\bh)]$ and $\pa_{\beta}\chi 
(\pa_{\alpha}\bh \cdot \pa_{\beta}\bh) = [\pa_{\beta}\chi (\pa_{\alpha}\bh \cdot 
\pa_{\beta}\bh)]$, up to gradients of the form $\pa_{\beta} l_{\alpha \beta}$ which 
 do not affect the relation between scale and conformal 
invariance~\cite{polchinski_npb_1988}.

Let us check that $V_{\alpha}$ cannot be reduced completely to the combination 
$V_{\alpha} = j_{\alpha} + \pa_{\beta}L'_{\alpha \beta}$ of a conserved current 
$j_{\alpha}$ and a total derivative.
If this was the case, $\pa_{\alpha} V_{\alpha}$ should reduce to the combination 
$\Op_{\rm red} + \pa_{\alpha} \pa_{\beta} l_{\alpha \beta}$ of a redundant operator 
$\Op_{\rm red}$ and a total second derivative.
Within dimensional regularization, candidates for $\Op_{\rm red}$ can be taken as 
linear combinations of operators proportional to the equations of motion $\bE$ and $E$ 
and, in order to match the power-counting dimension of $\pa_{\alpha} V_{\alpha}$, must 
have the form $f_{1}(h^{2}, \chi) (\bE \cdot \bh) + f_{2}(h^{2}, \chi) E$, where $f_{1}$ 
and $f_{2}$ are functions.
We checked from the explicit expression $V_{\alpha}$ that it is impossible to rewrite 
$\pa_{\alpha} V_{\alpha}$ as a combination of this type up to a total second derivative 
$\pa_{\alpha} \pa_{\beta} l_{\alpha \beta}$.
Since contributions to $V_{\alpha}$ do not renormalize, we expect that this result 
remains robust at the IR fixed point.
We are lead to the conclusion that the GCI model exhibits scale without conformal 
invariance.

Let us, then, investigate the scaling properties of the operators composing $V_{\alpha}$.
Since $P_{\mu, \alpha \beta}$ is not renormalized, it does not acquire anomalous 
exponents.
Therefore the naive dimension $\{P_{\mu, \alpha \beta}\} = 3 + 2\{\bh\} + \{\chi\} = 3 + 
2(D-4)/2 + 0 = D-1$ remains valid at the IR fixed point\textsuperscript{\ref{note1}}.
This scaling relation can also be understood in terms of the infrared dimensions of 
fields.
The renormalization relations discussed in Sec.~\ref{s:GCI}, $\bh = \sqrt{Z} \bh$, $\chi 
= Z^{-1} \chir$, imply that $\bh$ and $\chi$ scale in the long-wavelength limit with 
dimensionalities $\{\bh\} = (D - 4 + \eta)/2$ and $\{\chi\} = -\eta$.
The absence of divergences in the insertion of $P_{\mu, \alpha \beta}$ implies that the 
naive relation $\{V_{\alpha}\} = 3 + 2\{\bh\} + \{\chi\}$ remains valid in the IR and, in 
fact, it can be seen that the anomalous exponent $\eta$ cancels out leaving an exact 
canonical dimension.

The absence of nontrivial anomalous dimensions can be traced, as in the case of membrane 
theory, to the shift symmetries of the model.
These symmetries are manifested in momentum space as a special property of Feynman rules: 
for each external line connected to interaction vertices, it is always possible to 
factorize two powers of the corresponding momentum.
The result is a suppression of the degree of UV divergence~\cite{mauri_npb_2020, 
le-doussal_aop_2018}, which, in the power counting formula~\eqref{gci-power-counting} is 
expressed by the terms $-2 n - 2\ell$.
This explains why candidates for the virial current, which must have dimension $D-1$, 
arise naturally.

\section{Summary and conclusions}
To summarize, we analyzed two models for the scaling behavior of fluctuations in 
crystalline membranes: a widely-studied effective field theory based on elasticity and an 
alternative model, involving only scalar fields, which describes long-range 
phonon-mediated interactions between local Gaussian curvatures.
For both models, we argued that the infrared behavior is only scale-invariant: the 
asymptotic dilatation symmetry is not promoted to conformal invariance.
An analysis of the energy-momentum tensor of the two theories reveals, in both cases, the 
presence of non-trivial virial currents which, despite being non-conserved, maintain a 
scaling dimension equal to $D -1$, without corrections from interactions.
We traced the origin of this non-renormalization to the shift symmetries of the theory, 
which forbid the generation of several counterterms which would be allowed by a first 
power-counting analysis.
These results suggest a mechanism to elude a general reasoning according to which 
non-conserved currents with dimension $D-1$ are unlikely at generic interacting fixed 
points and thus, that conformal invariance should be an almost inevitable 
consequence of scale invariance in presence of interactions.
As a complementary analysis, in the case of the nonlinear elasticity theory of membranes, 
we present a simple argument, based only on the structure of symmetries, which suggests 
an inconsistency between conformal invariance and the invariance of the model under 
shifts.
The results derived in this paper are not in contradiction with general theorems and 
derivations on the relation between scale and conformal symmetries for two reasons.
First, we expect that the models investigated in this work are not reflection-positive.
Secondly, we studied fixed points in $D = 4 - \ve$, a dimension in which, to our 
knowledge, the connection between scaling and conformality is not yet firmly established.

\section*{Acknowledgements}
We thank S. Rychkov for stimulating discussions.
This work was supported by the Netherlands Organisation for Scientific Research (NWO) via 
the Spinoza Prize.

\appendix 
\section{Invariant composite operators in membrane theory}
\label{appA}

This appendix illustrates the renormalization of operators entering the expansion of the 
trace of the energy-momentum tensor.
Let us start by analyzing the set of composite fields
\begin{equation} \label{rotational-invariant-operators}
\begin{split}
& \Op_{1} = \frac{1}{2}(\pa^{2}\bh)^{2} + \lambda_{0} (u_{\alpha \alpha})^{2} + 2\mu_{0} 
u_{\alpha \beta} u_{\alpha \beta}~, \qquad \Op_{2} = \frac{1}{2}\lambda_{0} 
(u_{\alpha\alpha})^{2}~, 
\qquad \Op_{3} = \mu_{0}u_{\alpha \beta} u_{\alpha \beta}~, \\
& \Op_{4} = \pa^{2} u_{\alpha \alpha}~,  \qquad \Op_{5} = \pa_{\alpha}\pa_{\beta} 
u_{\alpha \beta}~, \qquad \Op_{6} = u_{\alpha \alpha}~,
\end{split}
\end{equation}
which are invariant under all symmetries of the theory, including translations in the 
embedding space $\bh \to \bh + \mathbf{B}$, $u_{\alpha} \to u_{\alpha} + B_{\alpha}$, and 
the linearized rotations in Eq.~\eqref{deformed-rotations}.
According to general renormalization theory~\cite{zinn-justin_qft}, the insertion of 
invariant operators of power-counting dimension $\Delta$ is renormalized by a linear 
combination of operators with the same symmetries and with dimension equal or lower to 
$\Delta$.
From the scaling of $\bh$ and $u_{\alpha}$ tree-level propagator, it follows that the 
power-counting dimension of a general operator of the schematic form $\pa^{k} \bh^{n} 
u^{\ell}$ is $k + n(D-4)/2 + \ell (D-2)/2$, which reduces to $k + \ell$ in the 
$\ve$-expansion at $D = 4 - \ve$.
The composite fields in Eq.~\eqref{rotational-invariant-operators} are a basis for the 
most general invariant operator with dimension $\leq 4$ and are, therefore, closed under 
renormalization.
It is possible to find a matrix $Z_{ij}$ of divergent coefficients such that bare and 
finite, renormalized operators, are related as $\Op_{i}(\bx) = Z_{ij} [\Op_{j}(\bx)]$.

In analogy with derivations in Ref.~\cite{brown_aop_1980}, it is possible to set strong 
constraints on renormalization by forming combinations which are a priori known to be 
finite and free of UV divergences.

The renormalization of $\Op_{1}$ can be fixed by the following argument.
The expression for a general correlation function $G^{(n, \ell)}$ in terms of a 
functional integral over $u_{\alpha}$ and $\bh$,
\begin{equation}
G^{(n, \ell)}_{i_{1}..i_{n} \alpha_{1}..\alpha_{\ell}}(\bx_{1}, .., \bx_{n}; \bx_{1}', 
.., \bx'_{\ell}) = \int [{\rm d}\bh] [{\rm d}u_{\alpha}] \lt\{{\rm e}^{-{\cal H}} 
h_{i_{1}}(\bx_{1})..h_{i_{n}}(\bx_{n}) u_{\alpha_{1}}(\bx_{1}') .. 
u_{\alpha_{\ell}}(\bx'_{\ell}) \rt\}~,
\end{equation}
must be invariant under change of variables.
If we choose a field redefinition $\bh \to (1 + \epsilon/2)\bh$, $u_{\alpha} \to (1 + 
\epsilon)u_{\alpha}$ the Hamiltonian ${\cal H}$ changes to first order by $\epsilon \int 
{\rm d}^{D}x \Op_{1}(\bx)$ while the string of fields in the correlator varies by an 
overall factor $(n/2 + \ell) \epsilon$.
Invariance of the functional integral then implies
\begin{equation} \label{O1-ward-identity}
\int {\rm d}^{D}x G^{(n, \ell)}_{\Op_{1}(\bx), i_{1}..i_{n} 
\alpha_{1}..\alpha_{\ell}}(\bx_{1}, .., \bx_{n}; \bx_{1}', .., \bx'_{\ell}) = 
\lt(\frac{n}{2} + \ell\rt) G^{(n, \ell)}_{i_{1}..i_{n} 
\alpha_{1}..\alpha_{\ell}}(\bx_{1}, .., \bx_{n}; \bx_{1}', .., \bx'_{\ell})~,
\end{equation}
where
\begin{equation}
G^{(n, \ell)}_{\Op_{1}(\bx), i_{1}..i_{n} \alpha_{1}..\alpha_{\ell}}(\bx_{1}, .., \bx_{n}; 
\bx_{1}', .., \bx'_{\ell}) = \int [{\rm d}\bh] [{\rm d}u_{\alpha}] \lt\{{\rm e}^{-{\cal 
H}} \Op_{1}(\bx) h_{i_{1}}(\bx_{1})..h_{i_{n}}(\bx_{n}) u_{\alpha_{1}}(\bx_{1}') .. 
u_{\alpha_{\ell}}(\bx'_{\ell}) \rt\}
\end{equation}
denotes correlation functions with $\Op_{1}(\bx)$ insertion.
From Eq.~\eqref{O1-ward-identity}, we see that $\int {\rm d}^{D}x G^{(n, 
\ell)}_{\Op_{1}(\bx)}$ is already finite after the renormalization of elementary fields, 
$\bh = \sqrt{Z} \bhr$, $u_{\alpha} = Z \ur_{\alpha}$, without the need of a new operator 
renormalization.
The only divergences in $\Op_{1}$ must be total derivatives, which vanish after space 
integration.
We thus conclude that $\Op_{1}$ can be renormalized as
\begin{equation}\label{renormalization-O1}
\Op_{1} = [\Op_{1}] + a_{1} \pa^{2}[u_{\alpha\alpha}] + b_{1} \pa_{\alpha}\pa_{\beta} 
[u_{\alpha \beta}]~,
\end{equation}
where $a_{1}$ and $b_{1}$ are divergent coefficients.

We can deduce additional constraints from the fact that derivatives of renormalized 
correlation functions with respect to $\lambdar$ and $\mur$ are 
finite~\cite{brown_aop_1980}.
Denoting as $G^{(n, \ell)}$ and $\tilde{G}^{(n, \ell)}$ bare and renormalized correlation 
functions with $n$ external $\bh$ fields and $\ell$ external $u_{\alpha}$ fields, we 
find, using Eq.~\eqref{renormalization-amplitudes-MS},
\begin{equation} \label{finiteness-relation}
\begin{split}
&\frac{\pa  \tilde{G}^{(n, \ell)}}{\pa \lambdar} = \frac{\pa}{\pa \lambdar} 
\lt(Z^{-\frac{n}{2} - \ell} G^{(n, \ell)}\rt) \\
& = -\lt(\frac{n}{2} + \ell\rt) \frac{\pa \ln Z}{\pa \lambdar} \tilde{G}^{(n, \ell)} + 
Z^{-\frac{n}{2} - \ell} \lt[\frac{\pa \ln \lambda_{0}}{\pa \lambdar}\Big|_{M, \mur} 
\frac{\pa }{\pa \ln \lambda_{0}} + \frac{\pa \ln \mu_{0}}{\pa \lambdar}\Big|_{M, \mur} 
\frac{\pa}{\pa \ln \mu_{0}}\rt] G^{(n, \ell)} = \text{finite}~.
\end{split}
\end{equation}
The derivatives $\pa/\pa \ln \lambda_{0}$ and $\pa/\pa \ln \mu_{0}$ generate, 
respectively, insertions of $-\int {\rm d}^{D}x \Op_{2}(\bx)$ and $-\int {\rm d}^{D}x 
\Op_{3}(\bx)$.
Moreover, as shown above, the counting factor $n/2 + \ell$ can be written via the 
insertion of $\int {\rm d}^{D}x \Op_{1}$.

As a result, Eq.~\eqref{finiteness-relation} is equivalent to
\begin{equation}\label{finiteness-relation-insertions}
\int {\rm d}^{D}x \lt[\frac{\pa \ln Z}{\pa \lambdar} \tilde{G}^{(n, \ell)}_{\Op_{1}(\bx)} 
+ \frac{\pa \ln \lambda_{0}}{\pa \lambdar}\Big|_{M, \mur} \tilde{G}^{(n, 
\ell)}_{\Op_{2}(\bx)} + \frac{\pa \ln \mu_{0}}{\pa \lambdar}\Big|_{M, \mur} 
\tilde{G}^{(n, \ell)}_{\Op_{3}(\bx)}\rt] = \text{finite}~,
\end{equation}
where $\tilde{G}^{(n, \ell)}_{\Op(\bx)}$ denotes correlation functions of renormalized 
fields with an insertion of the bare operator $\Op(\bx)$:
\begin{equation}
\tilde{G}^{(n, \ell)}_{\Op(\bx)} = \langle \Op(\bx) \hr_{i_{1}}(\bx_{1}) .. 
\hr_{i_{n}}(\bx_{n}) \ur_{\alpha_{1}}(\bx'_{1}) ..\ur_{\alpha_{\ell}} 
(\bx'_{\ell})\rangle~.
\end{equation}
Isolating operators from correlation functions and removing space integration, we can 
re-express Eq.~\eqref{finiteness-relation-insertions} as the statement that the 
combination
\begin{equation} \label{lambda-finiteness}
\frac{\pa \ln Z}{\pa \lambdar}\Big|_{M, \mur} \Op_{1}(\bx) + \frac{\pa \ln 
\lambda_{0}}{\pa \lambdar}\Big|_{M, \mur} \Op_{2}(\bx) + \frac{\pa \ln \mu_{0}}{\pa 
\lambdar}\Big|_{M, \mur} \Op_{3}(\bx)
\end{equation}
is finite up to total derivatives.
Assuming that amplitude, coupling, and operator renormalizations are all defined within 
the minimal subtraction scheme~\cite{zinn-justin_qft, brown_aop_1980}, this implies
\begin{equation} \label{lambda-MS}
\frac{\pa \ln Z}{\pa \lambdar}\Big|_{M, \mur} \Op_{1}(\bx) + \frac{\pa \ln 
\lambda_{0}}{\pa \lambdar}\Big|_{M, \mur} \Op_{2}(\bx) + \frac{\pa \ln \mu_{0}}{\pa 
\lambdar}\Big|_{M, \mur} \Op_{3}(\bx) = \frac{1}{\lambdar} [\Op_{2}(\bx)] + a_{\lambda} 
\pa^{2} [u_{\alpha\alpha}] + b_{\lambda} \pa_{\alpha} \pa_{\beta}[u_{\alpha \beta}]~,
\end{equation}
so that, up to the total-derivative terms, the right-hand side is equal to the 
tree-level contribution of the left hand side.
A consequence of Eq.~\eqref{lambda-MS} is that
\begin{equation}\label{derivative-renormalized-lambdar}
\frac{\pa \tilde{G}^{(n, \ell)}}{\pa \ln \lambdar}\Big|_{M, \mur} = -\int {\rm d}^{D}x 
\tilde{G}^{(n, \ell)}_{[\Op_{2}(\bx)]}~,
\end{equation}
where $\tilde{G}^{(n, \ell)}_{[\Op_{2}(\bx)]}$ is the correlation function of 
renormalized fields with insertion of the renormalized operator $[\Op_{2}(\bx)]$.
An analogue  relation was derived for scalar field theory in Ref.~\cite{brown_aop_1980}.

Identical arguments can be used to deduce that
\begin{equation} \label{mu-MS}
\frac{\pa \ln Z}{\pa \mur}\Big|_{M, \lambdar} \Op_{1}(\bx) + \frac{\pa \ln 
\lambda_{0}}{\pa \mur}\Big|_{M, \lambdar} \Op_{2}(\bx) + \frac{\pa \ln \mu_{0}}{\pa 
\mur}\Big|_{M, \lambdar} \Op_{3}(\bx) = \frac{1}{\mur} [\Op_{3}(\bx)] + a_{\mu} \pa^{2} 
[u_{\alpha\alpha}] + b_{\mu} \pa_{\alpha} \pa_{\beta}[u_{\alpha \beta}]~,
\end{equation}
a relation which follows from the finiteness of $\pa \tilde{G}^{(n, \ell)}/\pa \mur$.
A relation similar to Eq.~\eqref{derivative-renormalized-lambdar} holds:
\begin{equation}\label{derivative-renormalized-mur}
\frac{\pa \tilde{G}^{(n, \ell)}}{\pa \ln \mur}\Big|_{M, \lambdar} = -\int {\rm d}^{D}x 
\tilde{G}^{(n, \ell)}_{[\Op_{3}(\bx)]}~.
\end{equation}
As a particular case of Eqs.~\eqref{lambda-MS} and~\eqref{mu-MS}, let us take the linear 
combination $\beta_{\lambda}\times$\eqref{lambda-MS}$+\beta_{\mu}\times$\eqref{mu-MS}, 
where $\beta_{\lambda}$ and $\beta_{\mu}$ are the RG $\beta$-functions.
Using that~\cite{guitter_jpf_1989}
\begin{equation}
\beta_{\lambda} \frac{\pa \ln Z}{\pa \lambdar}\Big|_{M, \mur} + \beta_{\mu} \frac{\pa \ln 
Z}{\pa \mur}\Big|_{M, \lambdar} = \frac{\pa \ln Z}{\pa \ln M}\Big|_{\lambda_{0}, \mu_{0}} 
= \eta~,
\end{equation}
\begin{equation}
\beta_{\lambda} \frac{\pa \ln \lambda_{0}}{\pa \lambdar}\Big|_{M, \mur} + \beta_{\mu} 
\frac{\pa \ln \lambda_{0}}{\pa \mur}\Big|_{M, \lambdar}  = \frac{\pa \ln 
(\lambda_{0}/M^{\ve})}{\pa \ln M}\Big|_{\lambda_{0}, \mu_{0}} = -\ve~,
\end{equation}
and
\begin{equation}
\beta_{\lambda} \frac{\pa \ln \mu_{0}}{\pa \lambdar}\Big|_{M, \mur} + \beta_{\mu} 
\frac{\pa \ln \mu_{0}}{\pa \mur}\Big|_{M, \lambdar}  = \frac{\pa \ln 
(\mu_{0}/M^{\ve})}{\pa \ln M}\Big|_{\lambda_{0}, \mu_{0}} = -\ve~,
\end{equation}
we find
\begin{equation} \label{operators-beta-functions}
\ve \Op_{2} + \ve \Op_{3} = \eta \Op_{1} - \beta_{\lambda}/\lambdar [\Op_{2}] - 
\beta_{\mu}/\mur [\Op_{3}] + a \pa^{2} [u_{\alpha \alpha}] + b\pa_{\alpha}\pa_{\beta} 
[u_{\alpha \alpha}]
\end{equation}
with divergent coefficients $a$ and $b$.
This relation can be rewritten in a more explicit notation by setting $[\Op_{2}] = 
M^{\ve} \lambdar [(u_{\alpha\alpha})^{2}]/2$, $[\Op_{3}] = M^{\ve} \mur [u_{\alpha \beta} 
u_{\alpha \beta}]$.
In this basis, Eq.~\eqref{operators-beta-functions} becomes
\begin{equation}
\frac{\ve}{2} \lt(\lambda_{0} (u_{\alpha \alpha})^{2} + 2 \mu_{0} u_{\alpha \beta} 
u_{\alpha \beta} \rt) = \eta \Op_{1} - \frac{1}{2}\beta_{\lambda} M^{\ve} [(u_{\alpha 
\alpha})^{2}] - \beta_{\mu} M^{\ve} [u_{\alpha \beta} u_{\alpha \beta}] + a \pa^{2} 
[u_{\alpha \alpha}] + b\pa_{\alpha}\pa_{\beta} [u_{\alpha \alpha}]~.
\end{equation}

As a final remark, we note that Eqs.~\eqref{renormalization-O1},~\eqref{lambda-MS}, 
and~\eqref{mu-MS} imply that the operator $\Op_{6} = u_{\alpha \alpha}$ does not enter 
the renormalization of $\Op_{1}$, $\Op_{2}$, and $\Op_{3}$.
This is due to the use of dimensional regularization, implicit in the derivations above.
This regularization scheme automatically removes ultraviolet divergences of power-law 
type, implying that operators of dimension 4 do not mix under renormalization with 
operators of dimension $2$.

With results derived above, it is possible to show that the Ward identity for broken 
dilatation invariance is equivalent to the RG equations~\eqref{RG-equations}. (
An analogue result was derived for scalar field theory in Ref.~\cite{brown_aop_1980, 
paulos_npb_2016}).
Away from fixed points, the dilatation current $S_{\alpha} = x_{\beta} T_{\alpha \beta} - 
V_{\alpha}$ is not conserved: the RG flow functions $\beta_{\lambda}$ and $\beta_{\mu}$ 
act as sources for the violation of the conservation law of $S_{\alpha}$
\begin{equation}
\pa_{\alpha} S_{\alpha} = \frac{\beta_{\lambda}}{2}  M^{\ve} [(u_{\alpha \alpha})^{2}] + 
\beta_{\mu} M^{\ve} [u_{\alpha \beta} u_{\alpha \beta}] -x_{\beta} \lt(\bE \cdot 
\pa_{\beta}\bh + E_{\alpha}\pa_{\beta} u_{\alpha}\rt) - \frac{(\eta - \ve)}{2} \bE \cdot 
\bh - (1 + \eta - \ve) E_{\alpha} u_{\alpha}~.
\end{equation}
Renormalized correlation functions with insertions of $\pa_{\alpha}S_{\alpha}(\bx)$, 
which are relevant for the Ward identity, can be expressed more explicitly by using that 
the operators $\bE \cdot \pa_{\beta} \bh$, $E_{\alpha} \pa_{\beta}u_{\alpha}$, $\bE \cdot 
\bh$, $E_{\alpha} u_{\alpha}$, proportional to equations of motion, generate the contact 
terms~\cite{brown_aop_1980, paulos_npb_2016, de-polsi_jsp_2019}
\begin{equation} \label{contact-terms}
\begin{split}
\tilde{G}^{(n, \ell)}_{\bE(\bx) \cdot \bh(\bx)}(\bx_{1}, .., \bx_{n}; \bx'_{1}, .., 
\bx'_{\ell}) & = \sum_{p=1}^{n} \tilde{G}^{(n, \ell)}\delta(\bx - \bx_{p})(\bx_{m}, 
\bx'_{k})~,\\
\tilde{G}^{(n, \ell)}_{E_{\alpha}(\bx)u_{\alpha}(\bx)}(\bx_{m}, \bx'_{k}) & = 
\sum_{p=1}^{\ell} \delta(\bx - \bx'_{p})\tilde{G}^{(n, \ell)}(\bx_{m}, \bx'_{k})~,\\
\tilde{G}^{(n, \ell)}_{\bE(\bx) \cdot \pa_{\beta} \bh(\bx)}(\bx_{m}, \bx'_{k}) & = 
\sum_{p=1}^{n} \delta(\bx - \bx_{p}) \frac{\pa}{\pa x_{p\beta}}\tilde{G}^{(n, 
\ell)}(\bx_{m}, \bx'_{k})~,\\
\tilde{G}^{(n, \ell)}_{E_{\alpha}(\bx) \pa_{\beta} u_{\alpha}(\bx)}(\bx_{m}, \bx'_{k}) & 
= \sum_{p=1}^{\ell} \delta(\bx - \bx'_{p}) \frac{\pa}{\pa x'_{p \beta}}\tilde{G}^{(n, 
\ell)}(\bx_{m}, \bx'_{k})~,\\
\end{split}
\end{equation}
Using Eqs.~\eqref{derivative-renormalized-lambdar} 
and~\eqref{derivative-renormalized-mur}, and integrating over space, we obtain
\begin{equation}
\begin{split}
 \int {\rm d}^{D}x ~ \pa_{\alpha}\tilde{G}^{(n, \ell)}_{S_{\alpha}(\bx)} & 
=\Big[\frac{\pa}{\pa \ln \rho} + \beta_{\lambda} \frac{\pa}{\pa \lambdar}\Big|_{M, \mur} 
+ \beta_{\mu} \frac{\pa}{\pa \mur}\Big|_{M, \lambdar} + \frac{n}{2} (\eta - \ve) \\
& + \ell (1 + \eta - \ve)\Big] \tilde{G}^{(n, \ell)}(\rho \bx_{1}, .., \rho \bx_{n}; \rho 
\bx'_{1}, .., \rho 
\bx'_{\ell}) = 0
\end{split}
\end{equation}
a relation equivalent to the RG flow equation~\eqref{RG-equations}.

For completeness, we also discuss the composite field $u_{\alpha \beta}$. 
By symmetries and power counting its renormalization has the form
\begin{equation}\label{Z2}
u_{\alpha \beta} = Z_{2} [u_{\alpha \beta}] + \frac{1}{D}(Z_{2}' - Z_{2}) \delta_{\alpha 
\beta} 
[u_{\gamma \gamma}]~,
\end{equation}
where $Z_{2}$ and $Z'_{2}$ are divergent coefficients.
The factors $Z_{2}$ and $Z_{2}'$, moreover,are determined to all orders by the following 
argument.
Let us consider the stress field $\sigma_{\alpha \beta} = \lambda_{0} \delta_{\alpha 
\beta} u_{\gamma \gamma} + 2 \mu_{0} u_{\alpha \beta}$.
This composite operator can be viewed as the conserved current associated with the shift 
symmetry $u_{\alpha} \to u_{\alpha} + B_{\alpha}$ and it has a conservation law 
 $\pa_{\beta} \sigma_{\alpha \beta} = - E_{\alpha}$ which is identical, up to a 
sign, to the equations of motion of the $u_{\alpha}$ field.
By a general property, the renormalization of the equation of motion operator is dual to 
that of the corresponding field: since $u_{\alpha}$ renormalizes as $u_{\alpha} = Z 
\ur_{\alpha}$, then $Z E_{\alpha}$ is a finite operator.
It follows, as a result, that $Z(\pa_{\beta} \sigma_{\alpha \beta})$ is finite.
However, this also implies that $Z \sigma_{\alpha \beta}$ is finite by itself, because 
any divergence in $Z \sigma_{\alpha \beta}$ would inevitably appear in the derivative.
To see this more precisely, note that the infinite part $Z\sigma^{\rm div}_{\alpha 
\beta}$ of $Z \sigma_{\alpha \beta}$, if any, should be a linear combination of $u_{\alpha 
\beta}$ and $\delta_{\alpha \beta} u_{\gamma \gamma}$ satisfying the equation 
$\pa_{\alpha} (Z \sigma^{\rm div}_{\alpha \beta})=0$ identically.
It can be checked that the only possibility is $Z \sigma^{\rm div}_{\alpha \beta} = 0$ 
and, therefore, that the full tensor $Z \sigma_{\alpha \beta}$ is finite.
Using Eq.~\eqref{Z2} and Eq.~\eqref{renormalization-amplitudes-MS}, we see that the 
combinations of renormalization constants 
\begin{equation}
\frac{G_{\mu} Z_{2}}{Z}~, \qquad \frac{(G_{\lambda} + 2 G_{\mu}/D)}{Z}Z_{2}'
\end{equation}
are free of poles in $\ve$.
This implies that we can choose
\begin{equation}
Z_{2} = \frac{Z \mur}{G_{\mu}}~, \qquad Z_{2}' = \frac{Z (D \lambdar + 2 \mur)}{D 
G_{\lambda} + 2 G_{\mu}}~.
\end{equation}
The scaling dimensions of the scalar and traceless components of $u_{\alpha \beta}$ are 
then $\{u_{\alpha \alpha}\} = D - 2 + \eta_{2}'$ and $\{u_{\alpha \beta} - 
\delta_{\alpha \beta} u_{\gamma \gamma}/D\} = D - 2 + \eta_{2}$, where $\eta_{2}' = \pa 
\ln Z_{2}'/\pa \ln M|_{\lambda_{0}, \mu_{0}}$ and $\eta_{2} = \pa \ln Z_{2}/\pa \ln 
M|_{\lambda_{0}, \mu_{0}}$.
At the fixed point P$_{4}$ all components scale with the same dimension $\{u_{\alpha 
\beta}\} = 2 - \eta_{*}$.

\section{Renormalization of non-invariant currents}
\label{appB}

Besides invariant operators, expansion of the trace $T_{\alpha \alpha}$ includes the 
vector fields $u_{\alpha} u_{\beta \beta}$ and $u_{\beta} u_{\alpha \beta}$, which break 
the shift symmetry $u_{\alpha} \to u_{\alpha} + B_{\alpha}$ and the linearized 
embedding-space rotational symmetry.
This appendix shows that these vectors are non-renormalized, up to total derivatives.

As a first step, it is convenient to analyze the tensor $J_{\gamma, \alpha \beta} = 
u_{\gamma} \sigma_{\alpha \beta}$, where $\sigma_{\alpha \beta} = \lambda_{0} 
\delta_{\alpha \beta} u_{\delta \delta} + 2 \mu_{0} u_{\alpha \beta}$ is the stress 
field, which is also the conserved current associated with the symmetry under $u_{\alpha} 
\to u_{\alpha} + B_{\alpha}$.
A priori, the renormalization of $u_{\gamma} \sigma_{\alpha \beta}$ involves the mixing 
of all composite fields of dimension $3$ symmetric under $\alpha \leftrightarrow \beta$ 
and invariant under $\bh \to \bh + \mathbf{B}$.
(In dimensional regularization there is no mixing with operators of lower dimension).
Renormalization is however simplified by the following considerations.
Taking the derivative $\pa_{\alpha}J_{\gamma, \alpha \beta} = - E_{\beta}u_{\gamma} + 
\pa_{\alpha} u_{\gamma} \sigma_{\alpha \beta}$ gives the sum of two simple terms.
The first, $-E_{\beta}(\bx) u_{\gamma}(\bx)$, vanishes with equations of motion and acts, 
when inserted in a correlation function, as the generator of the infinitesimal field 
redefinition $u_{\alpha}(\bx') \to u_{\alpha}(\bx') + \epsilon \delta_{\alpha \beta} 
\delta(\bx - \bx') u_{\gamma}(\bx)$.
This transformation, being linear, can be equivalently represented in terms of 
renormalized fields as $\ur_{\alpha}(\bx') \to \ur_{\alpha}(\bx') + 
\epsilon \delta_{\alpha \beta} \delta(\bx - \bx') \ur_{\gamma}(\bx')$, a change of 
variables which preserves the finiteness of correlation functions.
It follows that insertions of $-E_{\beta} u_{\gamma}$ in renormalized functions is 
finite, and does not require renormalization.
It is, in fact, a general property that operators of the form $E_{\phi} \phi$ are not 
renormalized~\cite{brown_aop_1980}. 
The second term in $\pa_{\alpha}J_{\gamma, \alpha \beta}$, $\pa_{\alpha}u_{\gamma} 
\sigma_{\alpha \beta}$, requires subtractions but, being invariant under shifts of the 
$u_{\alpha}$ field, it can only mix with composite fields which are symmetric under both 
$\bh \to \bh + \mathbf{B}$ and $u_{\alpha} \to u_{\alpha} + B_{\alpha}$.

We can thus conclude that the UV-divergent part $J^{\rm div}_{\gamma, \alpha \beta}$ of 
$J_{\gamma, \alpha \beta}$ must have the property that $\pa_{\alpha}J^{\rm div}_{\gamma, 
\alpha \beta}$ is invariant under shifts of all fields.
This, however, implies in turn that $J^{\rm div}_{\gamma, \alpha \beta}$ must be 
shift-invariant by itself.
To derive this result, let us denote as $\epsilon J^{\rm div}_{\mu, \gamma, \alpha 
\beta}$ the variation of $J^{\rm div}_{\gamma, \alpha \beta}$ under an infinitesimal 
uniform translation $u_{\alpha} \to u_{\alpha} + \epsilon \delta_{\alpha \mu}$.
By power counting it must be a field of dimension 2 and, therefore, must have the form
\begin{equation}\label{shift-variation}
J^{\rm div}_{\mu, \gamma, \alpha \beta} = a^{(1)}_{\rho \sigma, \mu, \gamma, \alpha 
\beta} M^{\ve} u_{\rho} u_{\sigma} + a^{(2)}_{\rho, \sigma, \mu, \gamma, \alpha \beta} 
\pa_{\rho}u_{\sigma} + \frac{1}{2} a^{(3)}_{\rho \sigma, \mu, \gamma, \alpha \beta} 
(\pa_{\rho} \bh\cdot \pa_{\sigma} \bh)~,
\end{equation}
where $a^{(1)}$, $a^{(2)}$, and $a^{(3)}$ are invariant tensors (linear combinations of 
products of Kronecker symbols).
At the same time, by the arguments above, it must satisfy the equation $\pa_{\alpha} 
J^{\rm div}_{\mu, \gamma, \alpha \beta} = 0$ identically.
It can be checked that the only possibility is $J^{\rm div}_{\mu, \gamma, \alpha \beta} = 
0$, which implies that $J^{\rm div}_{\gamma, \alpha \beta}$ is invariant under shifts.

\begin{figure}[h]
\centering
\includegraphics[scale=1]{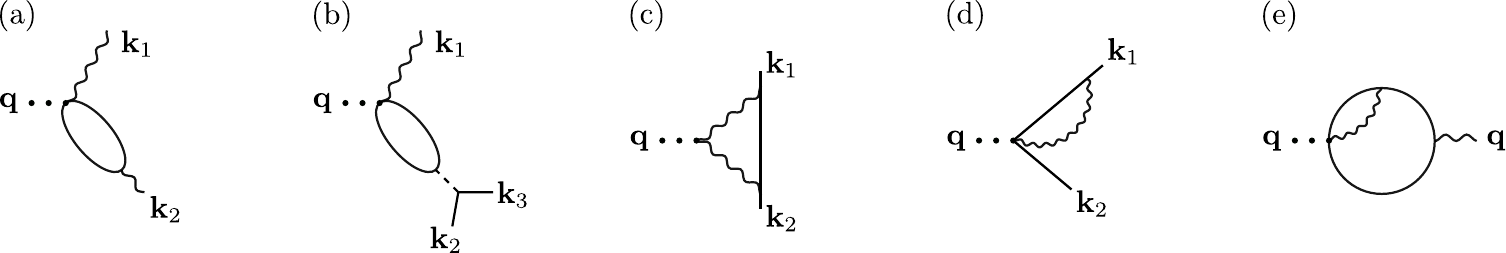}
\caption{\label{f:J-renormalization} Some of the first divergent 1PI diagrams with 
insertion of $J_{\gamma, \alpha \beta}$.
Dotted lines denote the operator insertion.
In diagrams (a) and (b), the undifferentiated field $u_{\gamma}$ is connected directly to 
external lines.
In diagrams (c), (d), and (e), instead, it enters as a loop line.}
\end{figure}

The conclusion of this argument is that any counterterm entering the renormalization of 
$J_{\gamma, \alpha \beta}$ must be a tensor of dimension 3 invariant under translations of 
both the $\bh$ and the $u_{\alpha}$ fields.
These tensors have the schematic form $\pa \pa u$ and $\pa \bh \cdot \pa \pa \bh$ and, 
since
\begin{equation}\label{total-derivative}
\pa_{\mu} \pa_{\nu} \bh \cdot \pa_{\rho} \bh = \frac{1}{2} \lt[\pa_{\nu} (\pa_{\mu}\bh 
\cdot \pa_{\rho} \bh) + \pa_{\mu} (\pa_{\nu}\bh\cdot \pa_{\rho} \bh) 
- \pa_{\rho}(\pa_{\mu}\bh \cdot \pa_{\nu}\bh)\rt]~,
\end{equation}
they can always be represented as total derivatives.
Therefore, general counterterms needed for the renormalization of $J_{\gamma, \alpha 
\beta}$ have the form
\begin{equation}
\frac{1}{2} B_{\rho, \mu \nu, \gamma, \alpha \beta} \pa_{\rho} (\pa_{\mu} \bh \cdot 
\pa_{\nu}\bh) + C_{\rho \sigma, \mu, \gamma, \alpha \beta} \pa_{\rho} \pa_{\sigma} 
u_{\mu}~,
\end{equation}
where $B_{\rho, \mu \nu, \gamma, \alpha \beta}$ and $C_{\rho \sigma, \mu, \gamma, \alpha 
\beta}$ are invariant tensors with divergent coefficients.
The renormalization of $J_{\gamma, \alpha \beta}$ in minimal subtraction can thus be 
written in the form
\begin{equation} \label{J-renormalization}
[J_{\gamma, \alpha \beta}] = J_{\gamma, \alpha \beta} + \frac{1}{2} B_{\rho, \mu \nu, 
\gamma, \alpha \beta} \pa_{\rho} (\pa_{\mu}\bh \cdot \pa_{\nu}\bh) + C_{\rho \sigma, \mu, 
\gamma, \alpha \beta} \pa_{\rho} \pa_{\sigma} u_{\mu}~.
\end{equation}
The final result for the renormalization of $J_{\gamma, \alpha \beta}$ has the following 
diagrammatic interpretation.
Among 1PI correlation functions with insertion of $J_{\gamma, \alpha \beta}=u_{\gamma} 
\sigma_{\alpha \beta}$, there are two types of divergent Feynman diagrams: the 
undifferentiated $u_{\gamma}$ field can be either connected to external legs or joined to 
loop lines (see Fig.~\ref{f:J-renormalization}).
In all diagrams of the second type, like (c), (d), and (e) of 
Fig.~\ref{f:J-renormalization}, it is possible to factorize one power of the momentum of 
each external solid and wiggly line, as it follows directly from the structure of the 
interaction vertices.
The corresponding divergences contribute to shift-invariant counterterms of the type $\pa 
\pa u$ and $\pa (\pa\bh \cdot\pa \bh)$ in Eq.~\eqref{J-renormalization}, but cannot 
generate renormalizations proportional to $J_{\gamma, \alpha \beta}$.

Counterterms of the same form of $J_{\gamma, \alpha \beta}$ can only arise from diagrams 
of the first type, like (a) and (b) in Fig.~\ref{f:J-renormalization}, which contribute 
to correlations which are not shift-invariant.
Since the undifferentiated $u_{\gamma}$ field is contracted with external lines, 
the loop part in this class of diagrams is entirely determined by the insertion of 
$\sigma_{\alpha \beta}$, whose renormalization was studied 
in appendix~\hyperref[appA]{A}.
The arguments above show that the UV divergences of $\lambda_{0}$ and $\mu_{0}$ are 
precisely cancelled to all orders by these loop contributions, so that $J_{\gamma, \alpha 
\beta}$ is finite (up to counterterms introduced in Eq.~\eqref{J-renormalization}).

Taking two independent traces over the components of $J_{\gamma, 
\alpha \beta}$ we finally obtain relations for the renormalization of the vector 
fields $u_{\alpha} u_{\beta \beta}$ and $u_{\beta} u_{\alpha \beta}$.
With a non-minimal renormalization choice, we can set
\begin{equation}
u_{\alpha} u_{\beta \beta} = \frac{M^{\ve}(D\lambdar + 2 \mur)}{D \lambda_{0} + 
2\mu_{0}}[u_{\alpha} u_{\beta \beta}] + b_{1}\pa_{\alpha} [\pa_{\beta}\bh\cdot 
\pa_{\beta}\bh] + b_{2} \pa_{\beta}[\pa_{\alpha}\bh \cdot \pa_{\beta}\bh] + b_{3} \pa^{2} 
[u_{\alpha}] + b_{4} \pa_{\alpha}\pa_{\beta}[u_{\beta}]~,
\end{equation}
\begin{equation}
\begin{split}
u_{\beta} u_{\alpha \beta} - \frac{1}{D} u_{\alpha} u_{\beta \beta} & = \frac{M^{\ve} 
\mur}{\mu_{0}} \lt\{[u_{\beta} u_{\alpha \beta}] - \frac{1}{D}[u_{\alpha} u_{\beta 
\beta}]\rt\}\\
& + b'_{1}\pa_{\alpha} [\pa_{\beta}\bh\cdot \pa_{\beta}\bh] + b'_{2} 
\pa_{\beta}[\pa_{\alpha}\bh \cdot \pa_{\beta}\bh] + b'_{3} \pa^{2} [u_{\alpha}] + b'_{4} 
\pa_{\alpha}\pa_{\beta}[u_{\beta}]~.
\end{split}
\end{equation}

\section{Energy-momentum tensor and operator renormalization in the GCI model}
\label{appC}

Starting from the explicit expression of its Hamiltonian, Eq.~\eqref{gci}, it can be 
shown that the GCI model admits the following symmetric energy-momentum tensor
\begin{equation}\label{energy-momentum-gci}
\begin{split}
T_{\alpha \beta} & = -\frac{1}{2}\delta_{\alpha \beta} (\pa^{2}\bh)^{2}
+ 2 \pa_{\alpha}\pa_{\beta}\bh \cdot \pa^{2}\bh - \pa_{\alpha}\bh \cdot 
\pa_{\beta}\pa^{2}\bh - \pa_{\beta}\bh \cdot \pa_{\alpha}\pa^{2}\bh\\
& +\frac{1}{D-1}\big\{\delta_{\alpha \beta} \pa_{\gamma}\bh \cdot 
\pa_{\gamma}\pa^{2}\bh + \delta_{\alpha \beta} \pa_{\gamma} \pa_{\delta} \bh \cdot 
\pa_{\gamma}\pa_{\delta}\bh + (D-2)\pa_{\gamma}\bh \cdot 
\pa_{\alpha}\pa_{\beta}\pa_{\gamma}\bh - D \pa_{\alpha}\pa_{\gamma}\bh \cdot 
\pa_{\beta}\pa_{\gamma}\bh\big\}\\
& -\frac{1}{2 Y_{0}}\delta_{\alpha \beta} (\pa^{2}\chi)^{2} + \frac{1}{Y_{0}} \big\{
2 \pa_{\alpha}\pa_{\beta}\chi\pa^{2}\chi - \pa_{\alpha}\chi \pa_{\beta}\pa^{2}\chi - 
\pa_{\beta}\chi \pa_{\alpha}\pa^{2}\chi\big\} \\
&+\frac{1}{(D-1)Y_{0}}\big\{\delta_{\alpha \beta} \pa_{\gamma}\chi 
\pa_{\gamma}\pa^{2}\chi + \delta_{\alpha \beta} \pa_{\gamma} \pa_{\delta} \chi 
\pa_{\gamma}\pa_{\delta}\chi + (D-2)\pa_{\gamma}\chi 
\pa_{\alpha}\pa_{\beta}\pa_{\gamma}\chi - D \pa_{\alpha}\pa_{\gamma}\chi 
\pa_{\beta}\pa_{\gamma}\chi\big\}\\
& -\frac{i}{2} \delta_{\alpha \beta} \big\{(\pa_{\gamma}\bh \cdot \pa_{\delta}\bh) 
\pa_{\gamma} \pa_{\delta} \chi  - (\pa_{\gamma}\bh \cdot \pa_{\gamma}\bh)  
\pa^{2}\chi\big\}   + i \big\{(\pa_{\beta}\bh \cdot \pa_{\gamma}\bh) \pa_{\alpha} 
\pa_{\gamma} \chi  + (\pa_{\alpha}\bh \cdot \pa_{\gamma}\bh)\pa_{\beta}\pa_{\gamma}\chi  
\\
&- (\pa_{\alpha}\bh \cdot \pa_{\beta}\bh) \pa^{2}\chi - (\pa_{\gamma}\bh \cdot 
\pa_{\gamma}\bh) \pa_{\alpha}\pa_{\beta}\chi \big\} + \frac{i}{2} \pa_{\gamma} 
\big\{(\pa_{\alpha}\bh \cdot \pa_{\beta}\bh) \pa_{\gamma}\chi - (\pa_{\alpha}\bh \cdot 
\pa_{\gamma}\bh) \pa_{\beta}\chi \\ &- (\pa_{\beta}\bh \cdot \pa_{\gamma}\bh) 
\pa_{\alpha}\chi + (\delta_{\alpha \gamma}\pa_{\beta}\chi + \delta_{\beta 
\gamma}\pa_{\alpha}\chi - 
\delta_{\alpha \beta} \pa_{\gamma}\chi) (\pa_{\delta} \bh \cdot \pa_{\delta}\bh) \big\}~,
\end{split}
\end{equation}
The identity for the trace, Eq.~\eqref{gci-T-renormalized}, can be derived from 
Eq.~\eqref{energy-momentum-gci} by some algebraic steps and by the following results for 
the renormalization of the invariant operators $(\pa^{2}\bh)^{2}$ and 
$(\pa^{2}\chi)^{2}/Y_{0}$.

Within dimensional regularization, symmetries and power counting imply that the set of 
composite operators
\begin{equation} \label{invariant-operators-gci}
\begin{split}
\Op_{1} &= \frac{1}{2} (\pa^{2}\bh)^{2} - \frac{1}{Y_{0}} (\pa^{2}\chi)^{2}~, \qquad 
\Op_{2} = \frac{1}{2Y_{0}} (\pa^{2}\chi)^{2}~, \\
\Op_{3} & = \frac{1}{2Y_{0}}((\pa^{2}\chi)^{2} - (\pa_{\alpha}\pa_{\beta}\chi 
\pa_{\alpha}\pa_{\beta}\chi)) = \frac{1}{2Y_{0}} \pa_{\alpha}\pa_{\beta} 
(\pa_{\alpha}\chi \pa_{\beta}\chi - \delta_{\alpha \beta}\pa_{\gamma}\chi 
\pa_{\gamma}\chi)~,\\
\Op_{4} & = \frac{1}{Y_{0}}\pa^{2}\pa^{2}\chi~, \qquad \Op_{5} = E = \frac{1}{Y_{0}} 
\pa^{2}\pa^{2}\chi + \frac{i}{2} ((\pa^{2}\bh)^{2} - (\pa_{\alpha}\pa_{\beta}\bh \cdot 
\pa_{\alpha}\pa_{\beta}\bh))
\end{split}
\end{equation}
is closed under renormalization.
The set~\eqref{invariant-operators-gci}, in fact, is a complete basis for all 
composite fields which are invariant under the symmetries of the GCI Hamiltonian 
(including the shifts $\bh \to \mathbf{A} + \mathbf{B}_{\alpha}x_{\alpha}$, $\chi \to 
\chi + A + B_{\alpha} x_{\alpha}$) and which have operator dimension 4 in the 
$\ve$-expansion.
A residual mixing with the softer field $\Op_{6} = \pa^{2}\chi$, which has dimension 2, 
is removed by dimensional regularization.

The operator $\Op_{4}$ is directly related to the elementary field $\chi$ and, therefore, 
has a simple multiplicative renormalization $\Op_{4} = Z Z^{-1}_{Y}[\Op_{4}]$.
Similarly $\Op_{5}$, which is equal to the equation of motion of the $\chi$ field, 
renormalizes in a multiplicative way as $\Op_{5} = Z [\Op_{5}]$.
We also note that the last three operators in the set~\eqref{invariant-operators-gci}, 
$\Op_{3}$, $\Op_{4}$, and $\Op_{5}$, are expressible as exact second derivatives of 
lower-dimensional fields.
In particular, this implies that $[\Op_{3}]$ is a linear combination of $\Op_{3}$, 
$\Op_{4}$, and $\Op_{5}$, without a mixing with $\Op_{1}$ and $\Op_{2}$. 

To study the renormalization of $\Op_{1}$ we note that, when integrated over all space, 
it is equivalent to the variation of the Hamiltonian under the infinitesimal rescaling 
$\bh \to (1 + \epsilon/2) \bh$, $\chi \to (1 - \epsilon) \chi$.
Therefore insertions of $\Op_{1}$ at zero momentum have the only effect to generate a 
factor $(n/2 - \ell)$ in front of correlation functions, where $n$ is the number of 
external $\bh$ fields and $\ell$ is the number of $\chi$ lines.
An immediate consequence is that $\Op_{1}$ is finite up to total-derivative operators 
which vanish at zero momentum.
Since the only total-derivative fields which can enter as counterterms are $\Op_{3}$, 
$\Op_{4}$, and $\Op_{5}$, we conclude that $\Op_{1}$ can be renormalized to 
all orders as $\Op_{1} = [\Op_{1}] + a_{1} [\Op_{3}] + b_{1} [\Op_{4}] + c_{1} 
[\Op_{5}]$, where $a_{1}$, $b_{1}$, and $c_{1}$ are divergent coefficients.

The renormalization of $\Op_{2}$ is constrained by the fact that the derivative of 
renormalized correlation functions with respect to $Y$ is finite.
Since, by Eq.~\eqref{gci-renormalization}, bare and renormalized correlation functions 
with $n$ external $\bh$ lines and $\ell$ external $\chi$ lines are related as 
$G^{(n, \ell)} = Z^{n/2 - \ell}\tilde{G}^{(n, \ell)}$, we obtain that
\begin{equation}
-\lt(\frac{n}{2} -\ell\rt) \frac{\pa \ln Z}{\pa Y}\Big|_{M} \tilde{G}^{(n, \ell)} +  
Z^{\ell - n/2} \frac{\pa \ln Y_{0}}{\pa Y}\Big|_{M}  \frac{\pa G^{(n, \ell)}}{\pa \ln 
Y_{0}}= \text{finite}~.
\end{equation}
The action of $\pa/\pa \ln Y_{0}$ on bare correlation functions generates insertion of 
$\Op_{2}$ at zero momentum.
The factor $(n/2 + \ell)$, moreover, can be represented via the zero-momentum insertion 
of $\Op_{1}$.

Using the relations
\begin{equation}
\frac{\pa \ln Z}{\pa Y}\Big|_{M} = \frac{\eta(Y)}{\beta(Y)}~, \qquad \frac{\pa \ln 
Y_{0}}{\pa Y}\Big|_{M} = - \frac{\ve}{\beta(Y)}~,
\end{equation}
we obtain that
\begin{equation}
\frac{\eta(Y)}{\beta(Y)}\Op_{1} + \frac{\ve}{\beta(Y)} \Op_{2} 
\end{equation}
is finite up to total derivatives.
It follows that the renormalization of $\Op_{2}$ has the form (in the minimal subtraction 
scheme)
\begin{equation}
\Op_{2} = -\frac{1}{\ve} \eta(Y) [\Op_{1}] -\frac{\beta(Y)}{\ve Y} [\Op_{2}] + a_{2} 
[\Op_{3}] + b_{2} [\Op_{4}] + b_{3} [\Op_{5}]~, 
\end{equation}
where $a_{2}$, $b_{2}$, and $c_{2}$ are new divergent coefficients.
Since $[\Op_{2}] = [(\pa^{2} \chi)^{2}]/(2 M^{\ve} Y)$ in minimal subtraction, we can 
rewrite this renormalization relation as
\begin{equation}
\frac{\ve}{2 Y_{0}} (\pa^{2}\chi)^{2} = - \eta(Y) [\Op_{1}] -
\frac{\beta(Y)}{2Y^{2}} M^{-\ve}[(\pa^{2} \chi)^{2}] + \ve a_{2} [\Op_{3}] + \ve b_{2} 
[\Op_{4}] + \ve b_{3} [\Op_{5}]~.
\end{equation}
As a further consequence, we note that
\begin{equation}
\frac{\pa \tilde{G}^{(n, \ell)}}{\pa \ln Y}\Big|_{M} = \frac{1}{2 M^{\ve} Y} \int {\rm 
d}^{D}x ~ \tilde{G}^{(n, \ell)}_{[(\pa^{2}\chi(\bx))^{2}]}~.
\end{equation}
This relation can be used to prove the equivalence between dilatation Ward identities 
and the RG equations~\eqref{RG-gci}, in analogy with Ref.~\cite{brown_aop_1980} and 
appendix~\hyperref[appA]{A}.


\begin{thebibliography}{10}
\expandafter\ifx\csname url\endcsname\relax
  \def\url#1{\texttt{#1}}\fi
\expandafter\ifx\csname urlprefix\endcsname\relax\def\urlprefix{URL }\fi
\expandafter\ifx\csname href\endcsname\relax
  \def\href#1#2{#2} \def\path#1{#1}\fi

\bibitem{poland_rmp_2019}
S.~R. D.~Poland, A.~Vichi, The conformal bootstrap: theory, numerical
  techniques, and applications, Rev. Mod. Phys. 91 (2019) 015002.
\newblock \href {https://doi.org/10.1103/RevModPhys.91.015002}
  {\path{doi:10.1103/RevModPhys.91.015002}}.

\bibitem{rychkov_epfl_2017}
V.~Rychkov, {EPFL} lectures on conformal field theory in $D \geq 3$ dimensions,
  {Springer Briefs} in {Physics}, Springer, 2017.
\newblock \href {https://doi.org/10.1007/978-3-319-43626-5}
  {\path{doi:10.1007/978-3-319-43626-5}}.

\bibitem{di-francesco_cft}
P.~Di~Francesco, P.~Mathieu, D.~S\'{e}n\'{e}chal, Conformal field theory,
  Springer, 1997.
\newblock \href {https://doi.org/https://doi.org/10.1007/978-1-4612-2256-9}
  {\path{doi:https://doi.org/10.1007/978-1-4612-2256-9}}.

\bibitem{nakayama_pr_2015}
Y.~Nakayama, Scale invariance vs. conformal invariance, Phys. Rep. 569 (2015).
\newblock \href {https://doi.org/10.1016/j.physrep.2014.12.003}
  {\path{doi:10.1016/j.physrep.2014.12.003}}.

\bibitem{zamolodchikov_jetp_1986}
A.~B. Zamolodchikov, {"Irreversibility"} of the flux of the renormalization
  group in a {2D} field theory, JETP Lett. 43 (1986) 730.

\bibitem{polchinski_npb_1988}
J.~Polchinski, Scale and conformal invariance in quantum field theory, Nucl.
  Phys. B 303 (1988) 226.
\newblock \href {https://doi.org/10.1016/0550-3213(88)90179-4}
  {\path{doi:10.1016/0550-3213(88)90179-4}}.

\bibitem{cardy_conformal}
J.~Cardy, \href{http://arxiv.org/abs/0807.3472}{Conformal field theory and
  statistical mechanics}, arXiv:0807.3472 (2008).
\newline\urlprefix\url{http://arxiv.org/abs/0807.3472}

\bibitem{jack_npb_1990}
I.~Jack, H.~Osborn, Analogs of the $c$-theorem for four-dimensional
  renormalisable field theories, Nucl. Phys. B 343 (1990) 647.
\newblock \href {https://doi.org/10.1016/0550-3213(90)90584-Z}
  {\path{doi:10.1016/0550-3213(90)90584-Z}}.

\bibitem{luty_jhep_2013}
M.~A. Luty, J.~Polchinski, R.~Rattazzi, The $a$-theorem and the asymptotics of
  4{D} quantum field theory, J. High Energy Phys. 01 (2013) 152.
\newblock \href {https://doi.org/10.1007/JHEP01(2013)152}
  {\path{doi:10.1007/JHEP01(2013)152}}.

\bibitem{fortin_jhep_2013}
J.-F. Fortin, B.~Grinstein, A.~Stergiou, Limit cycles and conformal invariance,
  J. High Energy Phys. 01 (2013) 184.
\newblock \href {https://doi.org/10.1007/JHEP01(2013)184}
  {\path{doi:10.1007/JHEP01(2013)184}}.

\bibitem{dymarsky_jhep_2015}
A.~Dymarsky, Z.~Komargodski, A.~Schwimmer, S.~Theisen, On scale and conformal
  invariance in four dimensions, J. High Energy Phys. 10~(171) (2015) 171.
\newblock \href {https://doi.org/10.1007/JHEP10(2015)171}
  {\path{doi:10.1007/JHEP10(2015)171}}.

\bibitem{dymarsky_jpa_2015}
A.~Dymarsky, A.~Zhiboedov, Scale-invariant breaking of conformal symmetry, J.
  Phys. A: Math. Theor. 48 (2015) 41FT01.
\newblock \href {https://doi.org/10.1088/1751-8113/48/41/41FT01}
  {\path{doi:10.1088/1751-8113/48/41/41FT01}}.

\bibitem{dymarsky_jhep_2016}
A.~Dymarsky, K.~Farnsworth, Z.~Komargodski, M.~A. Luty, V.~Prilepina, Scale
  invariance, conformality, and generalized free fields, J. High Energy Phys.
  02 (2016) 099.
\newblock \href {https://doi.org/10.1007/JHEP02(2016)099}
  {\path{doi:10.1007/JHEP02(2016)099}}.

\bibitem{nakayama_prd_2020}
Y.~Nakayama, Conformal invariance from scale invariance in nonlinear sigma
  models, Phys. Rev. D 102 (2020) 065018.
\newblock \href {https://doi.org/10.1103/PhysRevD.102.065018}
  {\path{doi:10.1103/PhysRevD.102.065018}}.

\bibitem{hull_npb_1986}
C.~M. Hull, P.~K. Townsend, Finiteness and conformal invariance in non-linear
  sigma models, Nucl. Phys. B 274 (1986) 349.
\newblock \href {https://doi.org/10.1016/0550-3213(86)90289-0}
  {\path{doi:10.1016/0550-3213(86)90289-0}}.

\bibitem{arutyunov_npb_2016}
G.~Arutyunov, S.~Frolov, B.~Hoare, R.~Roiban, A.~A. Tseytlin, Scale invariance
  of the $\eta$-deformed {AdS}$_{5} \times$ {S}$^{5}$ superstring, {T}-duality
  and modified type {II} equations, Nucl. Phys. B 903 (2016) 262.
\newblock \href {https://doi.org/10.1016/j.nuclphysb.2015.12.012}
  {\path{doi:10.1016/j.nuclphysb.2015.12.012}}.

\bibitem{riva_plb_2005}
V.~Riva, J.~Cardy, Scale and conformal invariance in field theory: a physical
  counterexample, Phys. Lett. B 622 (2005) 339.
\newblock \href {https://doi.org/10.1016/j.physletb.2005.07.010}
  {\path{doi:10.1016/j.physletb.2005.07.010}}.

\bibitem{ho_jhep_2008}
C.~M. Ho, Y.~Nakayama, Dangerous {Liouville} wave - exactly marginal but
  non-conformal deformation, J. High Energy Phys. 07 (2008) 109.
\newblock \href {https://doi.org/10.1088/1126-6708/2008/07/109}
  {\path{doi:10.1088/1126-6708/2008/07/109}}.

\bibitem{el-showk_npb_2011}
S.~El-Showk, Y.~Nakayama, S.~Rychkov, What {Maxwell} theory in $d \neq $ 4
  teaches us about scale and conformal invariance, Nucl. Phys. B 848 (2011)
  578.
\newblock \href {https://doi.org/10.1016/j.nuclphysb.2011.03.008}
  {\path{doi:10.1016/j.nuclphysb.2011.03.008}}.

\bibitem{nakayama_prd_2013}
Y.~Nakayama, Supercurrent, supervirial, and superimprovement, Phys. Rev. D 87
  (2013) 085005.
\newblock \href {https://doi.org/10.1103/PhysRevD.87.085005}
  {\path{doi:10.1103/PhysRevD.87.085005}}.

\bibitem{nakayama_prd_2017}
Y.~Nakayama, Interacting scale invariant but nonconformal field theories, Phys.
  Rev. D 95 (2017) 065016.
\newblock \href {https://doi.org/10.1103/PhysRevD.95.065016}
  {\path{doi:10.1103/PhysRevD.95.065016}}.

\bibitem{oz_epjc_2018}
Y.~Oz, On scale versus conformal symmetry in turbulence, Eur. Phys. K. C 78
  (2018).
\newblock \href {https://doi.org/10.1140/epjc/s10052-018-6147-8}
  {\path{doi:10.1140/epjc/s10052-018-6147-8}}.

\bibitem{nakayama_prd_2017b}
Y.~Nakayama, Euclidean {M}-theory background dual to a three-dimensional
  scale-invariant field theory without conformal invariance, Phys. Rev. D 95
  (2017) 046006.
\newblock \href {https://doi.org/10.1103/PhysRevD.95.046006}
  {\path{doi:10.1103/PhysRevD.95.046006}}.

\bibitem{nakayama_prd_2017c}
Y.~Nakayama, Topologically twisted renormalization group flow and its
  holographic dual, Phys. Rev. D 95 (2017) 066010.
\newblock \href {https://doi.org/10.1103/PhysRevD.95.066010}
  {\path{doi:10.1103/PhysRevD.95.066010}}.

\bibitem{li_epjc_2019}
Y.-Z. Li, H.~L\"{u}, H.-Y. Zhang, Scale invariance vs. conformal invariance:
  holographic two-point functions in {Horndeski} gravity, Eur. Phys. J. C 79
  (2019) 592.
\newblock \href {https://doi.org/10.1140/epjc/s10052-019-7096-6}
  {\path{doi:10.1140/epjc/s10052-019-7096-6}}.

\bibitem{schafer_jpa_1976}
L.~Sch\"{a}fer, Conformal covariance in the framework of {Wilson's}
  renormalization group approach, J. Phys. A: Math. Gen. 9 (1976) 377.
\newblock \href {https://doi.org/10.1088/0305-4470/9/3/008}
  {\path{doi:10.1088/0305-4470/9/3/008}}.

\bibitem{brown_aop_1980}
L.~S. Brown, Dimensional regularization of composite operators in scalar field
  theory, Ann. Phys. 126 (1980) 135.
\newblock \href {https://doi.org/10.1016/0003-4916(80)90377-2}
  {\path{doi:10.1016/0003-4916(80)90377-2}}.

\bibitem{paulos_npb_2016}
M.~F. Paulos, S.~Rychkov, B.~C. van Rees, B.~Zan, Conformal invariance in the
  long-range {Ising} model, Nucl. Phys. B 902 (2016) 246.
\newblock \href {https://doi.org/10.1016/j.nuclphysb.2015.10.018}
  {\path{doi:10.1016/j.nuclphysb.2015.10.018}}.

\bibitem{delamotte_pre_2016}
B.~Delamotte, M.~Tissier, N.~Wschebor, Scale invariance implies conformal
  invariance for the three-dimensional {Ising} model, Phys. Rev. E 93 (2016)
  012144.
\newblock \href {https://doi.org/10.1103/PhysRevE.93.012144}
  {\path{doi:10.1103/PhysRevE.93.012144}}.

\bibitem{de-polsi_jsp_2019}
G.~De~Polsi, M.~Tissier, N.~Wschebor, Conformal invariance and vector operators
  in the {$O(N)$} model, J. Stat. Phys. 177 (2019) 1089.
\newblock \href {https://doi.org/10.1007/s10955-019-02411-3}
  {\path{doi:10.1007/s10955-019-02411-3}}.

\bibitem{meneses_jhep_2019}
S.~Meneses, J.~Penedones, S.~Rychkov, J.~M. Viana Parente~Lopes, P.~Yvernay, A
  structural test for the conformal invariance of the critical 3d {Ising}
  model, J. High Energy Phys. 04 (2019) 115.
\newblock \href {https://doi.org/10.1007/JHEP04(2019)115}
  {\path{doi:10.1007/JHEP04(2019)115}}.

\bibitem{dietz_jhep_2013}
J.~A. Dietz, T.~R. Morris, Redundant operators in the exact renormalization
  group and in the {$f(R)$} approximation to asymptotic safety, J. High Energy
  Phys. 07 (2013) 64.
\newblock \href {https://doi.org/10.1007/JHEP07(2013)064}
  {\path{doi:10.1007/JHEP07(2013)064}}.

\bibitem{parisi_plb_1972}
G.~Parisi, Conformal invariance in perturbation theory, Phys. Lett. B 39 (1972)
  643.
\newblock \href {https://doi.org/10.1016/0370-2693(72)90020-2}
  {\path{doi:10.1016/0370-2693(72)90020-2}}.

\bibitem{pajer_jhep_2019}
E.~Pajer, D.~Stefanyszyn, Symmetric superfluids, J. High Energy Phys. 06 (2019)
  008.
\newblock \href {https://doi.org/10.1007/JHEP06(2019)008}
  {\path{doi:10.1007/JHEP06(2019)008}}.

\bibitem{nelson_statistical}
D.~R. Nelson, T.~Piran, S.~Weinberg (Eds.), Statistical mechanics of membranes
  and surfaces, 2nd Edition, World Scientific, Singapore, 2004.

\bibitem{bowick_pr_2001}
M.~J. Bowick, A.~Travesset, The statistical mechanics of membranes, Phys. Rep.
  344 (2001) 255.
\newblock \href {https://doi.org/10.1016/S0370-1573(00)00128-9}
  {\path{doi:10.1016/S0370-1573(00)00128-9}}.

\bibitem{katsnelson_graphene}
M.~I. Katsnelson, The physics of graphene, 2nd Edition, Cambridge University
  Press, Cambridge, 2020.

\bibitem{nelson_jpf_1987}
D.~R. Nelson, L.~Peliti, Fluctuations in membranes with crystalline and hexatic
  order, J. Physique 48~(7) (1987) 1085.
\newblock \href {https://doi.org/10.1051/jphys:019870048070108500}
  {\path{doi:10.1051/jphys:019870048070108500}}.

\bibitem{david_epl_1988}
F.~David, E.~Guitter, Crumpling transition in elastic membranes:
  renormalization group treatment, EPL 5 (1988) 709.
\newblock \href {https://doi.org/10.1209/0295-5075/5/8/008}
  {\path{doi:10.1209/0295-5075/5/8/008}}.

\bibitem{aronovitz_prl_1988}
J.~A. Aronovitz, T.~C. Lubensky, Fluctuations of solid membranes, Phys. Rev.
  Lett. 60 (1988) 2634.
\newblock \href {https://doi.org/10.1103/PhysRevLett.60.2634}
  {\path{doi:10.1103/PhysRevLett.60.2634}}.

\bibitem{aronovitz_jpf_1989}
J.~Aronovitz, L.~Golubovic, T.~C. Lubensky, Fluctuations and lower critical
  dimensions of crystalline membranes, J. Physique 50~(6) (1989) 609.
\newblock \href {https://doi.org/10.1051/jphys:01989005006060900}
  {\path{doi:10.1051/jphys:01989005006060900}}.

\bibitem{guitter_jpf_1989}
E.~Guitter, F.~David, S.~Leibler, L.~Peliti, Thermodynamical behavior of
  polymerized membranes, J. Physique 50 (1989) 1787.
\newblock \href {https://doi.org/10.1051/jphys:0198900500140178700}
  {\path{doi:10.1051/jphys:0198900500140178700}}.

\bibitem{kownacki_pre_2009}
J.-P. Kownacki, D.~Mouhanna, Crumpling transition and flat phase of polymerized
  phantom membranes, Phys. Rev. E 79 (2009) 040101(R).
\newblock \href {https://doi.org/10.1103/PhysRevE.79.040101}
  {\path{doi:10.1103/PhysRevE.79.040101}}.

\bibitem{gazit_pre_2009}
D.~Gazit, Structure of physical crystalline membranes within the
  self-consistent screening approximation, Phys. Rev. E 80 (2009) 041117.
\newblock \href {https://doi.org/10.1103/PhysRevE.80.041117}
  {\path{doi:10.1103/PhysRevE.80.041117}}.

\bibitem{bowick_prb_2017}
M.~J. Bowick, A.~Ko\ifmmode~\check{s}\else \v{s}\fi{}mrlj, D.~R. Nelson,
  R.~Sknepnek, Non-{Hookean} statistical mechanics of clamped graphene ribbons,
  Phys. Rev. B 95~(10) (2017) 104109.
\newblock \href {https://doi.org/10.1103/PhysRevB.95.104109}
  {\path{doi:10.1103/PhysRevB.95.104109}}.

\bibitem{le-doussal_aop_2018}
P.~Le~Doussal, L.~Radzihovsky, Anomalous elasticity, fluctuations and disorder
  in elastic membranes, Ann. Phys. 392 (2018) 340.
\newblock \href {https://doi.org/10.1016/j.aop.2017.08.033}
  {\path{doi:10.1016/j.aop.2017.08.033}}.

\bibitem{saykin_aop_2020}
D.~R. Saykin, I.~V. Gornyi, V.~Y. Kachorovskii, I.~S. Burmistrov, {Absolute
  Poisson's ratio and the bending rigidity exponent of a crystalline
  two-dimensional membrane}, Ann. Phys. 414 (2020) 168108.
\newblock \href {https://doi.org/10.1016/j.aop.2020.168108}
  {\path{doi:10.1016/j.aop.2020.168108}}.

\bibitem{mauri_npb_2020}
A.~Mauri, M.~I. Katsnelson, Scaling behavior of crystalline membranes: an
  $\epsilon$-expansion approach, Nucl. Phys. B 956 (2020) 115040.
\newblock \href {https://doi.org/10.1016/j.nuclphysb.2020.115040}
  {\path{doi:10.1016/j.nuclphysb.2020.115040}}.

\bibitem{coquand_pre_2020}
O.~Coquand, D.~Mouhanna, S.~Teber, Flat phase of polymerized membranes at
  two-loop order, Phys. Rev. E 101 (2020) 062104.
\newblock \href {https://doi.org/10.1103/PhysRevE.101.062104}
  {\path{doi:10.1103/PhysRevE.101.062104}}.

\bibitem{nakayama_aop_2016}
Y.~Nakayama, Hidden global conformal symmetry without {Virasoro} extension in
  theory of elasticity, Ann. Phys. 372 (2016) 392.
\newblock \href {https://doi.org/10.1016/j.aop.2016.06.010}
  {\path{doi:10.1016/j.aop.2016.06.010}}.

\bibitem{coquand_prb_2019}
O.~Coquand, Spontaneous symmetry breaking and the flat phase of crystalline
  membranes, Phys. Rev. B 100 (2019) 125406.
\newblock \href {https://doi.org/https://doi.org/10.1103/PhysRevB.100.125406}
  {\path{doi:https://doi.org/10.1103/PhysRevB.100.125406}}.

\bibitem{burmistrov_prb_2016}
I.~S. Burmistrov, I.~V. Gornyi, V.~Y. Kachorovskii, M.~I. Katsnelson, A.~D.
  Mirlin, Quantum elasticity of graphene: thermal expansion coefficient and
  specific heat, Phys. Rev. B 94 (2016) 195430.
\newblock \href {https://doi.org/10.1103/PhysRevB.94.195430}
  {\path{doi:10.1103/PhysRevB.94.195430}}.

\bibitem{burmistrov_aop_2018}
I.~S. Burmistrov, V.~Y. Kachorovskii, I.~V. Gornyi, A.~D. Mirlin, Differential
  {Poisson}'s ratio of a crystalline two-dimensional membrane, Ann. Phys. 396
  (2018) 119.
\newblock \href {https://doi.org/10.1016/j.aop.2018.07.009}
  {\path{doi:10.1016/j.aop.2018.07.009}}.

\bibitem{zinn-justin_qft}
J.~Zinn-Justin, Quantum field theory and critical phenomena, 4th Edition, Vol.
  113 of International series of monographs on physics, Oxford University
  Press, 2002.

\bibitem{parisi_sft}
G.~Parisi, Statistical field theory, Frontiers in Physics, Addison-Wesley,
  1988.

\bibitem{sun_pnas_2012}
K.~Sun, A.~Souslov, X.~Mao, T.~C. Lubensky, Surface phonons, elastic response,
  and conformal invariance in twisted kagome lattices, Proc. Natl. Acad. Sci.
  U.S.A 109 (2012) 12369.
\newblock \href {https://doi.org/10.1073/pnas.1119941109}
  {\path{doi:10.1073/pnas.1119941109}}.

\bibitem{xing_pre_2003}
X.~Xing, R.~Mukhopadhyay, T.~Lubensky, L.~Radzihovsky, Fluctuating nematic
  elastomer membranes, Phys. Rev. E 68 (2003) 021108.
\newblock \href {https://doi.org/10.1103/PhysRevE.68.021108}
  {\path{doi:10.1103/PhysRevE.68.021108}}.

\bibitem{arici_jmp_2018}
F.~Arici, D.~Becker, C.~Ripken, F.~Saueressig, W.~D. van Sluijlekom, Reflection
  positivity in higher derivative scalar theories, J. Math. Phys. 59 (2018)
  082302.
\newblock \href {https://doi.org/10.1063/1.5027231}
  {\path{doi:10.1063/1.5027231}}.

\bibitem{pisarski_prd_1983}
R.~D. Pisarski, Soluble theory with massive ghosts, Phys. Rev. D 28~(10) (1983)
  2547.
\newblock \href {https://doi.org/10.1103/PhysRevD.28.2547}
  {\path{doi:10.1103/PhysRevD.28.2547}}.

\bibitem{goon_jhep_2016}
G.~Goon, K.~Hinterbichler, A.~Joyce, M.~Trodden, Aspects of {Galileon}
  non-renormalization, J. High Energy Phys. 11 (2016) 100.
\newblock \href {https://doi.org/10.1007/JHEP11(2016)100}
  {\path{doi:10.1007/JHEP11(2016)100}}.

\bibitem{brust_jhep_2017}
C.~Brust, K.~Hinterbichler, Free {$\Box^{k}$} scalar conformal field theory, J.
  High Energy Phys. 02 (2017) 066.
\newblock \href {https://doi.org/http://dx.doi.org/10.1007/JHEP02(2017)066}
  {\path{doi:http://dx.doi.org/10.1007/JHEP02(2017)066}}.

\end{thebibliography}

\end{document}